\documentclass{aa}
\usepackage{natbib,graphicx,color,psfrag,amsmath}
\usepackage[varg]{txfonts}
\bibpunct{(}{)}{,}{a}{}{,}
\newcommand{\msun}{\ensuremath{M_{\odot}}}
\newcommand{\lsun}{\ensuremath{L_{\odot}}}
\newcommand{\rsun}{\ensuremath{R_{\odot}}}

\newcommand{\Teff}{\ensuremath{T_{\rm eff}}}
\newcommand{\vinf}{\ensuremath{v_{\infty}}}
\newcommand{\mdot}{\ensuremath{\dot{M}}}

\newcommand{\Mdu}{\ensuremath{\cdot 10^{-6}\,M_{\odot} {\rm yr}^{-1}}}
\newcommand{\mdu}{\ensuremath{10^{-6}\,M_{\odot} {\rm yr}^{-1}}}

\newcommand{\beq}{\begin{equation}}
\newcommand{\eeq}{\end{equation}}
\newcommand{\beqa}{\begin{eqnarray}}
\newcommand{\eeqa}{\end{eqnarray}}

\newcommand{\nbeq}{\begin{equation*}}
\newcommand{\neeq}{\end{equation*}}

\newcommand{\eu}{\ensuremath{{\rm e}}}

\newcommand{\kms}{\ensuremath{{\rm km}\,{\rm s}^{-1}}}
\newcommand{\seconds}{\ensuremath{{\rm s}}}

\newcommand{\dd}{{\rm d}}

\newcommand{\Rstar}{\ensuremath{R_{\ast}}}

\newcommand{\Mstar}{\ensuremath{M_{\ast}}}
\newcommand{\Lstar}{\ensuremath{L_{\ast}}}

\newcommand{\Trad}{\ensuremath{T_{\rm rad}}}

\newcommand{\vrot}{\ensuremath{v_{\rm rot}}}

\newcommand{\chibar}{\ensuremath{\bar{\chi}}}

\newcommand{\Jbar}{\ensuremath{\bar J}}

\newcommand\ie{\hbox{i.e.}}
\newcommand\eg{\hbox{e.g.}}

\definecolor{orange}{rgb}{1.,0.5,0.}
\newcommand{\vecown}[1]{\ensuremath{\vec{#1}}}
\newcommand{\matown}[1]{\ensuremath{\vec{#1}}}

\newcommand{\alo}{\ensuremath{\Lambda^{\rm (A)}}}
\newcommand{\alom}{\ensuremath{\matown{\Lambda}^\ast}}
\newcommand{\unitym}{\ensuremath{\matown{1}}}

\newcommand{\epsc}{\ensuremath{\epsilon_{\rm C}}}
\newcommand{\epsl}{\ensuremath{\epsilon_{\rm L}}}
\newcommand{\scont}{\ensuremath{S_{\rm C}}}
\newcommand{\sline}{\ensuremath{S_{\rm L}}}

\newcommand{\kcont}{\ensuremath{k_{\rm C}}}
\newcommand{\kline}{\ensuremath{k_{\rm L}}}

\newcommand{\chic}{\ensuremath{\chi_{\rm C}}}
\newcommand{\chil}{\ensuremath{\chi_{\rm L}}}
\newcommand{\chith}{\ensuremath{\chi_{\rm Th}}}

\newcommand{\profile}{\ensuremath{\Phi_x}}

\newcommand{\nx}{\ensuremath{N_{\rm x}}}
\newcommand{\ny}{\ensuremath{N_{\rm y}}}
\newcommand{\nz}{\ensuremath{N_{\rm z}}}

\newcommand{\nomega}{\ensuremath{N_{\rm \Omega}}}

\newcommand{\vmin}{\ensuremath{v_{\rm min}}}

\newcommand{\xobs}{\ensuremath{x_{\rm obs}}}
\newcommand{\xcmf}{\ensuremath{x_{\rm cmf}}}

\newcommand{\indx}[1]{\ensuremath{{#1}}}
\newcommand{\indxu}{\ensuremath{{\rm u}}}
\newcommand{\indxp}{\ensuremath{{\rm p}}}
\newcommand{\indxd}{\ensuremath{{\rm d}}}
\newcommand{\indxc}{\ensuremath{{\rm c}}}
\newcommand{\xii}{\ensuremath{x_{i}}}
\newcommand{\ximo}{\ensuremath{x_{i-1}}}
\newcommand{\xipo}{\ensuremath{x_{i+1}}}
\newcommand{\fii}{\ensuremath{f_{i}}}
\newcommand{\fimo}{\ensuremath{f_{i-1}}}
\newcommand{\fipo}{\ensuremath{f_{i+1}}}
\newcommand{\dxi}{\ensuremath{\Delta x_{i}}}
\newcommand{\dxipo}{\ensuremath{\Delta x_{i+1}}}

\newcommand{\Jnu}{\ensuremath{J_{\rm \nu}}}
\newcommand{\Bnu}{\ensuremath{B_{\rm \nu}}}

\newcommand{\vthfid}{\ensuremath{v_{\rm th}^{\rm \ast}}}
\newcommand{\ddopfid}{\ensuremath{\Delta \nu_{\rm D}^{\rm \ast}}}

\newcommand{\sigmae}{\ensuremath{\sigma_{\rm e}}}
\newcommand{\sigmab}{\ensuremath{\sigma_{\rm B}}}
\newcommand{\nel}{\ensuremath{n_{\rm e}}}

\newcommand{\Rpole}{\ensuremath{R_{\rm p}}}
\newcommand{\Req}{\ensuremath{R_{\rm eq}}}
\newcommand{\Teffeq}{\ensuremath{T_{\rm eff, eq}}}
\newcommand{\Teffpole}{\ensuremath{T_{\rm eff, p}}}

\newcommand{\clatitude}{\ensuremath{\Theta}}
\newcommand{\azimuth}{\ensuremath{\Phi}}

\newcommand{\Cul}{\ensuremath{C_{\rm ul}}}
\newcommand{\Aul}{\ensuremath{A_{\rm ul}}}

\defcitealias{Hennicker2018}{paper I}

\newcommand{\paramon}{\ensuremath{\omega}}

\newcommand{\mrerr}[1]{\mkern 1.5mu\overline{\mkern-1.5mu#1\mkern-1.5mu}\mkern 1.5mu}

\newcommand{\nconv}{\ensuremath{N_{\rm iter}}}

\begin{document}
\title{A 3D short-characteristics method for continuum and line scattering
  problems in the winds of hot stars}

\author{L. Hennicker\inst{1}, J. Puls\inst{1}, N. D. Kee\inst{2}, and J. O. Sundqvist\inst{2}
}

\institute{LMU M\"unchen, Universit\"atssternwarte, Scheinerstr. 1, 81679
M\"unchen, Germany, \email{levin@usm.uni-muenchen.de}
           \and
           Instituut voor Sterrenkunde, KU Leuven, Celestijnenlaan 200D,
           3001 Leuven, Belgium
}

\date{Received 27 August 2019; Accepted 8 October 2019}

\abstract 
{Knowledge about hot, massive stars is usually inferred from quantitative
  spectroscopy. To analyse non-spherical phenomena, the existing 1D codes must
  be extended to higher dimensions, and corresponding tools need to be
  developed.}
{We present a 3D radiative transfer code that is capable of calculating
  continuum and line scattering problems in the winds of hot stars. By
  considering spherically symmetric test models, we discuss potential error
  sources, and indicate advantages and disadvantages by comparing
  different solution methods. Further, we analyse the ultra-violet (UV)
  resonance line formation in the winds of rapidly rotating O stars.}
{We consider both a (simplified) continuum model including scattering and
  thermal sources, and a UV resonance line transition approximated by a
  two-level-atom. We applied the short-characteristics (SC) method, using linear
  or monotonic B\'ezier interpolations, for which monotonicity is of prime
  importance, to solve the equation of radiative transfer on a non-uniform
  Cartesian grid. To calculate scattering dominated problems, our solution
  method is supplemented by an accelerated $\Lambda$-iteration scheme using newly developed
  non-local operators.}
{For the spherical test models, the mean relative error of the source function
  is on the $5-20\,\%$ percent level, depending on the applied interpolation
  technique and the complexity of the considered model. All calculated line
  profiles are in excellent agreement with corresponding 1D solutions. Close
  to the stellar surface, the SC methods generally perform better than a 3D
  finite-volume-method; however, they display specific problems in searchlight-beam
  tests at larger distances from the star. The predicted line profiles from
  fast rotating stars show a distinct behaviour as a function of rotational
  speed and inclination. This behaviour is tightly coupled to the wind
  structure and the description of gravity darkening and stellar surface
  distortion.}
{Our SC methods are ready to be used for quantitative analyses of UV resonance
  line profiles. When calculating optically thick continua, both SC methods
  give reliable results, in contrast to the alternative finite-volume method.}
\keywords{Radiative transfer -- Methods: numerical -- Stars: massive -- Stars:
  rotation -- Stars: winds, outflows}
\titlerunning{3D short-characteristics method in the winds from OB stars}
\authorrunning{L. Hennicker et al.}
\maketitle
%
%
\section{Introduction} 
\label{sec:intro}
Understanding of hot, massive stars is a basic prerequisite for interpreting
fundamental properties of our Universe.  Already during their lifetimes, such
stars influence the evolution of galaxies via feedback of ionizing radiation
and radiatively driven winds, and they enrich the interstellar medium (ISM)
with metals. But also, their deaths are of large impact. Depending on initial
mass and mass loss, OB stars either explode as supernovae or end up as `heavy'
stellar-mass black holes \citep{Heger03}. Obviously, supernova explosions
enrich the ISM with metals even further. Additionally, the associated shock
fronts possibly trigger star formation, resulting in a new generation of
(massive) stars.

With the advent of gravitational wave (GW) observations (\eg~the black hole
merger GW150914 observed at the advanced Laser Interferometric
Gravitational-Wave Observatory (aLIGO), \citealt{Abbott16}), the formation of
heavy stellar-mass black holes becomes of key interest.  Since massive stars
are frequently found to be members of multiple star systems (see,
\eg~\citealt{Mason2009}, \citealt{Sana13}), they might explain the occurrence
of GW events in the correct mass range. At least the formation of heavy black
holes from single-star evolution, however, requires comparatively moderate
mass loss rates (\eg~in low metalicity environments, or mass-loss quenching by
magnetic fields, \citealt{Petit17}, \citealt{Keszthelyi2017}).

Current knowledge of OB stars is inferred from quantitative spectroscopy, that
is, by comparing observed spectra with synthetic ones, the latter being
obtained from numerically modelling their stellar atmospheres (photosphere +
wind). State of the art atmospheric modelling is performed by assuming
spherical symmetry (\eg~{\sc CMFGEN}: \citealt{hilliermiller98}; {\sc
  PHOENIX}: \citealt{Haus92}; {\sc PoWR}: \citealt{Graf02}; WM-{\sc basic}:
\citealt{pauldrach01}; {\sc FASTWIND}: \citealt{Puls05} and
\citealt{rivero12}).

Several effects may alter the geometry of hot star atmospheres, affecting both
the photospheric and wind lines.  For instance, binary interaction (see,
\eg~\citealt{Vanbeveren1991} and \citealt{deMink13} for a discussion about
evolutionary aspects, and \citealt{Prilutskii1976}, \citealt{Cherepa1976},
\citealt{Stevens92} for the effects of colliding winds) influences the line
formation in such objects. Another example is the phenomenon of magnetic winds
(\eg~\citealt{udDoula02}, \citealt{udDoula08}, \citealt{Petit13}), for which
the resonance line formation has been extensively discussed in
\cite{Marcolino13}, \cite{Hennicker2018}, and \cite{DavidUraz2019}, for
instance. Additionally, (non-radial) pulsations may impact the geometry of hot
star atmospheres.

In this paper, we focus on (rapidly) rotating stars. In addition to polar
velocity components that affect the radiative line-driving, such stars and
their winds are influenced by centrifugal forces and gravity darkening. For a
detailed discussion, we refer to Sect.~\ref{sec:rotation}.

To analyse these effects, and to distinguish between different theories
(resulting in, \eg~prolate vs. oblate wind structures), multi-D radiative
transfer codes that include a treatment of (arbitrary) velocity fields are
required. In hot star winds, a key challenge is the implementation of
scattering processes. Such problems are most easily handled by an accelerated
$\Lambda$-iteration scheme, that is, by calculating the radiation field for
known sources and sink terms (the so-called formal solution), and iterating
the updated sources and sinks until convergence \citep{Cannon73}.

In order to obtain the formal solution in multi-D, the following two major
methods exist, each having specific advantages and disadvantages.  Firstly,
within the finite-volume method (FVM, see \citealt{Adam90} for the first
application in the context of radiative transfer problems), the calculation
volume is discretized into finite sub-volumes. All required physical values at
the cell centres are then considered as suitable averages within each
cell. This way, a relatively simple solution scheme can be derived, which,
however, is only of low order. Thus, the FVM breaks down for large optical
depths (\citealt{Hennicker2018}, hereafter
\citetalias{Hennicker2018}). Nevertheless, it is still useful for qualitative
interpretations (see, \eg~\citealt{Lobel08} for an application to corotating
interaction regions in the winds of rotating stars).  Secondly, within the
characteristics methods, the formal solution is obtained by exact integration
of the equation of radiative transfer along a ray.  One distinguishes between
the `long-characteristics' method (LC, \citealt{Jones73a},
\citealt{Jones73b}), and the 'short-characteristics' method (SC,
\citealt{Kunasz88}). The LC method follows a particular ray from the boundary
to the considered grid point, whereas the SC method considers rays only from
cell to cell. All required quantities at each upwind point are obtained from
interpolation. While the LC method is computationally more expensive than the
SC method, the latter introduces numerical errors from the interpolation
scheme. To date, few 3D codes that apply one or the other technique already
exist, such as \textsc{Phoenix/3D} (\citealt{Haus06} and other publications in
this series) and \textsc{IRIS} \citep{Ibgui13}, which, however, are either
proprietary, or do not include a suitable $\Lambda$-iteration scheme. For
additional information on other codes, we refer to Sect.~1 of
\citetalias{Hennicker2018}.

In this paper, we present a newly developed 3D SC code, which is capable of
treating arbitrary velocity fields, and includes an accelerated
$\Lambda$-iteration scheme to handle continuum and line scattering problems.
In Sect.~\ref{sec:eqrt}, we discuss the basic assumptions of the code. The
numerical methods are described in Sect.~\ref{sec:numerical_methods}. In
Sect.~\ref{sec:err_analysis}, we perform a detailed error analysis by
comparing our 3D results for spherically symmetric atmospheres with
corresponding 1D solutions. Additionally, we present comparisons with the 3D
finite-volume method from \citetalias{Hennicker2018}, when appropriate. As a
first application to non-spherical winds, we calculated UV resonance line
profiles for different models of rotating stars in Sect.~\ref{sec:rotation},
and discuss the implications. Our conclusions are summarized in
Sect.~\ref{sec:conclusions}.
%
%
\section{Basic assumptions}
\label{sec:eqrt}
In this paper, we use the same basic assumptions as in
\citetalias{Hennicker2018}, that is we aim to solve the time-independent
equation of radiative transfer (\eg~\citealt{mihalasbook78}), formulated in
the observer's frame:
\beq
\label{eq:eqrt}
   \vecown{n} \vecown{\nabla} I_\nu(\vecown{r}, \vecown{n}) =
   \chi_\nu(\vecown{r}, \vecown{n}) \bigl(S_\nu(\vecown{r}, \vecown{n}) -
   I_\nu(\vecown{r}, \vecown{n}) \bigr) \,.
\eeq
$I_\nu$ describes the specific intensity for a given frequency $\nu$ and
direction $\vecown{n}$, and $\chi_\nu$ and $S_\nu$ are the opacities (in units
of ${\rm cm}^{-1}$) and source functions for continuum and line processes. For
our following tests, we either consider a pure continuum case with source
function
\beq \label{eq:scont}
\scont = (1-\epsc) \Jnu + \epsc \Bnu \,,
\eeq
and thermalization parameter $\epsc$, or a line (within an optically thin
continuum) approximated by a two-level-atom (TLA). A generalization to a
multitude of lines, coupled via corresponding rate equations, is one of our
next objectives. Within the TLA approach, the line source function reads:
\beqa \label{eq:sline}
\sline &=& (1-\epsl) \Jbar + \epsl B \\
\epsl &=& \dfrac{\epsilon'}{1+\epsilon'} , \quad
\epsilon'=\dfrac{\Cul}{\Aul}\Biggl[1-\exp\Bigl(-\frac{h \nu}{k_{\rm B} T} \Bigr)
  \Biggr] \,.
\eeqa
\Jbar~is the `scattering integral' (see Eq.~\eqref{eq:mintbar}), and \Cul~and
\Aul~describe the collisional rate and Einstein coefficient for spontaneous
emission, respectively. In this paper, $\epsl$ is considered as input
parameter. The profile function \profile ~is approximated by a Doppler
profile:
\beq
\label{eq:profile}
\profile = \dfrac{1}{\sqrt{\pi}\delta} \exp \Biggl[-\Bigl(\dfrac{\xcmf}{\delta} \Bigr)^2\Biggr] =
\dfrac{1}{\sqrt{\pi}\delta} \exp \Biggl[- \Bigl(\dfrac{\xobs - \vecown{n}
    \cdot \vecown{V}}{\delta} \Bigr)^2\Biggr] \,,
\eeq
where \xcmf~and \xobs~denote the frequency shift from line centre in the
comoving and observer's frame, in units of a fiducial Doppler width,
$\ddopfid=\nu_{\rm 0} \vthfid/c$.  $\vecown{V}$ is the local velocity vector
in units of \vthfid, and
\beq
\label{eq:vmicro}
\delta = \dfrac{1}{\vthfid} \sqrt{\dfrac{2 k_{\rm B} T}{m_{\rm A}} + v_{\rm
    micro}^2}
\eeq
is the ratio between the local thermal velocity (accounting for
micro-turbulent velocities) and the fiducial thermal velocity. This
description of the profile function allows for a depth-independent frequency
grid (see also \citetalias{Hennicker2018}).

Finally, we express the continuum and line opacities in terms of the
Thomson-opacity, $\chith = \nel \sigmae$, and depth-independent input
parameters, \kcont, \kline:
\beqa
\label{eq:opac}
\chic &=& \kcont \cdot \chith \\
\label{eq:opal}
\chil &=& \kline \cdot \chith \cdot \profile =: \bar{\chi}_{\rm L} \profile  \,,
\eeqa
where the line-strength \kline ~is related to the frequency integrated opacity
$\bar{\chi}_{\rm 0}=\ddopfid \bar{\chi}_{\rm L}=\kline\ddopfid\chith$
(\eg~\citetalias[Eqs.~(10) and (12)]{Hennicker2018}, and with
$[\bar{\chi}_{\rm 0}]={\rm (cm\,s)}^{-1}$).
In the following, we present (efficient) numerical tools for solving the
coupled equations \eqref{eq:eqrt} and \eqref{eq:scont}/\eqref{eq:sline}.
Since a direct solution is computationally prohibitive for 3D atmospheres due
to limited memory capacity, we apply an accelerated $\Lambda$-iteration (ALI)
scheme that overcomes the convergence problems of the classical
$\Lambda$-iteration for optically thick, scattering dominated atmospheres by
using an appropriate approximate $\Lambda$-operator (ALO).
%
%
\section{Numerical methods} 
\label{sec:numerical_methods}
The most time-consuming part of the ALI scheme consists of calculating the
formal solution. In \citetalias{Hennicker2018}, we have shown that a formal
solution obtained using the FVM suffers from various numerical inaccuracies
related to numerical diffusion and to the order of accuracy, the latter
influencing the solution particularly in the optically thick regime. To avoid
these errors, we implement an integral method along short
characteristics. When compared with a long-characteristics solution scheme,
the computation time becomes reduced by roughly a factor of $N/2$, with $N$
the number of spatial grid points (per dimension), since the LC method
integrates the equation of radiative transfer along the complete path from the
boundary to a considered grid point. Thus, LC methods become feasible only on
massively parallelized architectures.

We follow the same approach as in \citetalias{Hennicker2018}, and solve the
equation of radiative transfer on a non-uniform, 3-dimensional Cartesian
grid. In contrast to curvilinear coordinate systems, the direction vector
$\vecown{n}$ then becomes constant with respect to the spatial grid, thus
avoiding the otherwise required angular interpolation of upwind intensities
and a complicated bookkeeping of intensities (and corresponding integration
weights) for different directions. Furthermore, the sweep through the spatial
domain is considerably simplified, and can be performed grid point by grid
point along the $x$, $y$, and $z$ coordinates, since the intensities (required
for the upwind interpolation) are always known on the previous grid layer. To
enable a straightforward implementation of non-monotonic velocity fields, we
use the observer's frame formulation.

SC methods have been successfully implemented already for 3D non-LTE (NLTE)
radiative transfer problems in cool stars (\eg~\citealt{Vaeth1994},
\citealt{Leenaarts09}, \citealt{Hayek2010}, \citealt{Holzreuter2012}). These
codes, however, are mostly designed for planar geometries, and only account
for subsonic and slightly supersonic velocity fields. For scattering problems
including highly supersonic velocity fields, there exist, to our knowledge,
only the 2D codes by \cite{Dullemond2000} (planar/spherical),
\cite{vanNoort2002} (planar/spherical/cylindrical), \cite{Georgiev06} and
\cite{Zsargo2006} (spherical). The only 3D SC code including arbitrary
velocity fields, \textsc{IRIS} \citep{Ibgui13}, has also been formulated for
planar geometries, and lacks the implementation of a $\Lambda$-iteration
scheme thus far. As has been shown in all these studies, the final performance
of the SC method crucially depends on the choice of the applied interpolation
schemes.
%
%
\subsection{The discretized radiative transfer equation along a ray}
\label{subsec:sc}
The equation of radiative transfer along a given direction can be written as
\beq \label{eq:eqrtray}
\dfrac{\dd I}{\dd \tau} = S - I \,,
\eeq
where $\dd \tau :=\chi \dd s$ is the optical-depth increment along a ray
segment $\dd s$. Here and in the following, we suppress the notation for the
frequency dependence, and explicitly distinguish between continuum and line
only when appropriate.  Eq.~\eqref{eq:eqrtray} is integrated along a ray
propagating through a current grid point $\indxp$ with Cartesian grid indices
$(\indx{ijk})$ and corresponding upwind point $\indxu^{(\indx{ijk})}$. The
geometry for a 3D Cartesian grid is shown in the upper panel of
Fig.~\ref{fig:sc_cell_interp}. In the following, upwind and downwind
quantities corresponding to a considered grid point $(ijk)$ are indicated by
$q_{\indxu}^{(\indx{ijk})}$ and $q_{\indxd}^{(\indx{ijk})}$, while local
quantities are denoted either as $q_{\indxp}^{(\indx{ijk})}$ or simply
$q_{\indx{ijk}}$. For a given ray segment, we then obtain:
\beqa
\nonumber
I_{\indx{ijk}} &=& I_{\indxu}^{(\indx{ijk})} \eu^{-(\tau_{\indxp} - \tau_{\indxu})}
+\int_{\tau_{\indxu}}^{\tau_{\indxp}}\eu^{-(\tau_{\indxp} - \tau)} S(\tau) \dd
\tau \\ \label{eq:eqrt2}
 &=& I_{\indxu}^{(\indx{ijk})} \eu^{-\Delta\tau_{\indxu}} + \eu^{-\Delta\tau_{\indxu}}
\int_0^{\Delta\tau_{\indxu}}\eu^{t} S(t + \tau_{\indxu}) \dd t  \,,
\eeqa
with upwind optical-depth increment $\Delta
\tau_{\indxu}:=\tau_{\indxp}-\tau_{\indxu}\geq 0$, and
$t:=\tau-\tau_{\indxu}$. For the SC solution scheme, the location of the
reference point, $\tau=0$, plays no role, since only the optical-depth
increments, $\Delta \tau_{\indxu}$ and $\Delta \tau_{\indxd}$ (see below), are
required.
To calculate the source contribution, the source function is commonly
approximated by first- or second-order polynomials (\citealt{Kunasz88},
\citealt{vanNoort2002}), B\'ezier curves (\citealt{Hayek2010},
\citealt{Holzreuter2012}, \citealt{Auer2003}) or cubic Hermite splines
(\citealt{Ibgui13}).  While the 2nd- (and higher) order methods reproduce the
diffusion limit in optically thick media, they suffer from overshoots and need
to be monotonized with some effort to ensure that any interpolated quantity
remains positive between two given grid points. Monotonicity is usually
obtained by manipulating the interpolation scheme whenever overshoots
occur. Thus, the actual interpolation crucially depends on the specific
stratification of the considered quantity (\eg~the source function). The
$\Lambda$-operator then becomes non-linear, because its elements now
explicitly depend on the stratification of source functions (via corresponding
interpolation/integration coefficients). Within any $\Lambda$-iteration
scheme, this non-linearity can lead to oscillations.  In extreme cases,
`flip-flop situations' (\citealt[their Appendix A]{Holzreuter2012}) may occur,
which do not converge at all.

For the source contribution, we implement both a linear approximation as the
fastest and most stable method (monotonicity is always provided), and a
quadratic B\'ezier approximation (see Appendix \ref{app:bez}) for higher
accuracy\footnote{In this paper, different interpolation schemes are tested by
  considering simplified (though physically relevant) continuum and line
  scattering problems. We emphasize that our code will be further developed to
  enable the solution of more complex, multi-level problems in 3D. For such
  problems, highly accurate interpolation schemes are required to describe the
  variation of the mean intensities along a ray.}, which allows us to preserve
monotonicity in a rather simple way. The B\'ezier interpolation is constructed
from two given data points and one control point, the latter setting the slope
of the interpolating curve. The control point is located at the centre of the
data-points abscissae, with the ordinate calculated by accounting for the
information of a third data point to yield the parabola intersecting all three
data points. Whenever overshoots occur, the value of the control point will be
manipulated to ensure monotonicity (see Fig.~\ref{fig:bez}). The corresponding
formulation is given in Appendix \ref{app:bez},
Eqs.~\eqref{eq:bezleft_coeff}-\eqref{eq:bezleft_coeffc}. Applying these
equations to describe the behaviour of the mean intensities along the optical
path, and identifying the indices $(\indx{i-1})$, $(\indx{i})$, $(\indx{i+1})$
with the upwind, current, and downwind points, we find, after reordering for
the $t^0$, $t^1$, $t^2$ terms:
\beqa
\nonumber
 S(t+\tau_{\indxu}) &=& S_{\indxu}^{(\indx{ijk})} +
 \Biggl[\dfrac{(\paramon-2)}{\Delta\tau_{\indxu}}S_{\indxu}^{(\indx{ijk})} \\\nonumber
&+&
    \dfrac{(1-\paramon)\Delta\tau_{\indxu}+(2-\paramon)\Delta\tau_{\indxd}}{\Delta\tau_{\indxu}\Delta\tau_{\indxd}}S_{\indxp}^{(\indx{ijk})}
    + \dfrac{(\paramon-1)}{\Delta\tau_{\indxd}}S_{\indxd}^{(\indx{ijk})} \Biggr] \cdot
 t \\\nonumber
&+& \Biggl[
  \dfrac{(1-\paramon)}{\Delta\tau_{\indxu}^2}S_{\indxu}^{(\indx{ijk})} + 
  \dfrac{(\paramon-1)(\Delta\tau_{\indxu} +
    \Delta\tau_{\indxd})}{\Delta\tau_{\indxu}^2\Delta\tau_{\indxd}}S_{\indxp}^{(\indx{ijk})}
  \\\label{eq:source}
&+&
  \dfrac{(1-\paramon)}{\Delta\tau_{\indxu}\Delta\tau_{\indxd}}S_{\indxd}^{(\indx{ijk})}\Biggr]
 \cdot t^2 \,,
 \eeqa
with downwind optical-depth increment, $\Delta
\tau_{\indxd}=\tau_{\indxd}-\tau_{\indxp}\geq 0$. The parameter $\paramon$
defines the ordinate of the control point (Eq.~\eqref{eq:bezl_fc}). Within the
B\'ezier interpolation, we emphasize that $\paramon$ may explicitly depend on
$S_{\indxu}^{(\indx{ijk})}$, $S_{\indxp}^{(\indx{ijk})}$, and
$S_{\indxd}^{(\indx{ijk})}$ to ensure monotonicity, and not solely on the grid
spacing.  A major advantage of this parameterization is that we can globally
define a minimum allowed $\paramon$ that can be adapted during the iteration
process. The flip-flop situations discussed above can then be avoided by
gradually increasing $\paramon_{\min}$ towards unity ($\paramon\equiv1$
corresponds to a linear interpolation), that is, by suppressing the curvature of
the B\'ezier interpolation. This way, we can construct an always-convergent
iteration scheme, though with the drawback of using less accurate
interpolations.

Integrating Eq.~\eqref{eq:eqrt2} together with a source function described by
Eq.~\eqref{eq:source}, we obtain the discretized equation of radiative
transfer:
\beq
\label{eq:eqrt_disc}
I_{\indx{ijk}} = a_{\indx{ijk}} S_{\indxu}^{(\indx{ijk})} + b_{\indx{ijk}} S_{\indxp}^{(\indx{ijk})} + c_{\indx{ijk}}S_{\indxd}^{(\indx{ijk})} + d_{\indx{ijk}} I_{\indxu}^{(\indx{ijk})} \,,
\eeq
with
\begin{eqnarray*}
a_{\indx{ijk}} &:=& e_0 + \dfrac{\paramon-2}{\Delta\tau_{\indxu}}e_1 +
  \dfrac{1-\paramon}{\Delta \tau_{\indxu}^2}e_2 \\
b_{\indx{ijk}} &:=& \dfrac{(1-\paramon)\Delta\tau_{\indxu} +
    (2-\paramon)\Delta\tau_{\indxd}}{\Delta\tau_{\indxu}\Delta\tau_{\indxd}}e_1 +
  \dfrac{(\paramon-1)(\Delta\tau_{\indxu}+\Delta\tau_{\indxd})}{\Delta
    \tau_{\indxu}^2\Delta\tau_{\indxd}}e_2 \\
c_{\indx{ijk}} &:=& \dfrac{\paramon-1}{\Delta\tau_{\indxd}}e_1 +
  \dfrac{1-\paramon}{\Delta\tau_{\indxu}\Delta\tau_{\indxd}}e_2 \\
d_{\indx{ijk}} &:=& \eu^{-\Delta \tau_{\indxu}}\\
e_0 &:=& \eu^{-\Delta \tau_{\indxu}} \int_{0}^{\Delta \tau_{u}} \eu^t \dd t =
1-\eu^{-\Delta \tau_{\indxu}} \\
e_1 &:=& \eu^{-\Delta \tau_{\indxu}} \int_{0}^{\Delta \tau_{u}} t \eu^t \dd t =
\Delta \tau_{\indxu} - e_0 \\
e_2 &:=& \eu^{-\Delta \tau_{\indxu}} \int_{0}^{\Delta \tau_{u}} t^2 \eu^t \dd t =
\Delta \tau_{\indxu}^2 - 2e_1 \,.
\end{eqnarray*}
The calculation of the upwind and downwind $\Delta\tau$-steps proceeds
similarly, where now the opacity is integrated using the B\'ezier
interpolation.  Using Eqs.~\eqref{eq:bezl}, \eqref{eq:bezl_fc} for the upwind
interval, and Eqs.~\eqref{eq:bezr}, \eqref{eq:bezr_fc} for the downwind
interval, one easily obtains:
\beqa
\Delta \tau_{\indxu} &=& \int_{\indxu}^{\indxp} \chi(s) \dd s = \dfrac{\Delta
  s_{\indxu}}{3} (\chi_{\indxu} + \chi_{\indxc}^{[\indxu,\indxp]} + \chi_{\indxp}) \\
\Delta \tau_{\indxd} &=& \int_{\indxp}^{\indxd} \chi(s) \dd s = \dfrac{\Delta
  s_{\indxd}}{3} (\chi_{\indxp} + \chi_{\indxc}^{[\indxp,\indxd]} +
\chi_{\indxd}) \,,
\eeqa
where $\Delta s_{\indxu}$, $\Delta s_{\indxd}$ describe the path lengths of
the upwind and downwind intervals, respectively, and
$\chi_{\indxc}^{[\indxu,\indxp]}$, $\chi_{\indxc}^{[\indxp,\indxd]}$ refer to
the opacity at the control points in each interval.
%
%
\subsection{Grid refinement}
\label{subsec:refine}
Since the opacity of a line transition depends on the velocity field through
the Doppler effect, regions of significant opacity may become spatially
confined in a highly supersonic wind with strong acceleration. Thus, a grid
refinement along the short characteristic might be required to correctly
account for all so-called resonance zones. Because the profile function is
approximated by a Doppler profile and rapidly vanishes for $\lvert
\xcmf/\delta\rvert \gtrsim 3$, a resonance zone is here defined by a region
where $\xcmf/\delta\in[-3,3]$.

A numerically sufficient condition to resolve all such resonance zones along a
given ray is to demand that $\lvert \Delta \xcmf \rvert / \delta = \lvert
\Delta V_{\rm proj}\rvert / \delta \lesssim 1/3$ if a resonance zone lies in
between the points $[\indxu^{(ijk)},\indxp]$, where $\Delta V_{\rm proj}$ is
the projected velocity step along the ray in units of \vthfid. Assuming a
linear dependence of the projected velocities on the ray coordinate $s$, this
condition directly translates to an equidistant refined spatial grid along the
ray. For short ray segments (as is mostly the case within our calculations),
neglecting the second-order (curvature) terms of the projected velocity
influences the solution only weakly. The line source function on the refined
grid is obtained by B\'ezier interpolation in $s$-space
(Eqs.~\eqref{eq:bezleft_coeff}-\eqref{eq:bezleft_coeffb}):
\beq
S_{\rm L}(s_{\indx{\ell}}) = S_{\indx{\ell}} = \tilde{a}_{\indx{\ell}}S_{\indxu}^{(\indx{ijk})} +
\tilde{b}_{\indx{\ell}}S_{\indxp}^{(\indx{ijk})} + \tilde{c}_{\indx{\ell}}S_{\indxd}^{(\indx{ijk})} \,,
\eeq
where the index $\indx{\ell}$ refers to the points on the refined grid, and
$\indxu^{(\indx{ijk})}, \indxp^{(\indx{ijk})}, \indxd^{(\indx{ijk})}$ describe
the original geometry of the short characteristic. $\chibar_{\rm L}$ is
obtained analogously, and the required $\Delta \tau_{\indx{\ell}}$ steps are
calculated with the trapezoidal rule, for simplicity. Contrasted to the
Sobolev method (which also assumes a linear velocity law along the ray
segment, \eg~\citealt{Rybicki1978}), our grid refinement procedure explicitly
accounts for variations of the opacity and the source function.

Using Eq.~\eqref{eq:eqrt_disc} for the inter-grid points, such that the
(local) upwind, current, and downwind quantities are now described by the
indices $[\indx{\ell-1},\indx{\ell},\indx{\ell+1}]$, we obtain:
\beqa
I_{\indx{\ell}} = I_{\indx{\ell}-1} \eu^{-\Delta \tau_{\indx{\ell}}} &+& \Bigl(a_{\indx{\ell}}
\tilde{a}_{\indx{\ell}-1} + b_{\indx{\ell}} \tilde{a}_{\indx{\ell}} + c_{\indx{\ell}} \tilde{a}_{\indx{\ell}+1}\Bigr) S_{\indxu}^{(\indx{ijk})} \\
  &+& \Bigl(a_{\indx{\ell}} \tilde{b}_{\indx{\ell}-1} + b_{\indx{\ell}} \tilde{b}_{\indx{\ell}} + c_{\indx{\ell}} \tilde{b}_{\indx{\ell}+1}\Bigr) S_{\indxp}^{(\indx{ijk})} \\
  &+& \Bigl(a_{\indx{\ell}} \tilde{c}_{\indx{\ell}-1} + b_{\indx{\ell}} \tilde{c}_{\indx{\ell}} + c_{\indx{\ell}} \tilde{c}_{\indx{\ell}+1}\Bigr) S_{\indxd}^{(\indx{ijk})} \\
 :=  I_{\indx{\ell}-1} \eu^{-\Delta \tau_{\indx{\ell}}} &+&\tilde{\alpha}_{\indx{\ell}}
S_{\indxu}^{(\indx{ijk})} + \tilde{\beta}_{\indx{\ell}} S_{\indxp}^{(\indx{ijk})} +
\tilde{\gamma}_{\indx{\ell}} S_{\indxd}^{(\indx{ijk})} \,.
\eeqa
For a number of $N_{\rm ref}$ refinement points (including the upwind and
current point) within the interval
$[\indxu^{(\indx{ijk})},\indxp^{(\indx{ijk})}]$, the intensity at point
$\indxp^{(\indx{ijk})}$ is finally given by:
\beqa
\nonumber
I_{\indx{ijk}} &=& I_{\indxu}^{(\indx{ijk})} \eu^{-\sum_{m=2}^{N_{\rm ref}}\Delta \tau_{m}} + S_{\indxu}^{(\indx{ijk})} \sum_{m=2}^{N_{\rm ref}} \tilde{\alpha}_{m}
\eu^{-\sum_{n=m+1}^{N_{\rm ref}}\Delta \tau_{n}} \\ \label{eq:eqrt_disc2}
 &+& S_{\indxp}^{(\indx{ijk})} \sum_{m=2}^{N_{\rm ref}} \tilde{\beta}_{m}
\eu^{-\sum_{n=m+1}^{N_{\rm ref}}\Delta \tau_{n}} + S_{\indxd}^{(\indx{ijk})} \sum_{m=2}^{N_{\rm ref}} \tilde{\gamma}_{m}
\eu^{-\sum_{n=m+1}^{N_{\rm ref}}\Delta \tau_{n}} \,,
\eeqa
where the upwind and current points always correspond to the indices $m=1$ and
$m=N_{\rm ref}$, respectively, and the sum over $m$ is performed over $N_{\rm
  ref}-1$ intervals.  The discretized radiative transfer equation for the
refined grid obviously has the same form as for the standard short
characteristic (Eq.~\eqref{eq:eqrt_disc}), with different coefficients though.
%
%
\subsection{Upwind and downwind interpolations}
\label{subsec:interp2d}
\begin{figure}[!ht]
%
\psfrag{auu}{\LARGE ${\rm A_u}$}
\psfrag{buu}{\LARGE ${\rm B_u}$}
\psfrag{cuu}{\LARGE ${\rm C_u}$}
\psfrag{duu}{\LARGE ${\rm D_u}$}
\psfrag{euu}{\LARGE ${\rm E_u}$}
\psfrag{fuu}{\LARGE ${\rm F_u}$}
\psfrag{guu}{\LARGE ${\rm G_u}$}
\psfrag{huu}{\LARGE ${\rm H_u}$}
\psfrag{iuu}{\LARGE ${\rm I_u}$}
\psfrag{juu}{\LARGE ${\rm J_u}$}
\psfrag{kuu}{\LARGE ${\rm K_u}$}
\psfrag{luu}{\LARGE ${\rm L_u}$}
\psfrag{muu}{\LARGE ${\rm M_u}$}
\psfrag{nuu}{\LARGE ${\rm N_u}$}
\psfrag{ouu}{\LARGE ${\rm O_u}$}
\psfrag{puu}{\LARGE ${\rm P_u}$}
\psfrag{quu}{\LARGE ${\rm Q_u}$}
\psfrag{ruu}{\LARGE ${\rm R_u}$}
\psfrag{suu}{\LARGE ${\rm S_u}$}
\psfrag{nnu}{\LARGE $\vecown{n}$}
%
\psfrag{add}{\LARGE ${\rm A_d}$}
\psfrag{bdd}{\LARGE ${\rm B_d}$}
\psfrag{cdd}{\LARGE ${\rm C_d}$}
\psfrag{ddd}{\LARGE ${\rm D_d}$}
\psfrag{edd}{\LARGE ${\rm E_d}$}
\psfrag{fdd}{\LARGE ${\rm F_d}$}
\psfrag{gdd}{\LARGE ${\rm G_d}$}
\psfrag{hdd}{\LARGE ${\rm H_d}$}
\psfrag{idd}{\LARGE ${\rm I_d}$}
\psfrag{jdd}{\LARGE ${\rm J_d}$}
\psfrag{kdd}{\LARGE ${\rm K_d}$}
\psfrag{ldd}{\LARGE ${\rm L_d}$}
\psfrag{mdd}{\LARGE ${\rm M_d}$}
\psfrag{ndd}{\LARGE ${\rm N_d}$}
\psfrag{odd}{\LARGE ${\rm O_d}$}
\psfrag{pdd}{\LARGE ${\rm P_d}$}
\psfrag{qdd}{\LARGE ${\rm Q_d}$}
\psfrag{rdd}{\LARGE ${\rm R_d}$}
\psfrag{sdd}{\LARGE ${\rm S_d}$}
\psfrag{nnd}{\LARGE $\vecown{n}$}
\psfrag{ijk}{\LARGE \bf ${\rm p}^{(\indx{ijk})}$}
%
\psfrag{euc}{\Huge ${\rm E_u}$}
\psfrag{fuc}{\Huge ${\rm F_u}$}
\psfrag{nuc}{\Huge ${\rm N_u}$}
\psfrag{ouc}{\Huge ${\rm O_u}$}
\psfrag{huc}{}
\psfrag{iuc}{}
\psfrag{suc}{}
\psfrag{sdc}{}
\psfrag{edc}{}
\psfrag{fdc}{}
\psfrag{ndc}{\Huge ${\rm N_d}$}
\psfrag{odc}{\Huge ${\rm O_d}$}
\psfrag{hdc}{\Huge ${\rm H_d}$}
\psfrag{idc}{\Huge ${\rm I_d}$}
\psfrag{x}{\huge $\alpha x$}
\psfrag{y}{\huge $\beta y$}
\psfrag{z}{\huge $\gamma z$}
\psfrag{nn}{\huge $\vecown{n}$}
\psfrag{uu}{\Huge $\indxu^{(\indx{ijk})}$}
\psfrag{pp}{\Huge $\indxp^{(\indx{ijk})}$}
\psfrag{dd}{\Huge $\indxd^{(\indx{ijk})}$}
\centering
\resizebox{0.8\hsize}{!}{
   \begin{minipage}{\hsize}
      \resizebox{\hsize}{!}{\includegraphics{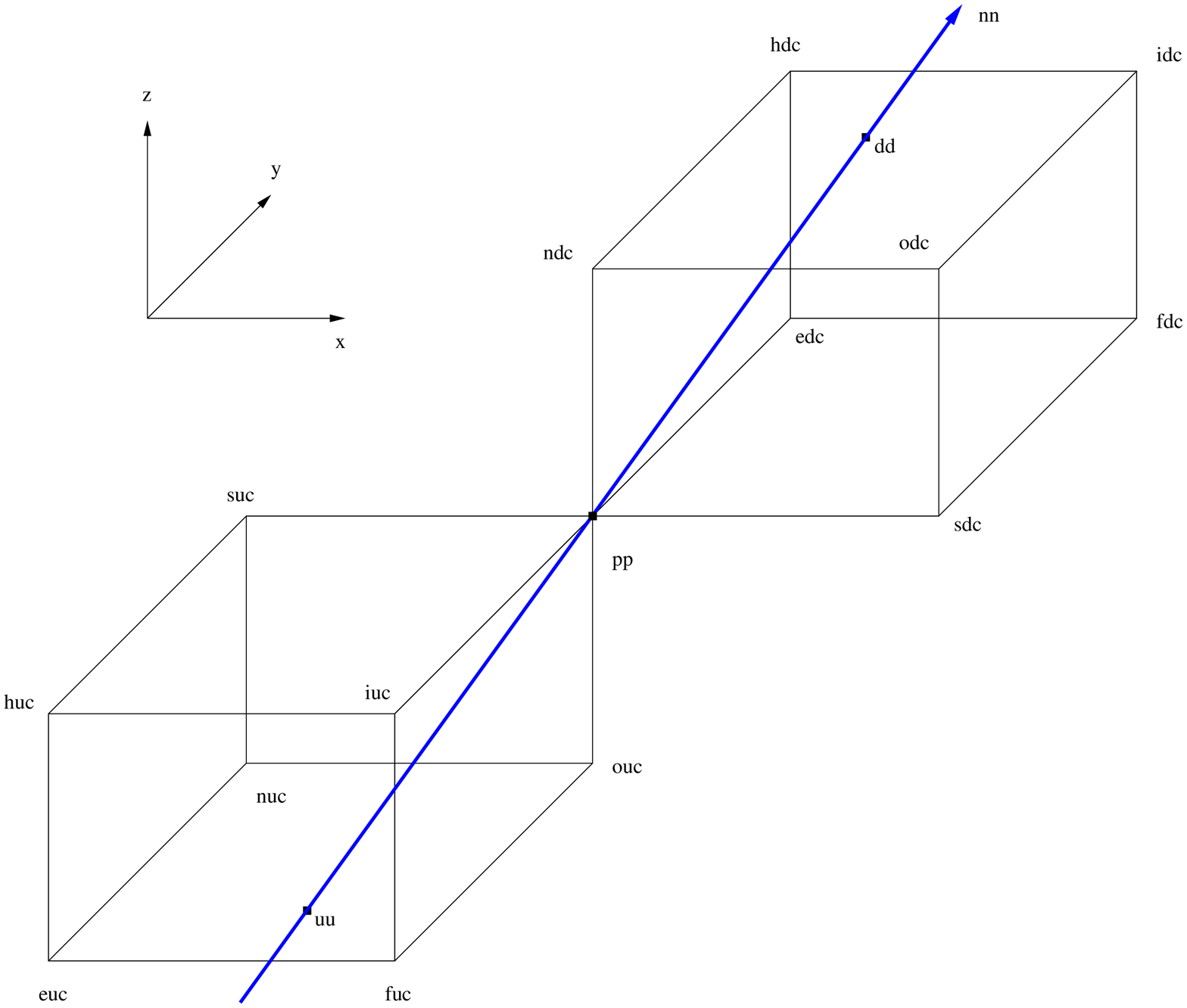}}
   \end{minipage}
}
\\
\resizebox{0.8\hsize}{!}{
   \centering
   \begin{minipage}{\hsize}
      \resizebox{\hsize}{!}{\includegraphics{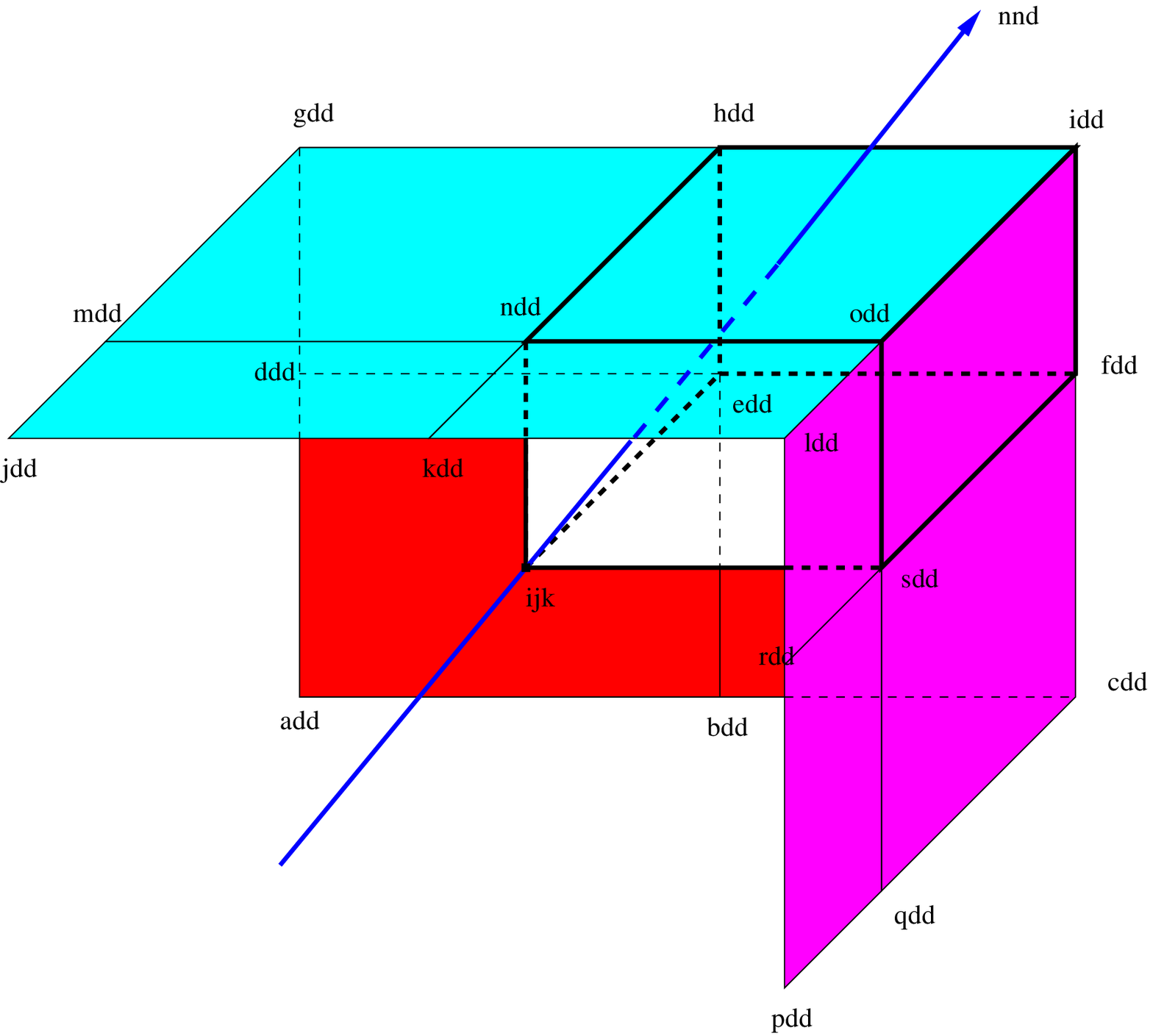}}
   \end{minipage}
}
\\
\resizebox{0.8\hsize}{!}{
   \centering
   \begin{minipage}{\hsize}
      \resizebox{\hsize}{!}{\includegraphics{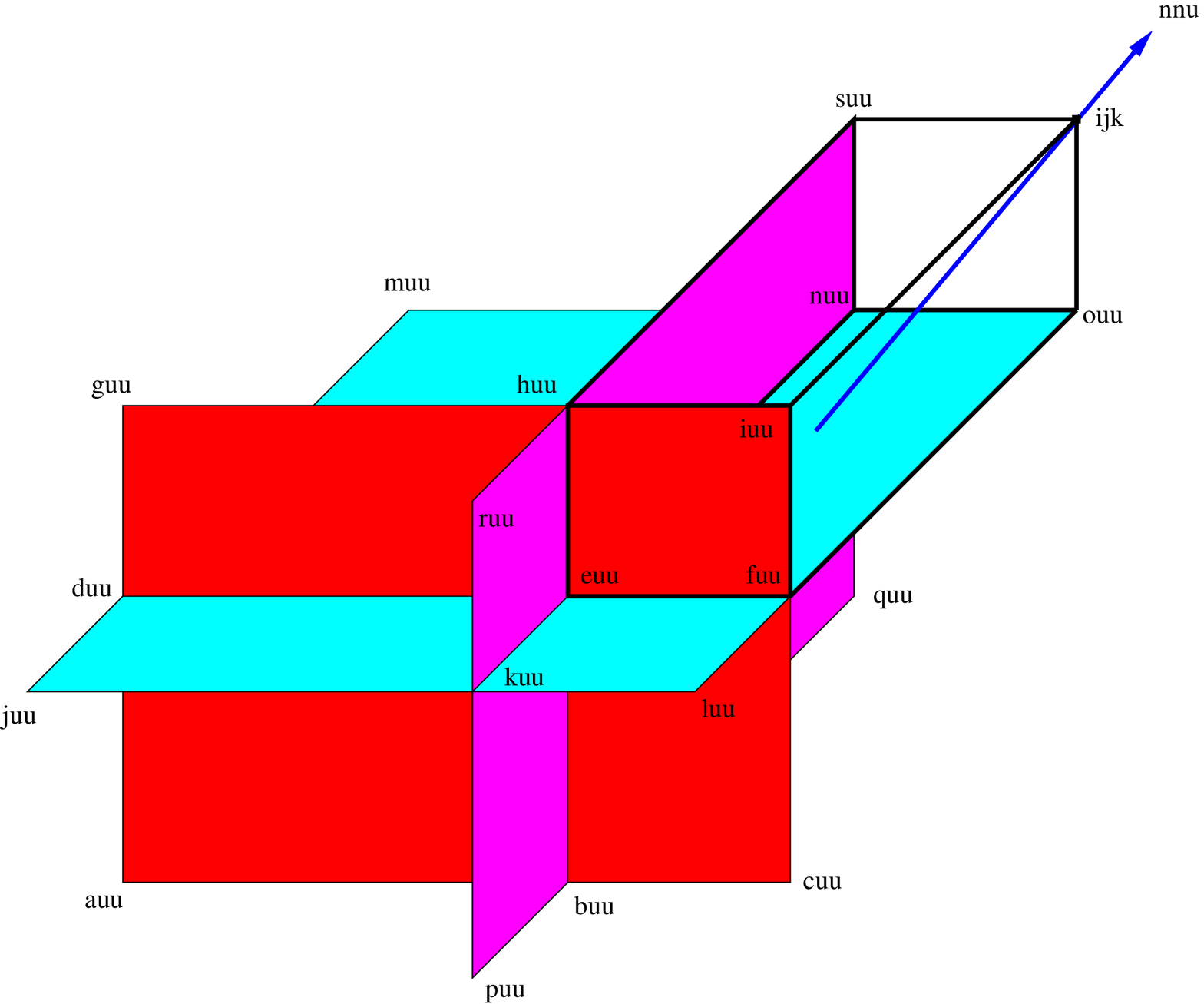}}
   \end{minipage}
}
\caption{Geometry of the SC method for a particular ray with
  direction $\vecown{n}$ propagating from the upwind point
  $\indxu^{(\indx{ijk})}$ to a considered grid point
  $\indxp^{(\indx{ijk})}$. The downwind point $\indxd^{(\indx{ijk})}$ is
  required to set the slope of a B\'ezier curve representing the opacities and
  source functions along the ray. The middle and lower panel display all
  possible downwind and upwind intersection surfaces for a short
  characteristic at a grid point $\indxp^{(\indx{ijk})}$. For rays
  intersecting the $xy$-, $xz$-, and $yz$-planes, the 2D B\'ezier interpolation
  is obtained from given quantities at grid points located in the cyan, red,
  and magenta shaded surfaces, respectively. The coordinate system is
  indicated at the upper left, where $\alpha$, $\beta$, $\gamma$ determine the
  direction of the coordinate-axes and are defined in
  Sect.~\ref{subsec:interp2d}.}
\label{fig:sc_cell_interp}
\end{figure}
To solve the discretized equation of radiative transfer, the opacities
$\chi_{\rm{C} (\indxu,\indxd)}$, $\chibar_{\rm{L} (\indxu,\indxd)}$, source
functions $S_{\rm{C} (\indxu,\indxd)}$, $S_{\rm{L} (\indxu,\indxd)}$, and
velocity vectors $\vecown{V}_{(\indxu,\indxd)}$, are required at the upwind
and downwind points, together with the incident intensity, $I_{\indxu}$. We
emphasize that the subscript ${\rm C}$ describes continuum quantities, and
should not be confused with the subscript ${\rm c}$ denoting the control
points of the interpolation scheme.

All required quantities are obtained from a 2D B\'ezier interpolation (see
Appendix~\ref{app:bez2d}) on the surfaces that intersect with a given ray. The
intersection surfaces depend on the considered direction and the size of the
upwind and downwind grid cells. For a given direction
\beq
\vecown{n} = 
\begin{pmatrix}
n_x \\ n_y \\ n_z
\end{pmatrix}
=
\begin{pmatrix}
\sin\theta \cos\phi \\ \sin\theta\sin\phi \\ \cos\theta
\end{pmatrix} \,,
\eeq
where $\theta$ is the co-latitude (measured from the Cartesian $z$-axis), and
$\phi$ is the azimuth (measured from the $x$-axis), the distances from a
considered grid point to the neighbouring $xy$-, $xz$-, and $yz$-planes are
calculated from trigonometry and yield:
\begin{align*}
\Delta s_{xy}^{(\indxu)} & = \dfrac{z_{\indx{k}}-z_{\indx{k-\gamma}}}{n_z}  & 
\Delta s_{xy}^{(\indxd)} & = \dfrac{z_{\indx{k+\gamma}}-z_{\indx{k}}}{n_z}  \\
\Delta s_{xz}^{(\indxu)} & = \dfrac{y_{\indx{j}}-y_{\indx{j-\beta}}}{n_y}  & 
\Delta s_{xz}^{(\indxd)} & = \dfrac{y_{\indx{j+\beta}}-y_{\indx{j}}}{n_y} \\
\Delta s_{yz}^{(\indxu)} & = \dfrac{x_{\indx{i}}-x_{\indx{i-\alpha}}}{n_x} & 
\Delta s_{yz}^{(\indxd)} & = \dfrac{x_{\indx{i+\alpha}}-x_{\indx{i}}}{n_x} \,,
\end{align*}
with $\alpha, \beta, \gamma$ set to $\pm 1$ for direction-vector components
$n_x,n_y,n_z \gtrless 0$, respectively. The intersection surface on the upwind
and downwind side are then found at the minimum of $\Delta
s_{xy}^{(\indxu,\indxd)}, \Delta s_{xz}^{(\indxu,\indxd)}, \Delta
s_{yz}^{(\indxu,\indxd)}$, and the corresponding coordinates are easily
calculated.

For each surface, the interpolation requires nine points within the
corresponding plane (see Fig.~\ref{fig:sc_cell_interp} and
Eq.~\eqref{eq:interp2d}). In each considered plane, we generally use grid
points running from (index-2) to (index) to determine upwind quantities, while
downwind quantities are calculated from (index-1) to (index+1). Such a
formulation greatly simplifies the calculation of the $\Lambda$-matrix
elements (see Appendix~\ref{app:alo}). In Fig.~\ref{fig:sc_cell_interp}, we
show an example for a ray intersecting the $xy$-plane at the upwind side. The
2D B\'ezier interpolation for the upwind point then consists of three 1D
B\'ezier interpolations along the $x$-axis using the points ${\rm (J_u, K_u,
  L_u)}$, ${\rm (D_u,E_u,F_u)}$, ${\rm (M_u,N_u,O_u)}$, followed by another 1D
B\'ezier interpolation along the $y$-axis at the upwind $x$-coordinates.  With
the 2D B\'ezier interpolation given by Eq.~\eqref{eq:interp2d}, we find for
each required quantity $q_{\indxu, \indxd}$:
\beqa
\nonumber
q_{\indxu}^{(\indx{ijk})} &=& w_{\rm A}^{(\indx{ijk})} q_{\indx{i-2\alpha, j-\beta,k-2\gamma}} + w_{\rm B}^{(\indx{ijk})}
q_{\indx{i-\alpha, j-\beta, k-2\gamma}} + w_{\rm C}^{(\indx{ijk})} q_{\indx{i, j-\beta, k-2\gamma}} \\
\nonumber
&+& w_{\rm D}^{(\indx{ijk})} q_{\indx{i-2\alpha, j-\beta,k-\gamma}} + w_{\rm E}^{(\indx{ijk})}
q_{\indx{i-\alpha, j-\beta, k-\gamma}} + w_{\rm F}^{(\indx{ijk})} q_{\indx{i, j-\beta, k-\gamma}} \\
\nonumber
&+& w_{\rm G}^{(\indx{ijk})} q_{\indx{i-2\alpha, j-\beta,k}} + w_{\rm H}^{(\indx{ijk})}
q_{\indx{i-\alpha, j-\beta, k}} + w_{\rm I}^{(\indx{ijk})} q_{\indx{i, j-\beta, k}} \\
\nonumber
&+& w_{\rm J}^{(\indx{ijk})} q_{\indx{i-2\alpha, j-2\beta,k-\gamma}} + w_{\rm K}^{(\indx{ijk})}
q_{\indx{i-\alpha, j-2\beta, k-\gamma}} + w_{\rm L}^{(\indx{ijk})} q_{\indx{i, j-2\beta, k-\gamma}} \\
\nonumber
&+& w_{\rm M}^{(\indx{ijk})} q_{\indx{i-2\alpha, j,k-\gamma}} + w_{\rm N}^{(\indx{ijk})}
q_{\indx{i-\alpha, j, k-\gamma}} + w_{\rm O}^{(\indx{ijk})} q_{\indx{i, j, k-\gamma}} \\
\nonumber
&+& w_{\rm P}^{(\indx{ijk})} q_{\indx{i-\alpha, j-2\beta,k-2\gamma}} + w_{\rm Q}^{(\indx{ijk})} q_{\indx{i-\alpha, j,
  k-2\gamma}}  \\
\label{eq:interp2du}
&+& w_{\rm R}^{(\indx{ijk})} q_{\indx{i-\alpha, j-2\beta,k}} + w_{\rm S}^{(\indx{ijk})} q_{\indx{i-\alpha, j,
  k}} + w_{\indx{ijk}} q_{\indx{ijk}} \\
\nonumber
q_{\indxd}^{(\indx{ijk})} &=& \tilde{w}_{\rm A}^{(\indx{ijk})} q_{\indx{i-\alpha, j+\beta,k-\gamma}} + \tilde{w}_{\indx{B}}^{(\indx{ijk})}
q_{\indx{i, j+\beta, k-\gamma}} + \tilde{w}_{\rm C}^{(\indx{ijk})} q_{\indx{i+\alpha, j+\beta, k-\gamma}} \\
\nonumber
&+& \tilde{w}_{\rm D}^{(\indx{ijk})} q_{\indx{i-\alpha, j+\beta,k}} + \tilde{w}_{\rm E}^{(\indx{ijk})}
q_{\indx{i, j+\beta, k}} + \tilde{w}_{\rm F}^{(\indx{ijk})} q_{\indx{i+\alpha, j+\beta, k}} \\
\nonumber
&+& \tilde{w}_{\rm G}^{(\indx{ijk})} q_{\indx{i-\alpha, j+\beta,k+\gamma}} + \tilde{w}_{\rm H}^{(\indx{ijk})}
q_{\indx{i, j+\beta, k+\gamma}} + \tilde{w}_{\rm I}^{(\indx{ijk})} q_{\indx{i+\alpha, j+\beta, k+\gamma}} \\
\nonumber
&+& \tilde{w}_{\rm J}^{(\indx{ijk})} q_{\indx{i-\alpha, j-\beta,k+\gamma}} + \tilde{w}_{\rm K}^{(\indx{ijk})}
q_{\indx{i, j-\beta, k+\gamma}} + \tilde{w}_{\rm L}^{(\indx{ijk})} q_{\indx{i+\alpha, j-\beta, k+\gamma}} \\
\nonumber
&+& \tilde{w}_{\rm M}^{(\indx{ijk})} q_{\indx{i-\alpha, j,k+\gamma}} + \tilde{w}_{\rm N}^{(\indx{ijk})}
q_{\indx{i, j, k+\gamma}} + \tilde{w}_{\rm O}^{(\indx{ijk})} q_{\indx{i+\alpha, j, k+\gamma}} \\
\nonumber
&+& \tilde{w}_{\rm P}^{(\indx{ijk})} q_{\indx{i+\alpha, j-\beta,k-\gamma}} + \tilde{w}_{\rm Q}^{(\indx{ijk})} q_{\indx{i+\alpha, j,
  k-\gamma}}  \\
&+& \tilde{w}_{\rm R}^{(\indx{ijk})} q_{\indx{i+\alpha, j-\beta,k}} + \tilde{w}_{\rm S}^{(\indx{ijk})} q_{\indx{i+\alpha, j,
  k}}
\label{eq:interp2dd}
\,,
\eeqa
where the coefficients $w^{(\indx{ijk})}$ and $\tilde{w}^{(\indx{ijk})}$ refer
to the upwind and downwind interpolations corresponding to a considered point
$(\indx{ijk})$. Depending on the intersection surface, ten out of these 19
coefficients are set to zero.  For the upwind interpolation, we have already
included the local coefficient $(\indx{ijk})$, which is only required when
boundary conditions need to be specified (Sect.~\ref{subsec:boundary}). We
note that all (non-zero) interpolation coefficients may depend on the specific
values of a considered quantity at the given grid points, via the
interpolation parameter $\paramon$ to ensure monotonicity. As in
Sect.~\ref{subsec:sc}, also these monotonicity constraints result in
non-linear $\Lambda$-operators.
%
%
\subsection{Boundary conditions}
\label{subsec:boundary}
\begin{figure}[t]
\psfrag{n1}{$\vecown{n_1}$}
\psfrag{n2}{$\vecown{n_2}$}
\psfrag{n3}{$\vecown{n_3}$}
\psfrag{p}{\indxp (ijk)}
\psfrag{u1}{$\indxu_1$}
\psfrag{u2}{$\indxu_2$}
\psfrag{u3}{$\indxu_3$}
\psfrag{a}{$\indx{i-2,j,k-1}$}
\psfrag{b}{$\indx{i-1,j,k-1}$}
\psfrag{c}{$\indx{i,j,k-1}$}
\psfrag{d}{$\indx{i-1,j,k}$}
\resizebox{\hsize}{!}{\includegraphics{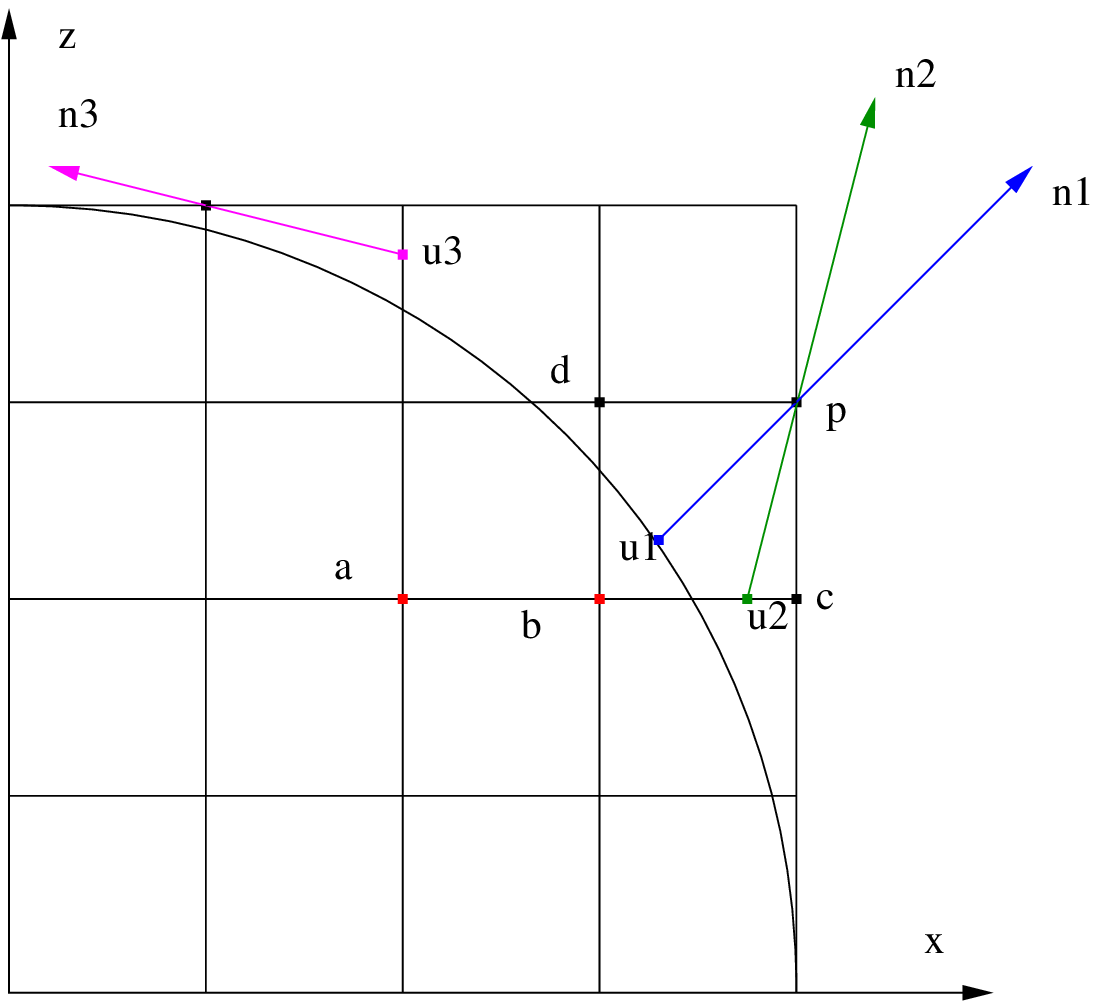}}
\caption{Boundary conditions for rays propagating in the $xz$-plane at
  $y$-level $(j)$ with three different directions $\vecown{n_1}$,
  $\vecown{n_2}$, $\vecown{n_3}$, and upwind points $\indxu_1$, $\indxu_2$,
  $\indxu_3$. For point $\indxu_1$, the intensity is set to
  $I_{\indxu}=I_{\indxc}^+$ and all remaining quantities are obtained by
  bilinear interpolation from points $(\indx{i-1,j,k-1})$, $(\indx{i,j,k-1})$,
  $(\indx{i-1,j,k})$, and $(\indx{ijk})$. The required quantities at point
  $\indxu_2$ are found from B\'ezier interpolation using the values at
  $(\indx{i-2,j,k-1})$, $(\indx{i-1,j,k-1})$, $(\indx{i,j,k-1})$. The
  (unknown) quantities inside the core are indicated by red dots, and need to
  be reasonably approximated (see text). For direction $\vecown{n}_3$, the
  unknown intensity inside the core, $I_{\indxc}^-$, is directed inwards. Such
  situations occur only for ray directions (nearly) parallel to the spatial
  grid, and thus are relatively seldom.}
\label{fig:boundary}
\end{figure}
Since the inner boundary is usually not aligned with the 3D Cartesian grid
(\eg~a spherical star at the origin), the upwind (and downwind) interpolations
need to be adapted near the stellar surface. For the upwind point, the
  following two situations may occur (see Fig.~\ref{fig:boundary} for an example in the
$xz$-plane). Firstly, the considered ray originates from the stellar surface
(direction $\vecown{n_1}$ in Fig.~\ref{fig:boundary}). In this case, we use a
core-halo approximation and set $I_{\indxu}=I_{\indxc}^+=\Bnu(\Trad)$, with
$I_{\indxc}^+$ the emergent intensity from the core, and $\Trad$ the radiation
temperature. Unless explicitly noted, we assume $\Trad=\Teff$ throughout this
work. All other quantities are obtained from trilinear interpolation using the
points $({\rm E_u,F_u,H_u,I_u,N_u,O_u,S_u,p}^{(\indx{ijk})})$ in
Fig.~\ref{fig:sc_cell_interp}, where representative estimates need to be
defined at the core points (those points that are located inside the star). In
hydrodynamic simulations, the analogue of these points are so-called `ghost
points'. Secondly, the considered ray originates from a plane spanned by grid
points that are partially located inside the star (direction $\vecown{n_2}$ in
Fig.~\ref{fig:boundary}). Then, the interpolation is performed as in
Sect.~\ref{subsec:interp2d}, using again representative estimates at the
core-points.

Inside the core, we define $I_{\indxc}^+=\sline=\scont=\Bnu(\Trad)$ and set
$I_{\indxc}^-$ and all velocity components to zero, where $I_{\indxc}^-$ is
the inward directed intensity which needs to be specified only in rare
situations (Fig.~\ref{fig:boundary}, direction $\vecown{n}_3$). The opacities
inside the star are found by extrapolation from the known values outside the
star. While this procedure certainly introduces errors (\eg~by
over- and underestimating the upwind source function in optically thin
and thick cases, respectively), it is still favourable to extrapolating all values
directly onto the stellar surface, mainly due to performance
reasons\footnote{For a given grid point, the number of neighbouring grid
  points that can be used for extrapolation is not a priori clear, and depends
  on the shape of the stellar surface, and the considered direction of the
  ray. Indeed, there are 64 special cases that would have to be implemented
  explicitly. This is computationally not feasible.}. In addition to the error
introduced by the predefined values inside the core, the calculation of
$\Delta s_{\rm r}$ is a certain issue, where $\Delta s_{\rm r}$ is the
distance of the current grid point to the stellar surface. Since the radiative
transfer near the stellar surface is (in most cases) very sensitive to the
path length of a considered ray, $\Delta s_{\rm r}$ needs to be known
exactly. Depending on the shape of the surface, $\Delta s_{\rm r}$ can be
calculated analytically, or needs to be determined numerically. A numerical
solution, however, might be time consuming and should be avoided when
possible. Downwind quantities are always calculated from
Eq.~\eqref{eq:interp2dd}, using the estimates at the core points as defined
above when necessary.
%
%
\subsection{Angular and frequency integration}
\label{subsec:integration}
\begin{table}
\begin{center}
\caption{Mean relative error (defined in Sect.~\ref{subsec:spherical_wind}) of
  the mean intensity for a zero-opacity model, obtained using the Lebedev,
  Gauss-Legendre, and trapezoidal integration method, the latter with nodes
  from \cite{Lobel08}. For $\mrerr{\Delta J}_{\rm ex}$ and $\mrerr{\Delta
    J}_{\rm SC}$, the incident intensities have been calculated exactly and
  from the SC method using linear upwind interpolations, respectively.}
\label{tab:mint_thin}
\begin{tabular}{c|c|c|c|c|c|c}
   & \multicolumn{2}{c|}{Trapez} & \multicolumn{2}{c|}{Legendre} &
\multicolumn{2}{c}{Lebedev} \\
\hline
\noalign{\vskip 0.5mm}
\nomega & 1037 & 2105 & 968 & 2048 & 974 & 2030 \\
$\mrerr{\Delta J_{\rm ex}}\,[\%]$ & 12.1 & 8.0 & 14.3 & 9.4 & 11.2 &  7.7 \\
$\mrerr{\Delta J_{\rm SC}}\, [\%]$ & 11.3 & 10.8 & 10.9 & 10.9 & 10.8 & 10.8
\end{tabular}
\end{center}
\end{table}
To obtain the mean intensity at each grid point in the atmosphere, we solve
the discretized equation of radiative transfer for many directions and
numerically integrate via:
\beq
J_{\indx{ijk}} = \dfrac{1}{4\pi} \int I_{\indx{ijk}} \dd \Omega = \sum_{\indx{l}} w_{\indx{l}} I_{\indx{ijk}}(\Omega_{\indx{l}}) \,,
\eeq
where $w_{\indx{l}}$ is the integration weight corresponding to a considered
direction $\Omega_{\indx{l}}=(\theta_{\indx{l}},\phi_{\indx{l}})$. The angular
integration is particular challenging for optically thin atmospheres, since in
such situations each (spatial) grid point is illuminated by the stellar
surface, and the distribution of intensities $I_{\indx{ijk}}(\theta,\phi)$
becomes a 2D step-function in the $\theta-\phi$-plane (if no upwind
interpolation errors were present). Depending on the considered position, the
shape of $I_{\indx{ijk}}(\theta,\phi)$ greatly varies. Thus, elaborate
integration methods are required to resolve the star \textbf{and} its edges at
any point of the atmosphere.

\cite{Lobel08} use the trapezoidal rule with a decreasing number of polar grid
points at higher latitudes to reasonably distribute the direction vectors on
the unit sphere. For the 3D FVM, we have shown in \citetalias{Hennicker2018}
that a Gauss-Legendre integration performs (slightly) better. However, the
corresponding directions are always clustered in certain regions since the
nodes of the Gauss-Legendre quadrature are fixed. Additionally, the
Gauss-Legendre integration should only be applied when the distribution of
intensities can be described by high order polynomials, that is, when
$I_{\indx{ijk}}(\theta,\phi)$ is smoothed out (\eg~by numerical diffusion).
When numerical diffusion errors are suppressed (\eg~by using elaborate upwind
interpolation schemes), the Gauss-Legendre integration should not be used (see
Table~\ref{tab:mint_thin}).

We have tested a multitude of other quadrature schemes, including trapezoidal
and (pseudo)-Gaussian rules on triangles, and the so-called Lebedev quadrature
(see, \eg~\citealt{Ahrens09}, \citealt{Beentjes2015}, and references therein).
The Lebedev quadrature is optimized to exactly integrate the spherical
harmonics up to a certain degree, with a (nearly) optimum distribution of
direction vectors on the unit sphere.  In Table \ref{tab:mint_thin}, we
summarize the errors for an optically thin atmosphere using different
integration methods. The incident intensities have been obtained exactly
(\ie~by setting $I_{\indx{ijk}}=I_{\rm c}$ and $I_{\indx{ijk}}=0$ for core and
non-core rays, respectively), or from the 3D SC method using linear
interpolations. Considering the SC solution scheme, the solution has only been
slightly improved (if at all) when doubling the angular grid resolution from
$N_\Omega\approx 1000$ to $N_\Omega \approx 2000$, for all applied integration
methods. We note that the mean relative error does not converge to zero due to
the upwind interpolation scheme (see Sect.~\ref{subsec:searchlight}). For the
exact solution of the optically thin radiative transfer, the
Lebedev-integration method performs best, and is therefore used within all our
calculations. When calculating line transitions, the location of resonance
zones depends on the considered direction. Thus, we generally use
$N_\Omega=2030$ direction vectors to ensure that no resonance zone has been
overlooked. The corresponding angular resolution is typical when calculating
3D radiative transfer problems in extended stellar
atmospheres. For instance, \cite{Lobel08} used $N_\Omega=6400$ within their 3D
finite-volume method.

To obtain the scattering integral, we apply the trapezoidal rule for the
frequency integration. The scattering integral then reads:
\beq
\label{eq:mintbar}
\bar{J}_{\indx{ijk}} = \dfrac{1}{4\pi} \int \dd\Omega \int_{\xobs^{\rm
    (min)}}^{\xobs^{\rm (max)}} \dd x I_{\indx{ijk}} \Phi_x^{(\indx{ijk})} \, =
\sum_{\indx{l}} w_{\indx{l}} \sum_{\indx{x}} w_{\indx{x}} I_{\indx{ijk}}\profile^{(\indx{ijk})},
\eeq
with $\xobs^{\rm (min)}$ and $\xobs^{\rm (max)}$ the required frequency shift
in the observer's frame obtained from the maximum absolute velocity occurring
in the atmosphere (see also \citetalias{Hennicker2018}), and $w_{\indx{x}}$
the corresponding frequency integration weight. To resolve the profile
function at each point in the atmosphere, we demand that $\lvert \Delta \xobs
\rvert / \delta \lesssim 1/3$. Since the profile function depends on the ratio
of fiducial to actual thermal width, the fiducial velocity should be set to
the minimum thermal velocity present in the atmosphere.
%
%
\subsection{$\Lambda$-iteration} \label{subsec:ali}
In Sect.~\ref{subsec:sc}, we have already noted that the $\Lambda$-operator
becomes non-linear due to monotonicity constraints (implemented by the
interpolation parameter $\paramon$). In this section, we present a suitable
workaround, beginning with a recapitulation of some fundamental ideas.
\subsubsection{$\Lambda$-matrix elements}
With the discretized equation of radiative transfer,
Eqs.~\eqref{eq:eqrt_disc}/\eqref{eq:eqrt_disc2}, and the upwind and downwind
quantities obtained from Eqs.~\eqref{eq:interp2du} and \eqref{eq:interp2dd},
the intensity at each spatial, angular, and frequency grid point can be
calculated for a given source function. We use the standard
$\Lambda$-formalism to write the formal solution of the intensity, mean
intensity, and scattering integral as:
\beqa
I &=& \Lambda_{\Omega,\nu}[S_{\rm C, L}] \\
J &=& \Lambda_{\nu}[\scont] \\
\label{eq:formal_sol}
\bar{J} &=& \Lambda [\sline] \,,
\eeqa
with subscripts $\Omega$ and $\nu$ defining the dependence of the
$\Lambda$-operator on direction and frequency, respectively. In the following,
we focus on the line transport. The continuum can be derived analogously. When
all interpolation parameters $\paramon$ have been determined (for a given
stratification of source functions and intensities), the $\Lambda$-operator is
an affine operator described by the $\matown{\Lambda}$-matrix and a constant
displacement vector $\vecown{\Phi}_{\rm B}$ representing the propagation of
boundary conditions (for a detailed discussion, see
\citetalias{Hennicker2018}, \citealt{Puls91}, and references therein). The
$\matown{\Lambda}$-matrix elements can then be obtained by:
\beq
\label{eq:lambda_elements}
\Lambda_{\indx{m,n}} = \bar{J}_{\indx{m}}(\vecown{S}_{\rm L} = \vecown{e}_{\indx{n}},
\vecown{\Phi}_{\rm B} = 0) \,,
\eeq
with the $\indx{n}$-th unit vector $\vecown{e}_{\indx{n}}$, and matrix indices
$\indx{m,n}$ related to the 3D indices $(\indx{i,j,k})$ by
\beq
\label{eq:index_conversion}
m=i+\nx(j-1)+\nx\ny(k-1) \,,
\eeq
where $\nx$ and $\ny$ denote the number of spatial grid points of the $x$ and
$y$ coordinate, respectively. Eq.~\eqref{eq:index_conversion} simply
transforms a data cube to a 1D array.  The $\indx{m,n}$-th
matrix-element describes the effect of a non-vanishing source function at grid
point $\indx{n}$ onto grid point $\indx{m}$. We emphasize that
Eq.~\eqref{eq:lambda_elements} holds only for pre-calculated interpolation
parameters $\paramon$, obtained from an already known stratification of source
functions.
\subsubsection{Accelerated $\Lambda$-iteration}
The classical $\Lambda$-iteration scheme is defined by calculating a formal
solution for a given source function using Eq.~\eqref{eq:formal_sol}, followed
by the calculation of a new source function by means of
Eq.~\eqref{eq:sline}. For optically thick, scattering dominated atmospheres,
however, this iteration scheme suffers from severe convergence problems (see
Fig.~\ref{fig:convergence} for the convergence behaviour of spherically
symmetric test models). To overcome these problems, we apply an accelerated
$\Lambda$-iteration scheme based on operator-splitting methods
(\citealt{Cannon73}). Within the ALI, the $\Lambda$-operator is written as
\beq
\label{eq:split}
\Lambda = \alo + (\Lambda - \alo) \,,
\eeq
where the first term is an appropriately chosen ALO acting on the new source
function, $S_{\rm L}^{(k)}$, and the second term acts on the previous one,
$S_{\rm L}^{(k-1)}$. For the converged solution, this scheme becomes an exact
relation. Using also, and in analogy to the exact $\Lambda$-operator, an
affine representation for the approximate one, $\alo\left[S\right]=\alom\cdot
\vecown{S} +\vecown{\Phi}_{\rm B}$ (cf. above), and evaluating $\alo$ at the
previous iteration step, $k-1$, we obtain:
\beqa
\nonumber
\vecown{S}^{(k)}_{\rm L} &=& \matown{\zeta} \cdot \bar{\vecown{J}}^{(k)} +
\vecown{\Psi} \\ \nonumber
&\approx&  \matown{\zeta} \cdot \alo_{k-1}[\vecown{S}_{\rm L}^{(k)}] + \matown{\zeta}
\cdot (\Lambda_{k-1}-\alo_{k-1}) [\vecown{S}_{\rm L}^{(k-1)}] + \vecown{\Psi}
\\ \nonumber
&=&
\matown{\zeta} \cdot \Bigl( \alom_{k-1} \vecown{S}_{\rm L}^{(k)} +
 \vecown{\Phi}_{\rm B}^{(k-1)} +
 \bar{\vecown{J}}^{(k-1)} - \alom_{k-1} \vecown{S}_{\rm L}^{(k-1)} - \vecown{\Phi}_{\rm B}^{(k-1)} \Bigl) \\\label{eq:ali_final}
&+& \vecown{\Psi} \,.
\eeqa
Here, the iteration indices $k-1$ and $k$ are indicated as sub- or
superscripts, $\matown{\zeta} := \unitym - \matown{\epsilon}_{\rm L}$ is a
diagonal matrix, and $\vecown{\Psi}:=\matown{\epsilon}_{\rm L} \cdot
\vecown{B}_{\rm \nu} (\vecown{T})$ is the thermal contribution vector. From
Eq.~\eqref{eq:ali_final}, it is obvious that the $\vecown{\Phi}_{\rm B}$ terms
cancel.  For multi-level atoms, we emphasize that $\Lambda$ and $\alo$ may
change within the ALI-cycle due to the variation of opacities (induced by the
subsequently updated occupation numbers). Furthermore, both operators might
also change even for the simplified TLA approach considered in this paper,
since the corresponding matrix elements depend on the source functions via the
interpolation parameters $\paramon_{k-1}$, $\paramon_{k}$. Rearranging terms,
we find:
\beq
\label{eq:ali}
 (\unitym - \matown{\zeta} \cdot \alom_{k-1}) \vecown{S}_{\rm L}^{(k)} = \matown{\zeta}
\cdot (\bar{\vecown{J}}^{(k-1)} - \alom_{k-1} \cdot \vecown{S}_{\rm L}^{(k-1)}) + \vecown{\Psi} \,.
\eeq
Eq.~\eqref{eq:ali} is solved to obtain a new source function $\vecown{S}_{\rm
  L}^{(k)}$ (see below). Since, however, $\alom_{k-1}$ has been optimized only
to ensure monotonicity in a specific step $k-1$ (based on source function
$\vecown{S}_{\rm L}^{(k-1)}$), the iteration scheme can oscillate due to
oscillations in $\alom_k$ and $\alom_{k-1}$. Even worse, the new source
function might become negative.
To overcome these problems, non-linear situations need to be avoided (by
providing almost constant $\alom$-matrices over subsequent iteration
steps). The following approach has proved to lead to a stable and convergent
scheme: We apply purely linear interpolations ($\paramon_{k-1}=\paramon_{k}=1$
and thus $\alom_{k}=\alom_{k-1}$) in the first four iteration steps to obtain
an already smooth stratification of source functions. Additionally, we
globally define a minimum allowed interpolation parameter and demand that
$\paramon > \paramon_{\rm min}$. Then, $\paramon$ becomes constant (namely
$\paramon=\paramon_{\rm min}$) in (most) critical situations, and again,
$\alom_k$ approaches $\alom_{k-1}$.  Whenever negative source functions or
oscillations occur within the iteration scheme, $\paramon_{\rm min}$ is
gradually increased to one. With this approach, we obtain an always convergent
iteration scheme, with a formal solution obtained by using linear
interpolations only in most challenging cases.
\subsubsection{Constructing the ALO}
The rate of convergence achieved by the accelerated $\Lambda$-iteration scheme
increases with the number of $\matown{\Lambda}$-matrix elements included in
the ALO. To minimize the computation time of the complete procedure, the
choice of the ALO is always a compromise between the number of matrix-elements
to be calculated, and the resulting convergence speed. While the inversion of
a diagonal ALO reduces to a simple division, a multi-band ALO needs to be
inverted with some more effort. When taking only the nearest neighbours into
account, however, the ALO becomes a sparse matrix, and can be efficiently
inverted by applying the Jacobi-Iteration for sparse systems (see
\citetalias{Hennicker2018}).  For 3D calculations, a multi-band ALO is
required to obtain a rapidly converging iteration scheme (see \citealt{Haus06}
and \citetalias{Hennicker2018} for solutions obtained with the LC method and
the FVM, respectively), and thus implemented within our 3D SC framework. To
calculate the corresponding $\matown{\Lambda}$-matrix elements (including all
upwind and downwind interpolations), we extend the procedure developed by
\cite{Olson87} and \cite{Kunasz88b}. A detailed derivation is given in
Appendix \ref{app:alo}.  Eqs.~\eqref{eq:alocoeff01}-\eqref{eq:alocoeff27}
correspond to the exact $\matown{\Lambda}$-matrix elements for a local point
and its 26 neighbours, and thus should give an excellent rate of convergence
when included in the ALO (see, \eg~the 26-neighbour ALO of
\textsc{Phoenix/3D}, \citealt{Haus06}). Furthermore, all elements can be
calculated in parallel to the formal solution. This property becomes important
when the ALO varies during the iteration scheme, that is, when applying monotonic
B\'ezier interpolations (as discussed above), or when accounting for
multi-level atoms (for which the occupation numbers and thus opacities might
change during the iteration)\footnote{For the simplified continuum and the TLA
  considered in this paper, the linear interpolation scheme is particularly
  advantageous in terms of computation time, since the corresponding ALO
  remains constant over all iteration steps, and therefore needs to be
  calculated only once.}.

In this paper, we analyse the convergence speed of the ALI for a
diagonal ALO given by Eq.~\eqref{eq:alocoeff14}, a `direct-neighbour' (DN)-ALO
given by Eqs.~\eqref{eq:alocoeff05}, \eqref{eq:alocoeff11},
\eqref{eq:alocoeff13}, \eqref{eq:alocoeff14}, \eqref{eq:alocoeff15},
\eqref{eq:alocoeff17}, \eqref{eq:alocoeff23}, and a `nearest-neighbour'
(NN)-ALO obtained from all
Eqs.~\eqref{eq:alocoeff01}-\eqref{eq:alocoeff27}. We note that only a moderate
improvement of the computation time can be expected when using the diagonal or
DN-ALO, since the diagonal and direct-neighbour elements depend on several
other neighbours through the inclusion of downwind interpolations. Since,
however, the downwind-integration weight is generally negative, neglecting
these terms will overestimate the considered matrix elements, possibly
resulting in a divergent iteration scheme. On the other hand, when using
purely linear interpolations (for the source contribution and upwind
interpolations), the calculation of the ALO is greatly simplified since all
coefficients $c_{\indx{ijk}}$ and $w_{\rm A}$, $w_{\rm B}$, $w_{\rm C}$,
$w_{\rm D}$, $w_{\rm G}$, $w_{\rm J}$, $w_{\rm K}$, $w_{\rm L}$, $w_{\rm M}$,
$w_{\rm P}$, $w_{\rm Q}$, $w_{\rm R}$ vanish. For third-order upwind/downwind
interpolations as used in IRIS (\citealt{Ibgui13}), the calculation of the ALO
coefficients becomes computationally prohibitive at some point.  Considering
both interpolation techniques used in this paper, the calculation of the
diagonal, DN-, and NN-ALO, in parallel to the formal solution requires 20\%,
30\%, and 40\% of the total computation time.

Finally, we have implemented the Ng-extrapolation (\citealt{Ng74},
\citealt{OAB86}) to improve the convergence speed further. As in
\citetalias{Hennicker2018}, the Ng-acceleration is applied in every fifth
iteration step in order to use independent extrapolations.
%
%
\subsection{Parallelization and timing}
\label{subsec:timing}
To minimize the computation time of our 3D code, we have implemented an
elaborate grid construction procedure using a non-uniform grid-spacing that
still enables a reasonably high spatial resolution (see
Appendix~\ref{subsec:grid_construction}). Furthermore, when calculating line
transitions, we have parallelized the code 
using \textsc{OpenMP} as in \citetalias{Hennicker2018}. The parallelization is
implemented over the frequency grid. We note that \textsc{OpenMP} creates a
local copy of the 3D arrays representing the intensity and the
(nearest-neighbour) $\matown{\Lambda}$-matrix. With 27
$\matown{\Lambda}$-matrix elements (per spatial grid point) included for the
ALO calculations, the (spatial) resolution becomes therefore memory limited. A
typical resolution of $\nx=\ny=\nz=93$, however, is still feasible, and gives
reasonable results. For the models calculated in
Sect.~\ref{subsec:spherical_wind}, typical computation times are $t_{\rm
  SC}^{\rm (linear)}\approx 2\,\rm{h}$ and $t_{\rm SC}^{\rm
  (B\acute{e}zier)}\approx 6\,\rm{h}$ per iteration when applying the 3D SC
methods on an \textsc{Intel Xeon X5650 (2.67 GHz)} machine with 16 CPUs. As a
reference, the FVM from \citetalias{Hennicker2018} `only' required roughly 50
minutes. A more meaningful comparison, however, is the computation time per
iteration, per CPU, and per angular and frequency grid point using the same
spatial grids for all methods. For an equidistant grid with $\nx=\ny=\nz=71$
grid points, we find computation times of $t_{\rm FVM}\approx 0.037 \seconds$,
$t_{\rm SC}^{\rm (linear)} \approx 0.138 \seconds$, and $t_{\rm SC}^{\rm
  (B\acute{e}zier)}\approx 0.448 \seconds$. Thus, the computation times of the
3D SC methods using linear/B\'ezier interpolations are increased by a factor
of roughly four/twelve, when compared to the 3D FVM. These differences
originate from the computationally more challenging upwind/downwind
interpolations, the integration of the discretized equation of radiative
transfer on (possibly) refined grids along a given ray, and from the
calculation of an ALO including 26 neighbouring elements (instead of the 6
direct neighbours as used for the 3D FVM).
%
%
\section{Spherically symmetric models} \label{sec:err_analysis}
With the numerical tools developed in Sect.~\ref{sec:numerical_methods}, we
are able to tackle 3D continuum and line scattering problems for arbitrary
velocity fields. In the following, we discuss the performance of the code when
applied to spherically symmetric test models. We compare the solutions
obtained from the 3D SC method using linear and B\'ezier interpolations
(hereafter denoted by SClin and SCbez, respectively), with those obtained from
the 3D FVM and from accurate 1D solvers\footnote{The 1D solution for the
  continuum transport is found from the Rybicki-algorithm (combined with the
  solution of the moment equations using variable Eddington factors, see,
  \eg~\citealt{mihalasbook78}). To calculate the line, a comoving-frame
  ray-by-ray solution scheme in pz-geometry is applied, ensuring convergence
  with a diagonal ALO.}. Although we are using the same models as in
\citetalias{Hennicker2018}, the FVM solutions might differ slightly from those
presented in this paper, since we are using a different grid, optimized for
the 3D SC method.
%
%
\subsection{Searchlight-beam test}\label{subsec:searchlight}
\begin{figure}[t]
\begin{minipage}{9cm}
   \resizebox{\hsize}{!}{\includegraphics{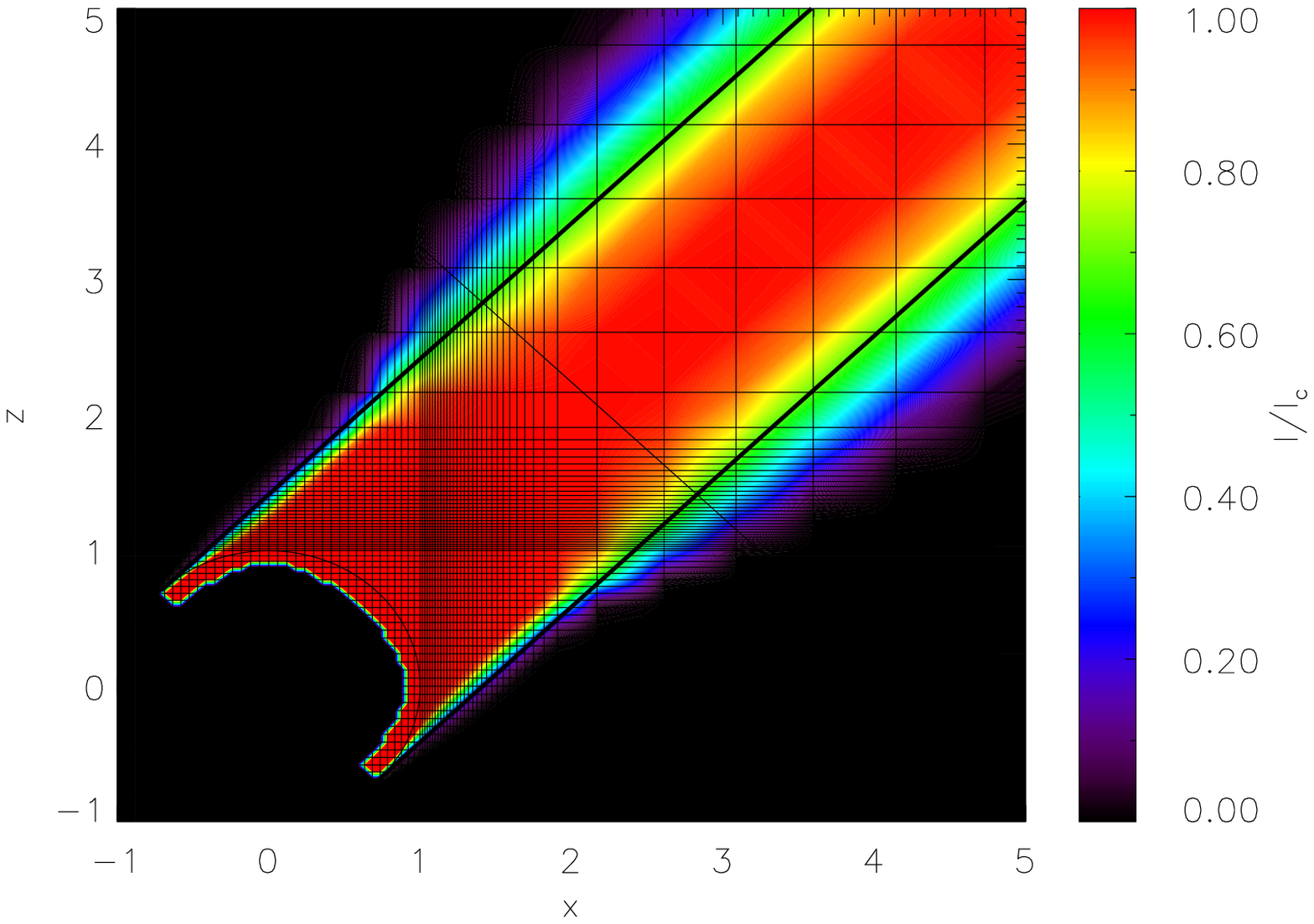}} 
\end{minipage}
\begin{minipage}{9cm}
   \resizebox{\hsize}{!}{\includegraphics{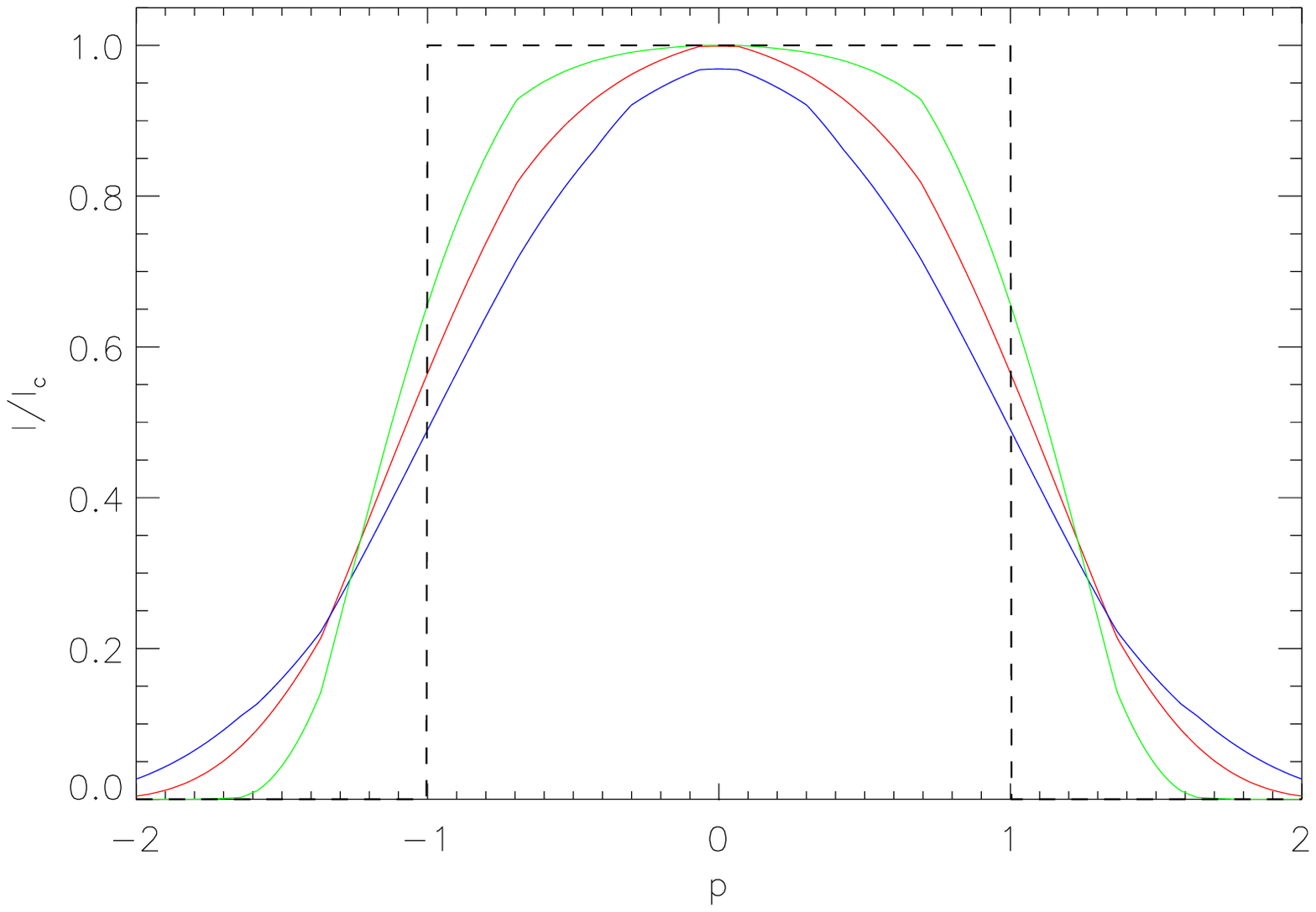}}
\end{minipage}
\begin{minipage}{9cm}
   \resizebox{\hsize}{!}{\includegraphics{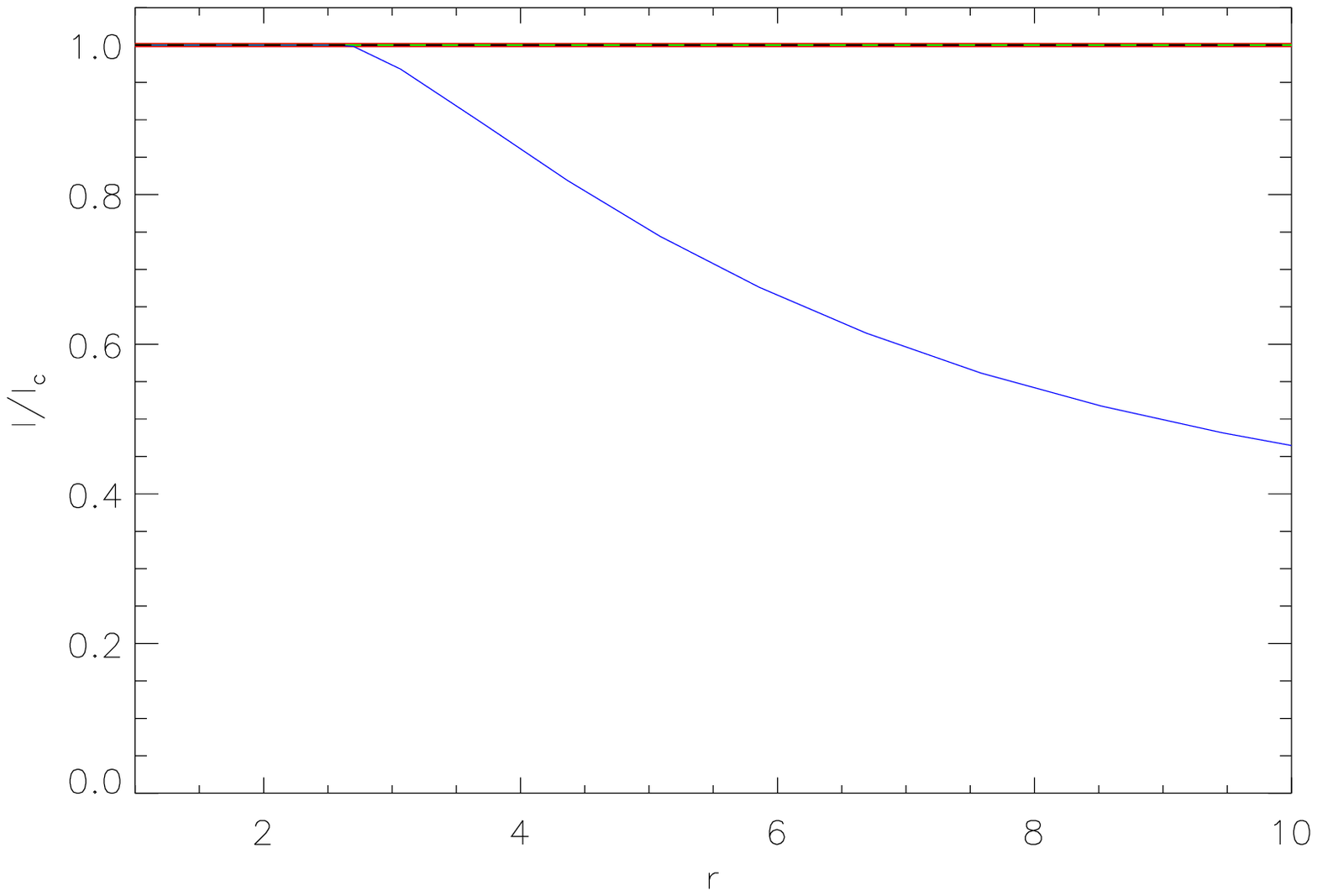}}
\end{minipage}
%
\caption{Searchlight-beam test for direction $\vecown{n}=(1,0,1)$ and a
  typical grid with $\nx=\ny=\nz=133$ grid points. Upper panel: Contour plot
  of the specific intensity as calculated with the SC method using B\'ezier
  interpolations in the $xz$-plane (cf.~\citetalias[Fig. 3]{Hennicker2018},
  for the finite-volume method). Middle panel: Specific intensity through the
  perpendicular area indicated by the straight line in the upper panel. The
  blue, red, and green profiles correspond to the FVM, SClin, and SCbez methods,
  respectively. The dashed line indicates the theoretical profile. Bottom
  panel: As middle panel, but along the centre of the searchlight beam. We
  note that the SC methods reproduce the exact solution at the centre of the
  beam, whereas the FVM solution decreases significantly for
  $r\gtrsim2.5\,\Rstar$.}
\label{fig:searchlight}
\end{figure}
\begin{figure}[t]
\resizebox{\hsize}{!}{\includegraphics{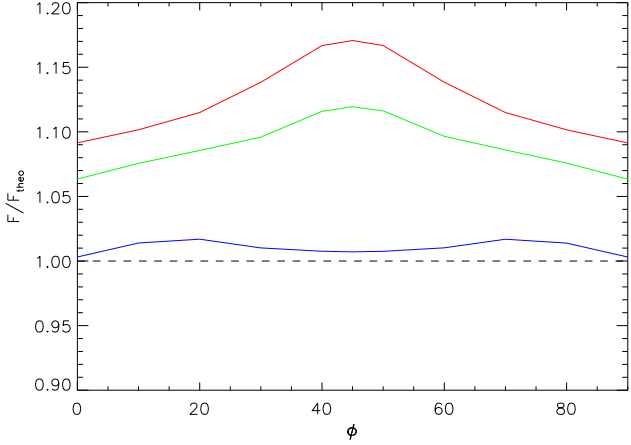}} 
%
%
\caption{Photon flux as a function of direction angle $\phi$ (with fixed
  $\theta=45^\circ$) through corresponding perpendicular areas, and with the
  opacity set to zero. The central distance of all areas to the stellar
  surface has been set to $2\,\Rstar$. The same colour coding as in
  Fig.~\ref{fig:searchlight} has been used.}
\label{fig:searchlight_fluxem}
\end{figure}
A first test of our 3D SC methods is the searchlight-beam test
(\eg~\citealt{Kunasz88}). Within this test, we set the opacity to zero and
consider the illumination of the atmosphere by a central star for a single
direction. For consistency (cf.~\citetalias{Hennicker2018}), the direction
vector corresponds to $\theta=45^\circ$, $\phi=0^\circ$. Since the discretized
equation of radiative transfer,
Eq.~\eqref{eq:eqrt_disc}/\eqref{eq:eqrt_disc2}, reduces to
$I_{\indx{ijk}}=I_{\indx{ijk}}^{(u)}$, this test extracts the effects of the
applied interpolation schemes for the upwind intensity.  The upper panel of
Fig.~\ref{fig:searchlight} shows the propagation of the specific intensity
scaled by $I_{\rm c}$ in the $xz$-plane. Due to the upwind interpolation, the
beam emerging from the stellar core becomes widened. These interpolation
errors are connected with numerical diffusion, and could only be avoided by
applying a LC method.  To obtain a quantitative measure of this effect, the
lower and middle panel of Fig.~\ref{fig:searchlight} display the specific
intensity along the given direction at the centre of the beam, and the
specific intensity through a circular area perpendicular to the ray direction
as a function of impact parameter $p$. The corresponding exact solutions are
given by a constant and rectangular function, respectively.

Along the beam centre, the SC methods perfectly reproduce the exact solution,
whereas the FVM solution decreases significantly due to the finite grid-cell
size (see \citetalias{Hennicker2018}). Considering the intensity through the
perpendicular area, both SC methods perform better than the FVM, with slight
advantages of the SCbez method when compared with the SClin method. Within the
3D SC methods, however, energy conservation is violated for our zero opacity
models, because the (nominal) specific intensity jumps from $I_{\rm c}^+$ to
zero for rays intersecting the stellar surface or not, due to the core-halo
approximation. As a consequence, almost all interpolations (and interpolation
schemes) overestimate the specific intensity\footnote{In contrast, the number
  of photons entering and leaving a given grid cell is (nearly) conserved
  within the FVM by definition. This statement, however, is not completely
  true for the FVM as formulated by \cite{Adam90}, \cite{Lobel08},
  \cite{Hennicker2018}, since all these authors apply an (averaged) upwind
  approximation.}. For optically thick models (where $I_{\rm c}^+$ at the core
plays a negligible or minor role), this effect should decrease though.  The
associated error can be quantified by calculating the corresponding flux,
that is, by integrating the specific intensity for a given direction over a
corresponding perpendicular area (defined as a circle with virtually infinite
radius). We emphasize that the flux as defined here constitutes the most
demanding test case, and should not be confused with the flux density (\ie~the
first moment of the specific intensity). Fig.~\ref{fig:searchlight_fluxem}
shows the resulting fluxes (normalized by the nominal value) for searchlight
beams with different directions defined by $\theta=45^\circ$ and $\phi \in
[0^\circ,90^\circ]$. For different directions $\phi$, the searchlight beams
propagate through different domains of the spatial grid with accordingly
different grid-cell sizes. Due to the distinct behaviour of numerical
diffusion errors within these domains, the total flux varies as a function of
$\phi$. Overall, the total fluxes for the SClin and SCbez methods are larger
than theoretically constrained, whereas the FVM gives (despite a small error)
reasonable results. This effect is largest in regions far from the star and
for diagonal directions. Thus, particularly in these regions, also the mean
intensities (for optically thin atmospheres) are expected to be
overestimated. The same problem arises when calculating line transitions,
since photons may freely propagate over large distances before a resonance
region is hit. Numerical diffusion errors can only be avoided by increasing
the grid resolution, or using higher order upwind-interpolation methods.
%
%
\subsection{Spherically symmetric stellar winds}\label{subsec:spherical_wind}
In this section, we test the performance of the 3D SC method when applied to
spherically symmetric, stationary atmospheres. For consistency, the same test
models as in \citetalias{Hennicker2018} have been calculated, with the wind
described by a $\beta$-velocity law and the continuity equation:
\beqa
\nonumber
v(r) &=& \vinf \Bigl(1 - b \dfrac{\Rstar}{r} \Bigr)^{\beta} \\\nonumber
b &=& 1-\Bigl(\dfrac{\vmin}{\vinf}\Bigr)^{1/\beta} \nonumber \\\nonumber
\rho(r) &=& \dfrac{\mdot}{4 \pi r^2 v(r)} \,.
\eeqa
For stellar and wind parameters, $\Rstar=19\,\rsun$, $\mdot=5\Mdu$, $\beta=1$,
$\vmin=10\,\kms$, $\vinf=2000\,\kms$, the density stratification and the
velocity field are completely determined.  For the considered scattering
problems ($\epsc=\epsl=10^{-6}$), effects of the temperature stratification
are negligible. The continuum and (frequency integrated) line opacities have
been calculated from Eqs.~\eqref{eq:opac} and \eqref{eq:opal}, with the electron
density derived for a completely ionized H/He plasma with helium abundance
$N_{\rm He}/N_{\rm H}=0.1$. We have calculated three different continuum
models by scaling the opacity with $\kcont=[1,10,100]$, respectively. These
models correspond to an optically thin, marginally optically thick, and
optically thick atmosphere, with radial optical depths
$\tau_r=[0.17,1.7,17]$. The line transport has been calculated for a weak,
intermediate, and strong line, with line-strengths $\kline=[1,10^3,10^5]$. To
minimize the computation time, we use a microturbulent velocity $v_{\rm micro}
= 100$~\kms~throughout this paper. Such a large velocity dispersion mimicks
the effects of multiply non-monotonic velocity fields resulting from the
line-driven instability (\citetalias{Hennicker2018} and references
therein). The atomic mass has been set to $m_{\rm A}=12\, m_{\rm p}$. Finally,
the emergent intensity from the stellar core is calculated from
$\Trad=\Teff=40\,{\rm kK}$.
%
%
\subsubsection{Convergence behaviour} \label{subsubsec:convergence}
\begin{figure*}[t]
\resizebox{\hsize}{!}{
   \begin{minipage}{0.5\hsize}
      \resizebox{\hsize}{!}{\includegraphics{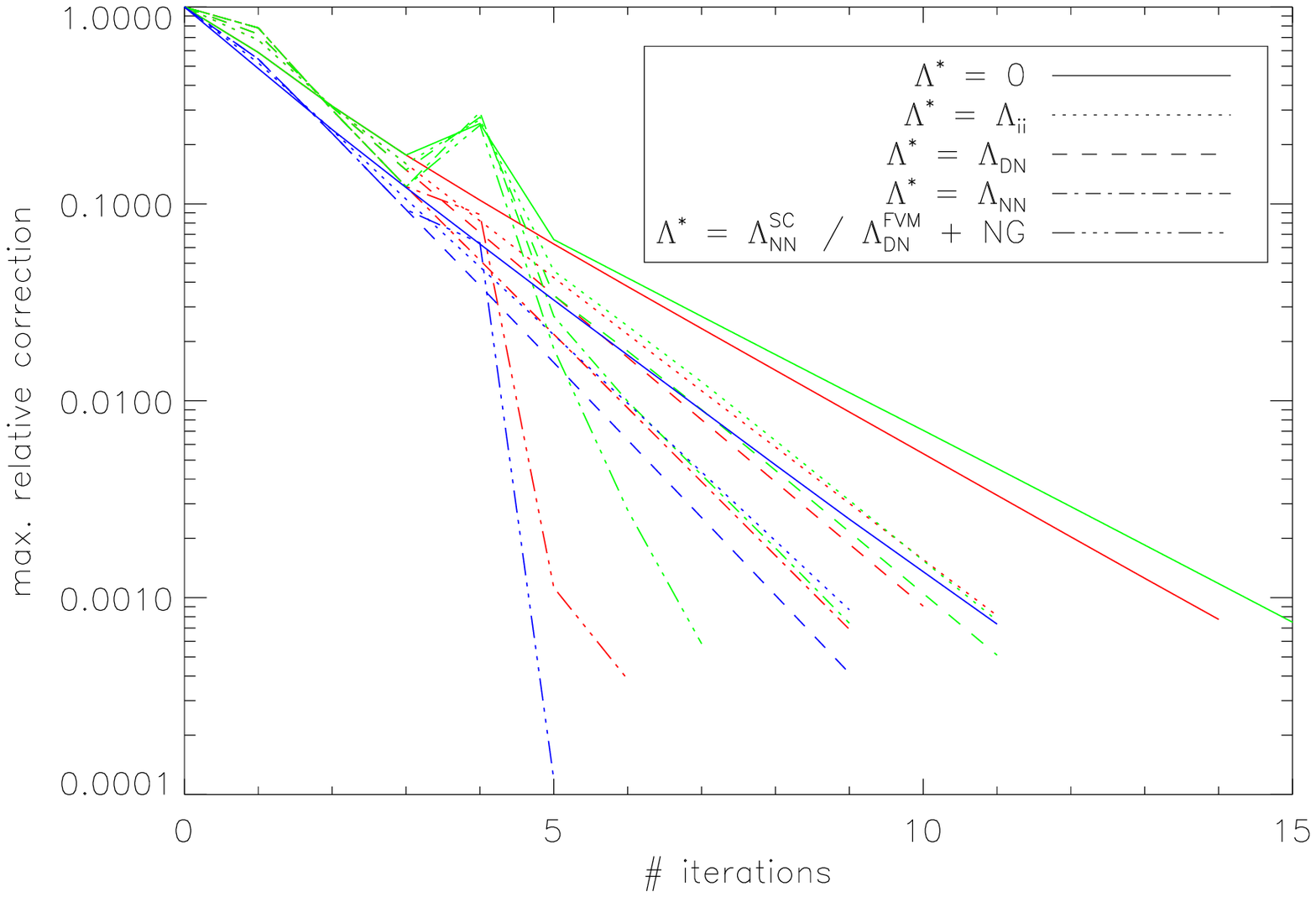}}
   \end{minipage}
   \begin{minipage}{0.5\hsize}
      \resizebox{\hsize}{!}{\includegraphics{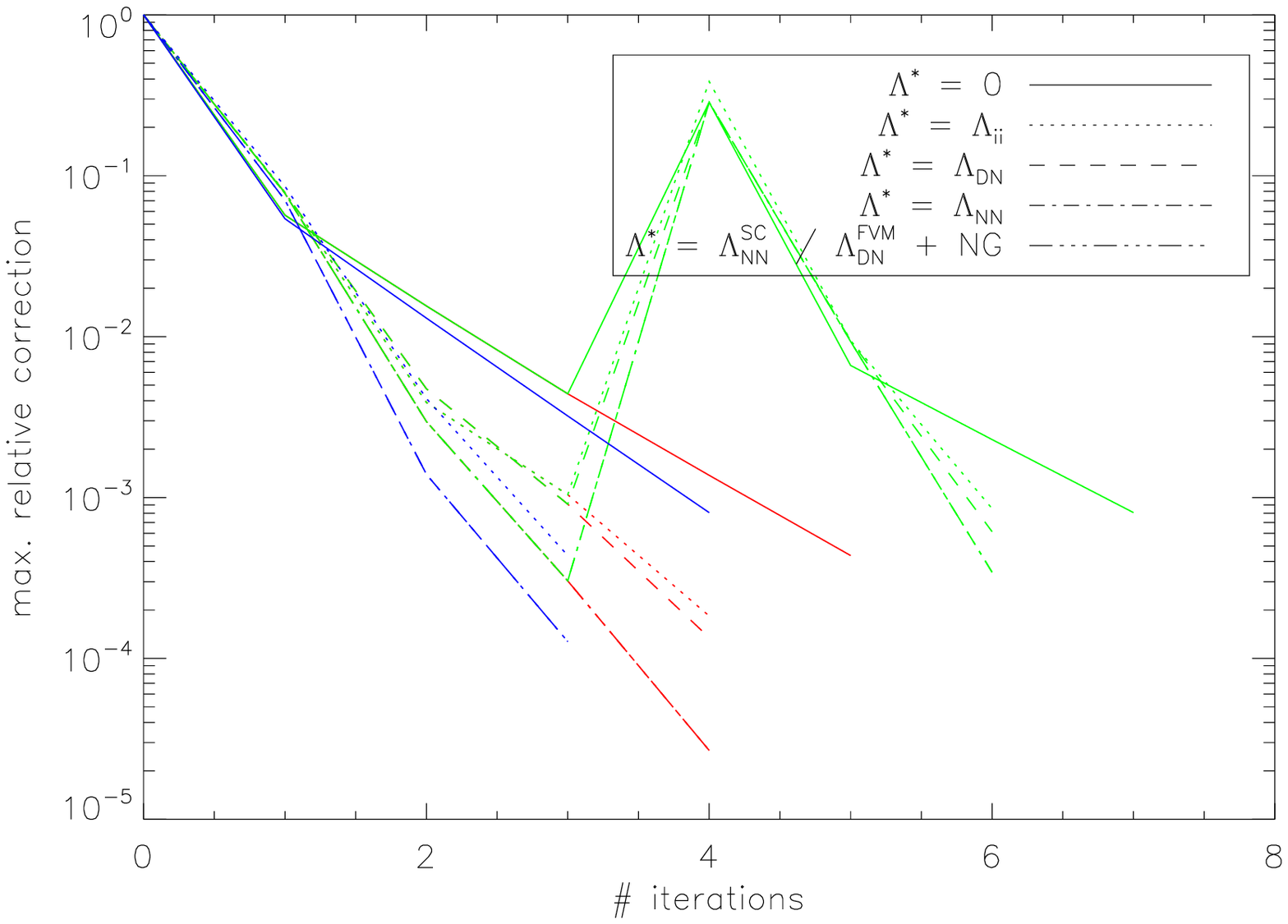}}
   \end{minipage}
}
\\
\resizebox{\hsize}{!}{
   \begin{minipage}{0.5\hsize}
      \resizebox{\hsize}{!}{\includegraphics{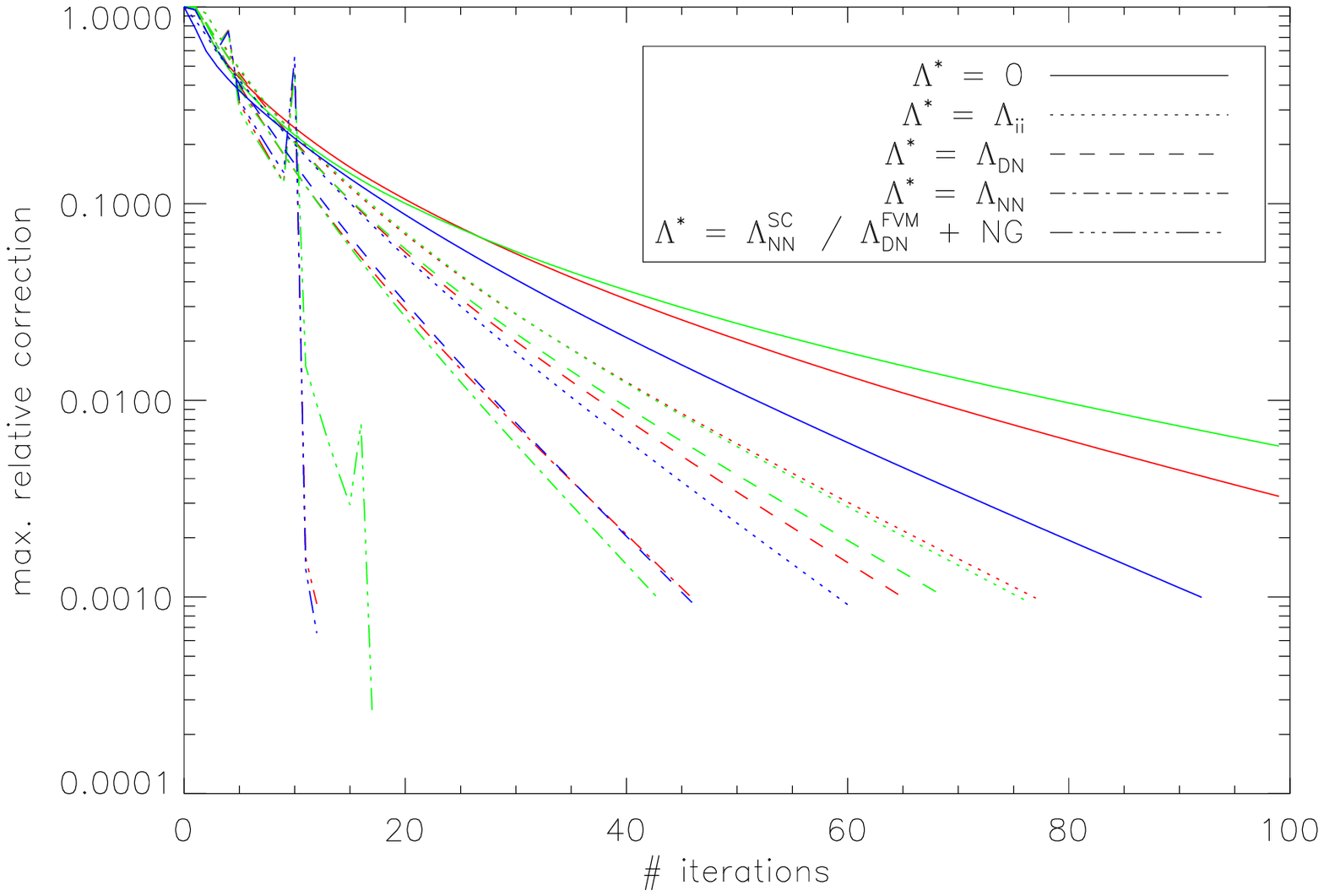}}
   \end{minipage}
   \begin{minipage}{0.5\hsize}
      \resizebox{\hsize}{!}{\includegraphics{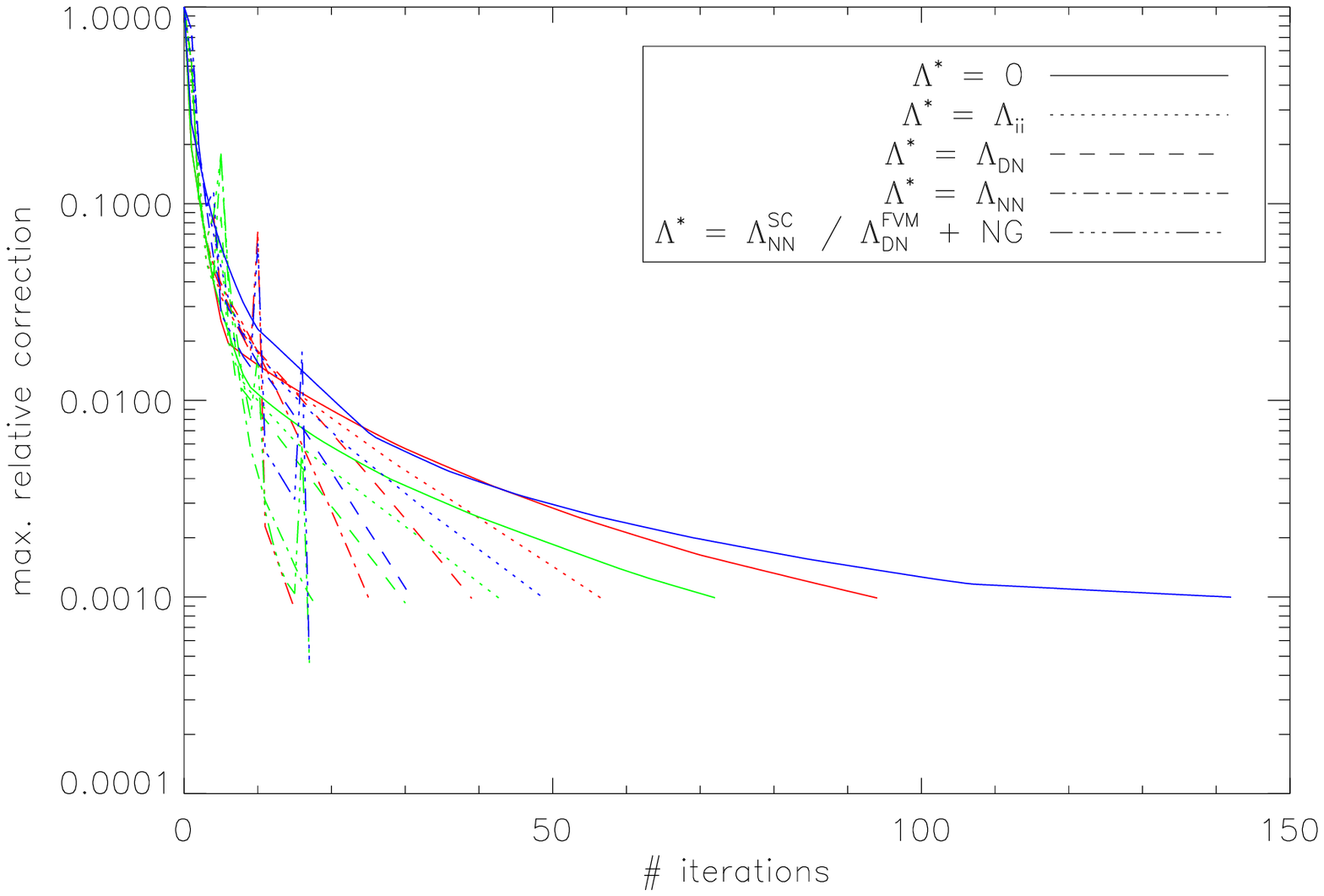}}
   \end{minipage}
}
%
%
\caption{Convergence behaviour of the standard spherically symmetric model
  calculated with the 3D FVM (blue) and the 3D SC method using linear (red) or
  B\'ezier (green) interpolations. The left and right panels show the
  convergence behaviour of the continuum and line transfer for
  $\epsc=\epsl=10^{-6}$, respectively. While the upper row displays the
  optically thin models with $\kcont=10$ and $\kline=10$, the lower row has
  been calculated using $\kcont=100$ and $\kline=10^5$. Different acceleration
  techniques have been applied, where the NN-ALO is implemented only for the
  SC method.}
\label{fig:convergence}
\end{figure*}
Fig.~\ref{fig:convergence} shows the maximum relative corrections of the mean
intensity (left panel) and line source function (right panel) after each
iteration step. Different methods (SClin, SCbez, and FVM), and different
acceleration techniques (classical $\Lambda$-iteration, and diagonal-,
direct-neighbour-, nearest-neighbour-ALO with the Ng-extrapolation switched on
or off) have been applied. We display the continuum and line calculations for
$\kcont=[10,100]$ and $\kline=[10,10^5]$, respectively, using $\nx=\ny=\nz=93$
spatial and $\nomega=974$ angular grid points (to save computation time when
calculating the slowly converging classical $\Lambda$-iteration). In the
following, we only discuss the convergence behaviour of the SClin and SCbez
methods, as the FVM has already been analysed in
\citetalias{Hennicker2018}. We usually stop the iteration scheme when the
maximum relative corrections become less than $10^{-3}$ between subsequent
iteration steps, emphasizing that a truly converged solution is only found
when the curve describing subsequent relative corrections is sufficiently
steep\footnote{For linearly convergent iteration schemes, the steepness of the
  convergence curve is described by the relative corrections in subsequent
  iterations steps, $\Delta_{\indx{k}}/\Delta_{\indx{k-1}} = \rm{const.} =:
  q$. To obtain a solution within a reasonable amount of computation time, we
  may demand that $q \lesssim 0.8$, corresponding to a reduction of relative
  errors by a factor of $10^{-3}$ every 30th iteration step.}. For instance, the
classical $\Lambda$-iteration (with $\alom=0$) fails to converge for strong
scattering lines (Fig.~\ref{fig:convergence}, lower right panel), since the
relative corrections become almost constant in each iteration step (`false
convergence', cf.~\citealt{hubenybook2014}).

Overall, and as expected, the number of iterations needed to obtain the
converged solution is decreasing with increasing number of matrix elements
used to define the ALO (see also \citealt{Hauschildt1994} and \citealt{Haus06}
for multi-band ALOs coupled to a 1D-SC and 3D-LC formal solution scheme,
respectively). In most cases, the convergence of the SClin is faster than that
of the SCbez method, because the interpolation scheme is intrinsically more
localized (with stronger weights assigned to local $\matown{\Lambda}$-matrix
elements). The FVM always performs best, since only the direct neighbours
directly influence the formal solution within this method.

For parameters $\kcont,\kline \leq 10$ (Fig.~\ref{fig:convergence}, upper
panels), all ALOs yield a converged solution within $\nconv \approx 10$
iteration steps. When applying the SCbez method, the first peak results from
switching the linear interpolations to B\'ezier interpolations at the fifth
iteration step. This peak is less pronounced for parameters $\kcont,\kline =
100, 10^5$, since the maximum relative corrections are still relatively large
in the first few iteration steps.

For the optically thick continuum model (Fig.~\ref{fig:convergence}, lower
left panel), the number of iterations until convergence is reduced from
$\nconv^{\rm diag}\approx75$ to $\nconv^{\rm DN}\approx 65$ and $\nconv^{\rm
  NN}\approx 45$ when using the diagonal, DN-, and NN-ALO within the SClin and
SCbez method, respectively. The Ng-extrapolation significantly reduces
$\nconv$ further, and is required to obtain the converged solution in
$\lesssim 20$ iteration steps.

For the strong line, the convergence behaviour becomes slightly improved,
because the radiative transfer is localized to (narrow) resonance
regions. Thus, already the diagonal and DN-ALO yield a solution within
$\nconv^{\rm diag}\approx50$ and $\nconv^{\rm DN}\approx 35$ iteration
steps. Again, the NN-ALO with the Ng-extrapolation scheme switched on performs
best, and reduces the number of iteration steps until convergence to $\lesssim
20$.

In total, we conclude that a NN-ALO together with the Ng-extrapolation is
required for the SClin and SCbez methods, in order to obtain a fast
convergence behaviour of the iteration scheme. This ALO also performs
excellently for extreme test-cases, that is, for optically thick continua and
strong lines in scattering dominated atmospheres.
%
%
\subsubsection{Continuum and line solution}
%
\begin{figure*}[t]
\resizebox{\hsize}{!}{
   \begin{minipage}{0.33\hsize}
      \resizebox{\hsize}{!}{\includegraphics{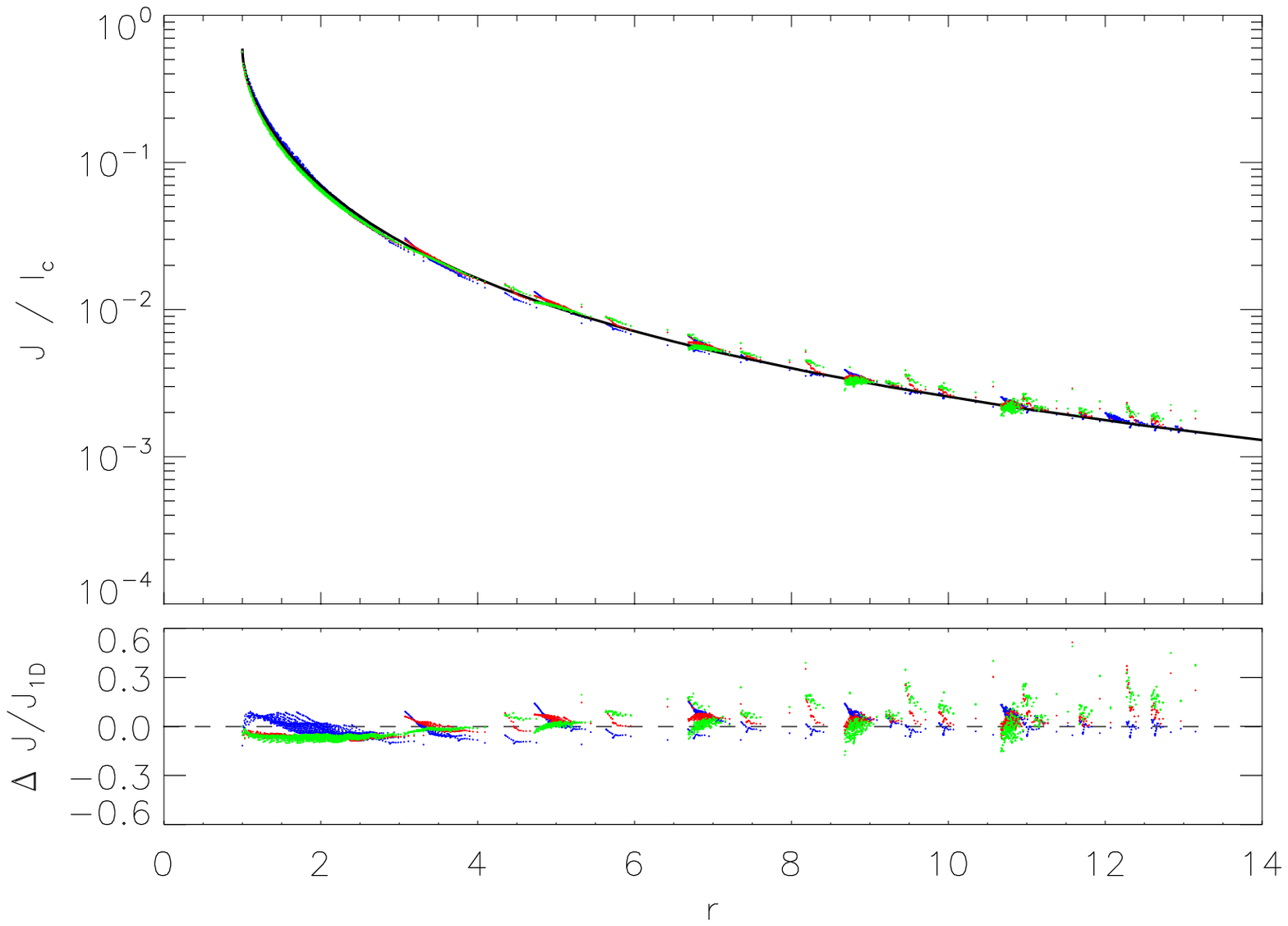}}
   \end{minipage}
   \begin{minipage}{0.33\hsize}
      \resizebox{\hsize}{!}{\includegraphics{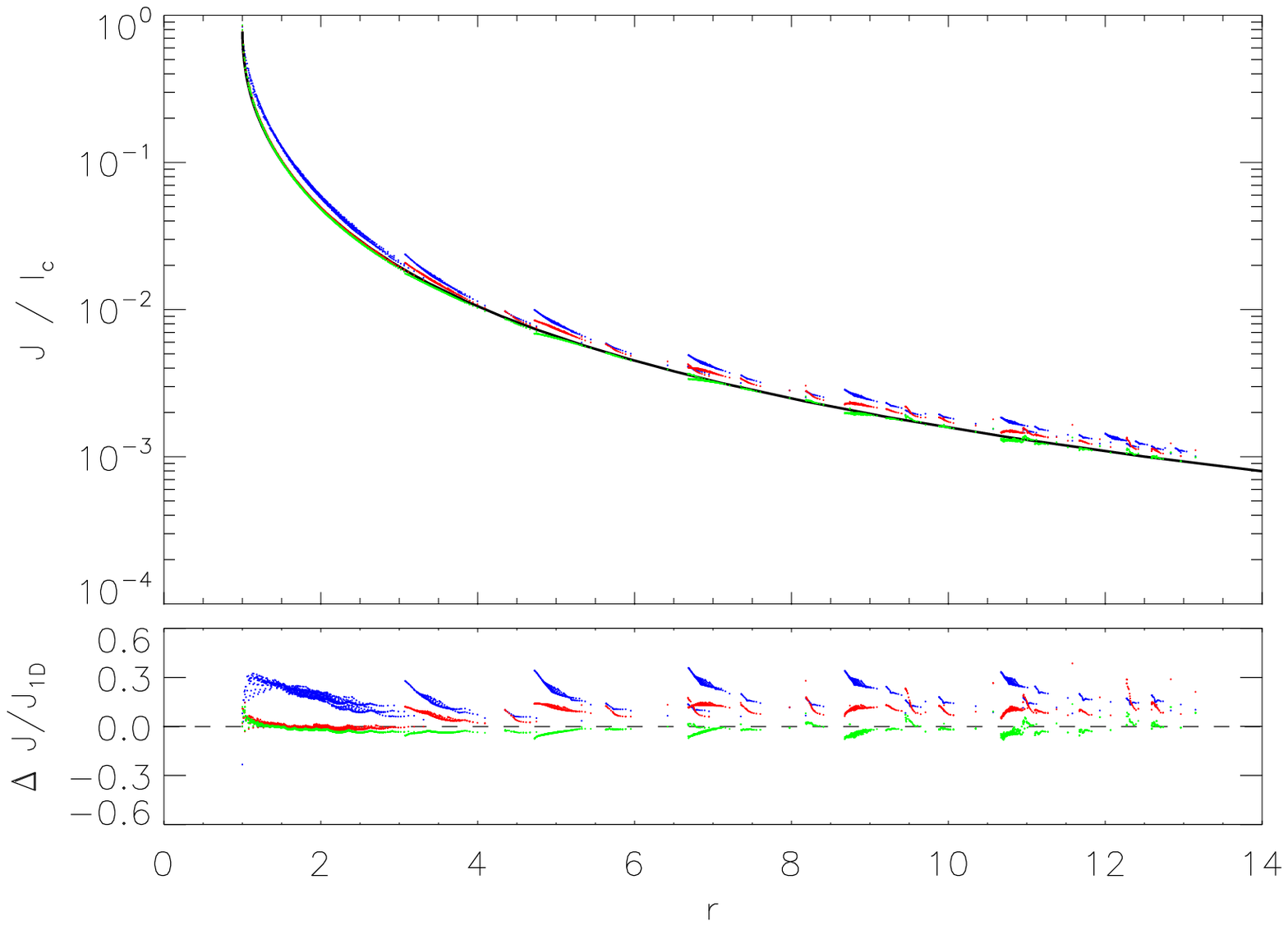}} 
   \end{minipage}
   \begin{minipage}{0.33\hsize}
     \resizebox{\hsize}{!}{\includegraphics{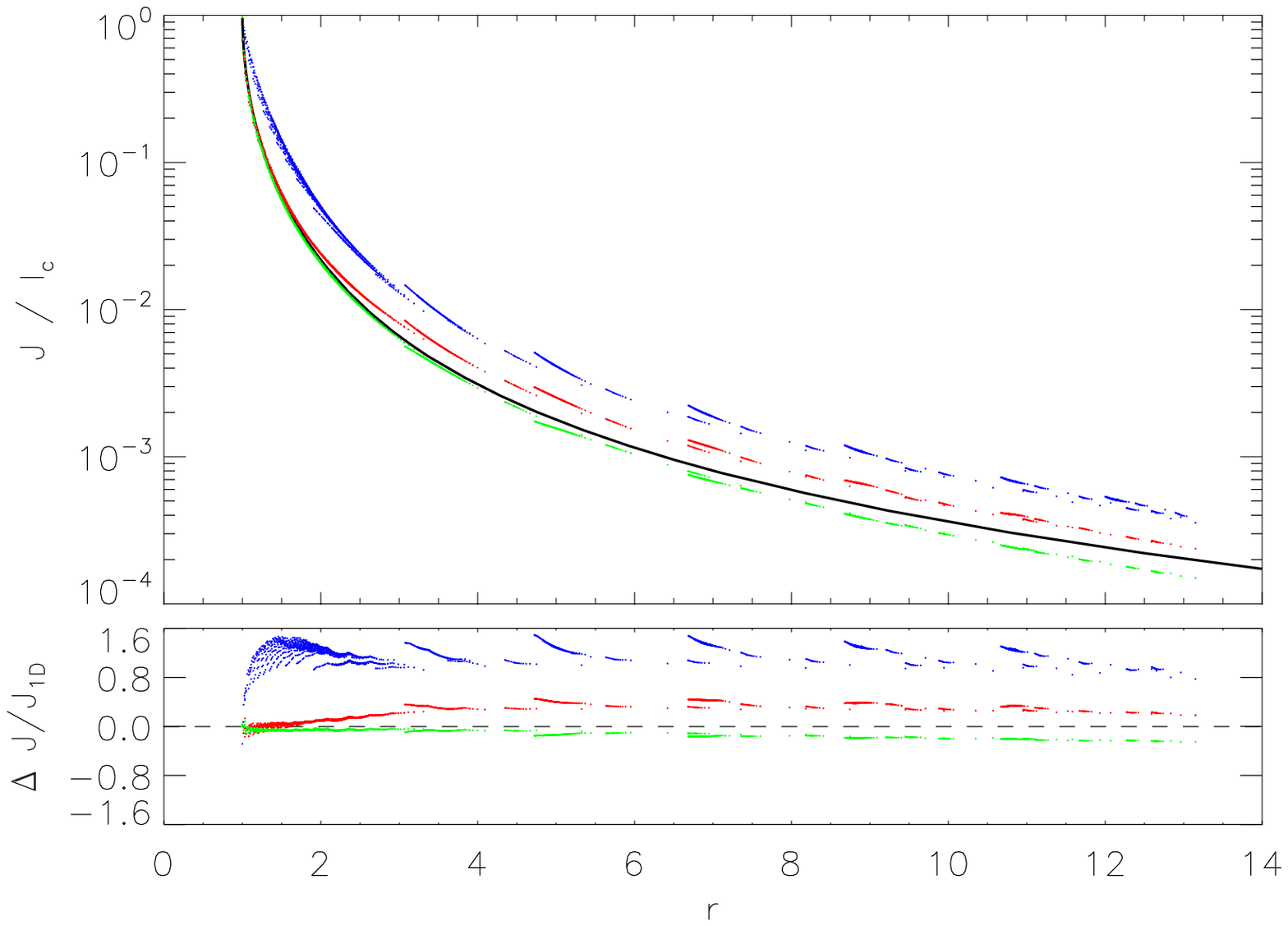}} 
   \end{minipage}
}
\\
\resizebox{\hsize}{!}{
   \begin{minipage}{0.33\hsize}
      \resizebox{\hsize}{!}{\includegraphics{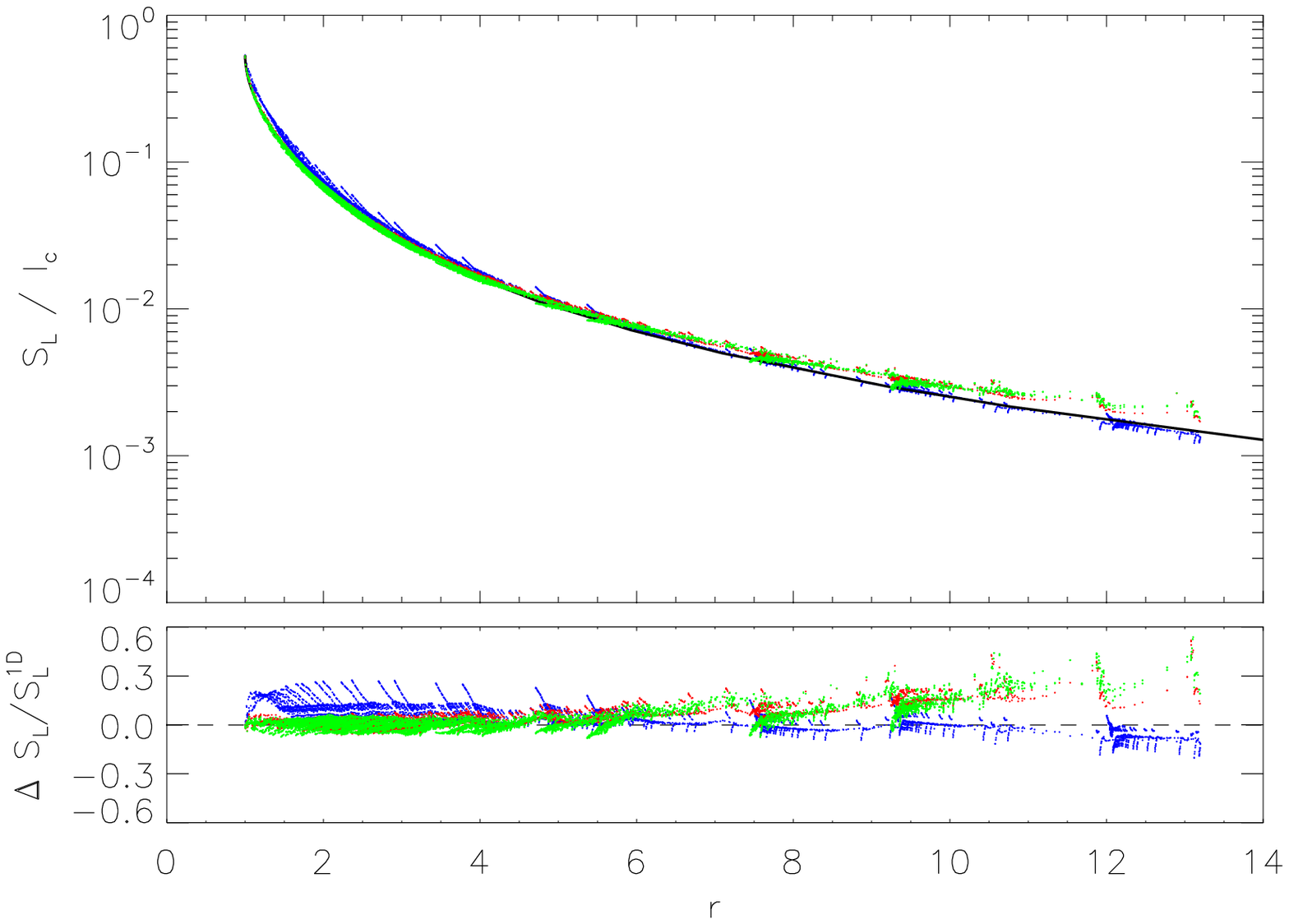}} 
   \end{minipage}
   \begin{minipage}{0.33\hsize}
      \resizebox{\hsize}{!}{\includegraphics{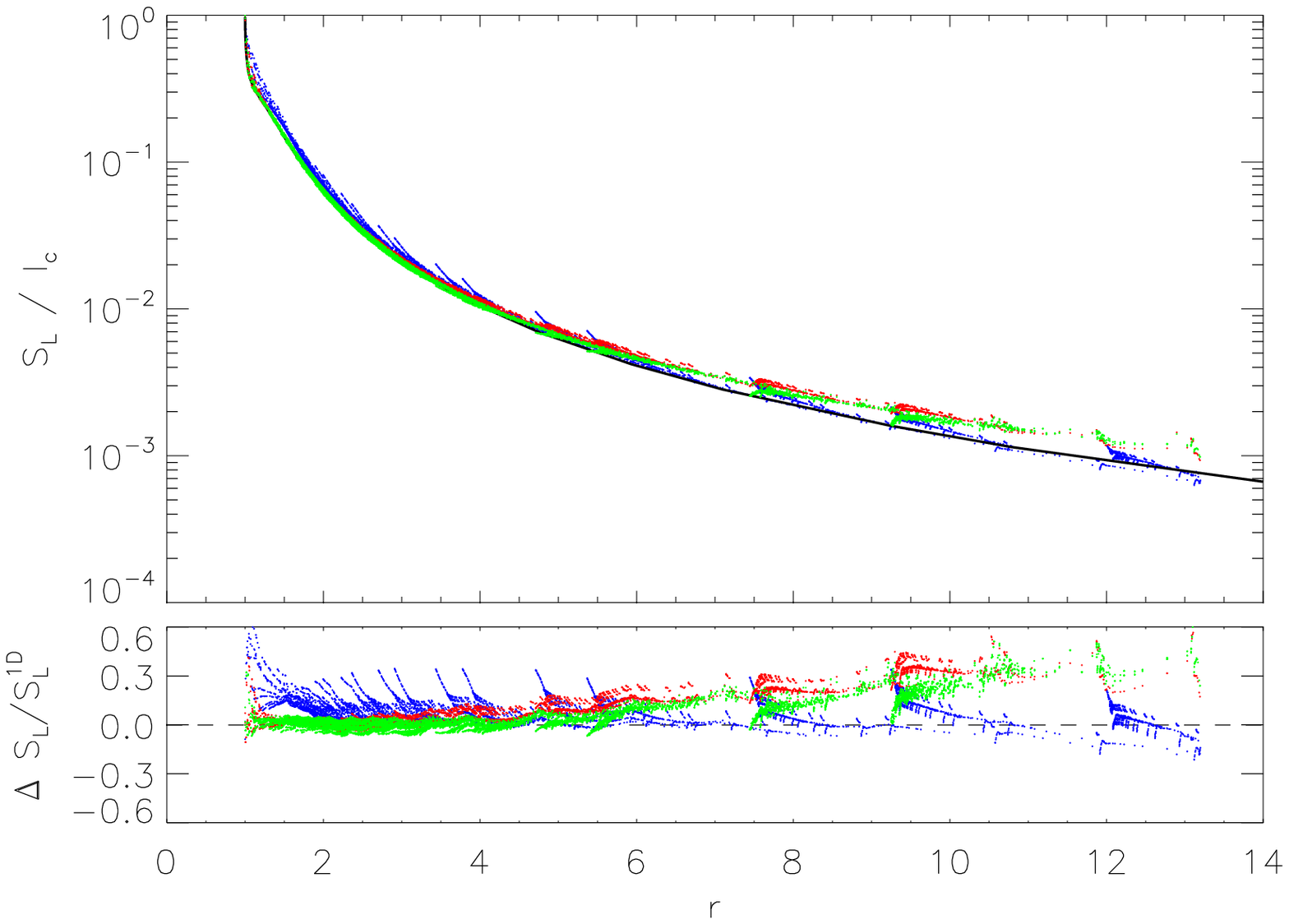}} 
   \end{minipage}
   \begin{minipage}{0.33\hsize}
      \resizebox{\hsize}{!}{\includegraphics{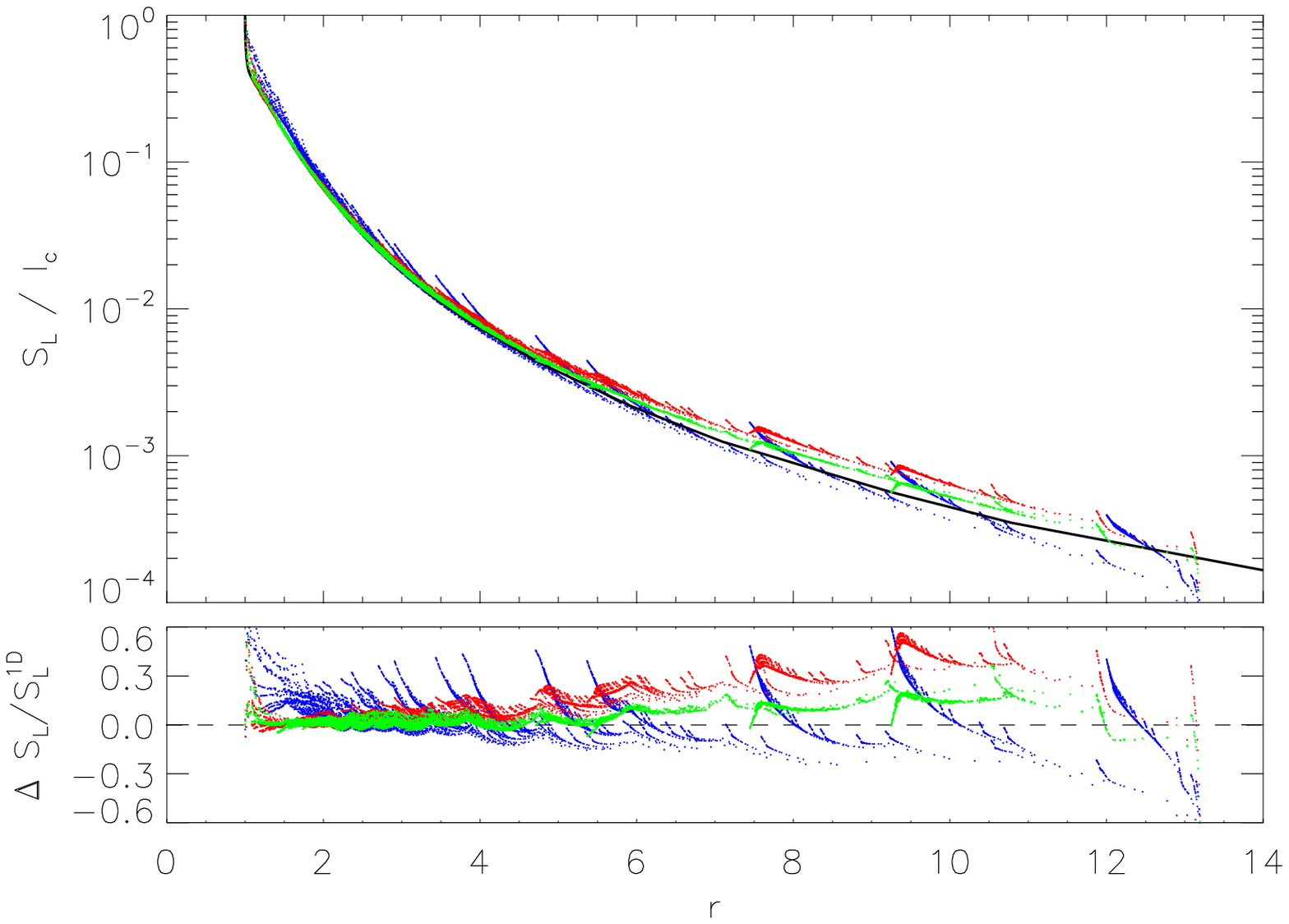}}
   \end{minipage}
}
%
\caption{Solutions for the standard spherically symmetric models as calculated
  with the 3D FVM (blue) and 3D SC methods using linear (red) or B\'ezier
  (green) interpolations, compared to an accurate 1D solution (black solid
  line). The dots represent the solutions at specific grid points (with
  different latitudes and longitudes), where only a subset of all grid points
  is displayed to compress the image. Corresponding errors are indicated at
  the bottom of each chart.  The top panel shows the mean intensity for the
  continuum transfer as a function of radius, with $\epsc=10^{-6}$, and
  $\kcont=[1,10,100]$ from left to right. The bottom panel shows the line
  source function with $\epsl=10^{-6}$, and $\kline=[10^0,10^3,10^5]$ from
  left to right.}
\label{fig:mint_odepth}
\end{figure*}
\begin{table}
\setlength{\tabcolsep}{0.5mm}
\begin{center}
\caption{Mean and maximum relative errors of the FVM and SClin, SCbez methods,
  when applied to spherically symmetric test models. The mean relative
  errors are listed for different regions with $r\in[\Rstar,3\,\Rstar]$,
  $r\in[3\,\Rstar,R_{\rm max}]$, and $r\in[\Rstar, R_{\rm max}]$, from top to
  bottom.}
\label{tab:mint_odepth}
\begin{tabular}{cccccccccc}
\multicolumn{5}{c}{$\mrerr{\Delta J}\, [\%]$ for $r\in[\Rstar,3\,\Rstar]$} & $\quad$
& \multicolumn{4}{c}{$\mrerr{\Delta \sline}\, [\%]$ for
  $r\in[\Rstar,3\,\Rstar]$} \\
\cline{1-5} \cline{7-10}
\noalign{\vskip 0.5mm}
\cline{1-5} \cline{7-10}
\noalign{\vskip 0.5mm}
\kcont & $\tau_r$ & FVM & SClin & SCbez & $\quad$ & \kline & FVM & SClin & SCbez \\
\cline{1-5} \cline{7-10}
\noalign{\vskip 0.5mm}
$10^0$ & $\,0.17\,$ & 3.4 & 6.2 & 6.6 & $\quad$ & $10^0$ & 8.7 & 2.0 & 2.4 \\
$10^1$ & $\,1.17\,$ & 18  & 1.2 & 2.2 & $\quad$ & $10^3$ & 9.4 & 2.0 & 2.6 \\
$10^2$ & $\,17.0\,$ & 120 & 9.7 & 5.7 & $\quad$ & $10^5$ & 10  & 3.3 & 2.2 \\
&&&&&&&&&\\
\multicolumn{5}{c}{$\mrerr{\Delta J}\, [\%]$ for $r\in[3\,\Rstar,R_{\rm max}]$} & $\quad$
& \multicolumn{4}{c}{$\mrerr{\Delta \sline}\, [\%]$ for $r\in[3\,\Rstar,R_{\rm max}]$} \\
\cline{1-5} \cline{7-10} 
\noalign{\vskip 0.5mm}
\cline{1-5} \cline{7-10}
\noalign{\vskip 0.5mm}
\kcont & $\tau_r$ & FVM & SClin & SCbez & $\quad$ & \kline & FVM & SClin & SCbez \\
\cline{1-5} \cline{7-10}
\noalign{\vskip 0.5mm}
$10^0$ & $\,0.17\,$ & 5.2 & 4.4 & 4.3 & $\quad$ & \makebox[3ex][c]{$10^0$} & \makebox[7ex][c]{4.8} & \makebox[7ex][c]{5.9} & \makebox[7ex][c]{4.7} \\    
$10^1$ & $\,1.17\,$ & 22  & 10  & 3.7 & $\quad$ & $10^3$ & 5.9 & 10  & 5.1  \\ 
$10^2$ & $\,17.0\,$ & 120 & 36  & 16  & $\quad$ & $10^5$ & 12  & 19  & 6.2 \\
&&&&&&&&& \\
\multicolumn{5}{c}{$\mrerr{\Delta J}\, [\%]$ for $r\in[\Rstar,R_{\rm max}]$} & $\quad$
& \multicolumn{4}{c}{$\mrerr{\Delta \sline}\, [\%]$ for $r\in[\Rstar,R_{\rm max}]$} \\
\cline{1-5} \cline{7-10} 
\noalign{\vskip 0.5mm}
\cline{1-5} \cline{7-10}
\noalign{\vskip 0.5mm}
\kcont & $\tau_r$ & FVM & SClin & SCbez & $\quad$ & \kline & FVM & SClin & SCbez \\
\cline{1-5} \cline{7-10}
\noalign{\vskip 0.5mm}
$10^0$ & $\,0.17\,$ & 4.4 & 5.2 & 5.4 & $\quad$ & $10^0$ & 6.3 & 4.5 & 3.9 \\
$10^1$ & $\,1.17\,$ & 20  & 6.0 & 3.0 & $\quad$ & $10^3$ & 7.2 & 7.0 & 4.2 \\
$10^2$ & $\,17.0\,$ & 120 & 2 3 & 11  & $\quad$ & $10^5$ & 11  & 13  & 4.7 \\
 & & & & & & & & & \\
\multicolumn{5}{c}{$\Delta J_{\rm max}\, [\%]$ for $r\in[\Rstar,R_{\rm max}]$} & $\quad$
& \multicolumn{4}{c}{$\Delta S_{\rm L, \,max}\, [\%]$ for $r\in[\Rstar,R_{\rm max}]$} \\
\cline{1-5} \cline{7-10} 
\noalign{\vskip 0.5mm}
\cline{1-5} \cline{7-10}
\noalign{\vskip 0.5mm}
\kcont & $\tau_r$ & FVM & SClin & SCbez & $\quad$ & \kline & FVM & SClin & SCbez \\
\cline{1-5} \cline{7-10}
\noalign{\vskip 0.5mm}
$10^0$ & $\,0.17\,$ & 17  & 51 & 49 & $\quad$ & $10^0$ & 27 & 46 & 45 \\
$10^1$ & $\,1.17\,$ & 36  & 39 & 14 & $\quad$ & $10^3$ & 73 & 50 & 45 \\
$10^2$ & $\,17.0\,$ & 150 & 46 & 24 & $\quad$ & $10^5$ & 70 & 65 & 55
%
%
\end{tabular}
\end{center}
\end{table}
In the following, we discuss the errors resulting from the upwind and downwind
interpolations, and from the integration of the discretized radiative transfer
equation. We apply the NN-ALO together with the Ng-extrapolation to ensure
convergence.  When calculating the line, we used $\nx=\ny=\nz=93$ spatial grid
points. Since the continuum transport has been calculated at only one frequency
point, we applied a higher grid resolution with $\nx=\ny=\nz=133$ for such
problems. Fig.~\ref{fig:mint_odepth} shows the continuum and line solutions
together with corresponding relative errors obtained for the spherically
symmetric model when calculated with the FVM, SClin, and SCbez methods, and
compared to the `exact' 1D solution. The mean and maximum relative errors are
shown for different regions in Table~\ref{tab:mint_odepth}, where the mean and
maximum relative errors of any quantity are defined throughout this paper by
\beqa
\nonumber
\mrerr{\Delta q} &:=& \dfrac{1}{N} \sum_{i=1}^N \dfrac{\lvert q_i-q_i^{\rm (exact)}\rvert}{q_i^{\rm(exact)}} \\\nonumber
\Delta q_{\rm max} &:=& \max\limits_{\forall i \in [1,N]} \dfrac{\lvert q_i-q_i^{\rm
    (exact)}\rvert}{q_i^{\rm (exact)}} \,,
\eeqa
with $N$ the number of grid points within the considered region.

For the continuum models, the solutions obtained from the 3D SC methods are
superior to the solution obtained from the FVM in most cases. Particularly
near the stellar surface (at $r\lesssim 3\,\Rstar$), both SC methods are in
good agreement with the 1D solution (see Fig.~\ref{fig:mint_odepth}, top
panel, and bottom of each chart for the radial dependence of the relative
errors). When considering the most challenging problem of optically thick,
scattering dominated atmospheres, the mean relative errors of the SClin and
SCbez method for the complete calculation region are on the $20$- and
$10\,\%$-level, respectively. For such models, the FVM breaks down due to the
order of accuracy (see \citetalias{Hennicker2018}), and a (high order) SC
method is indeed required to solve the radiative transfer with reasonable
accuracy. For marginally optically thick continua, the mean relative errors of
the SClin and SCbez methods are on the $5\,\%$-level and below. While the FVM
allows for a qualitative interpretation of the radiation field for such
models, the SC methods should be used for quantitative discussions.  The
optically thin model calculations give mean relative errors of the order of
$5\,\%$ for all methods, with the maximum relative error being lowest for the
FVM. Since, additionally, the computation times of the SClin and SCbez methods
are typically highest (see Sect.~\ref{subsec:timing}), the FVM is to be
preferred when calculating optically thin continua. We note that all errors
originate from the interplay between upwind and downwind interpolations of
opacities, source functions, and intensities, and from the integration of the
discretized radiative transfer equation. Numerical diffusion and the
associated violation of energy conservation influences the converged solution
particularly in the optically thin regime.

The mean relative errors for the line transition are of the order of
$5-10\,\%$, with increasing accuracy from strong to weak lines, and slight
advantages of the SCbez method when compared to the SClin method.  The radial
stratification of relative errors for each considered line is shown in the
bottom panel of Fig.~\ref{fig:mint_odepth}, bottom of each chart. While the
FVM gives largest errors near the stellar surface (at $r\lesssim 3\,\Rstar$),
both SC techniques are in excellent agreement with the 1D solution in such
regions.  At larger radii, however, the SC solutions are generally
overestimated when compared to the 1D solution due to numerical diffusion
errors. The distinct behaviour of the applied solution schemes in different
atmospheric regions finally determines the quality of emergent flux profiles.
%
%
\subsubsection{Emergent flux profile}
\begin{figure}[t]
\resizebox{\hsize}{!}{\includegraphics{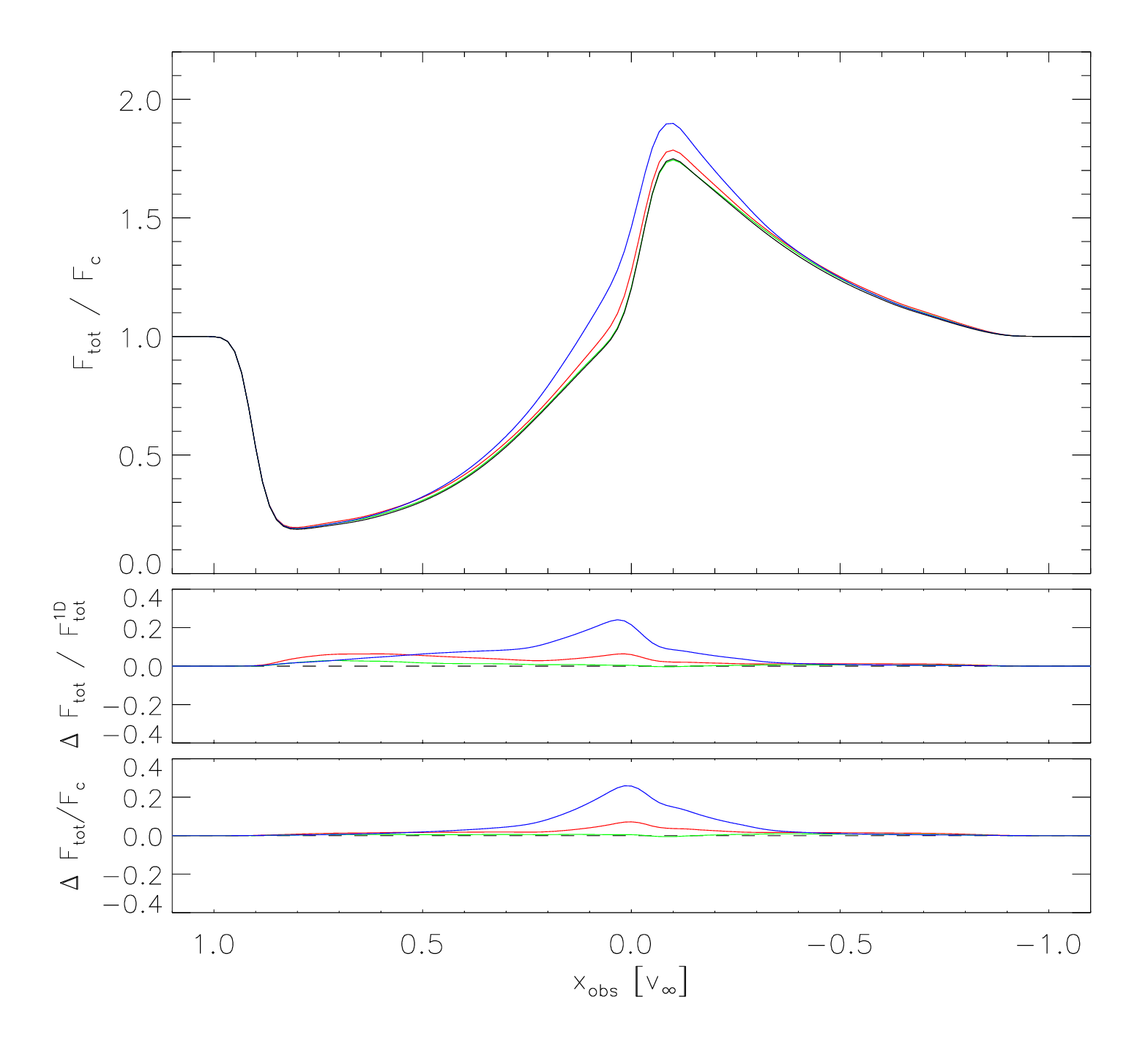}}
\\
\resizebox{\hsize}{!}{\includegraphics{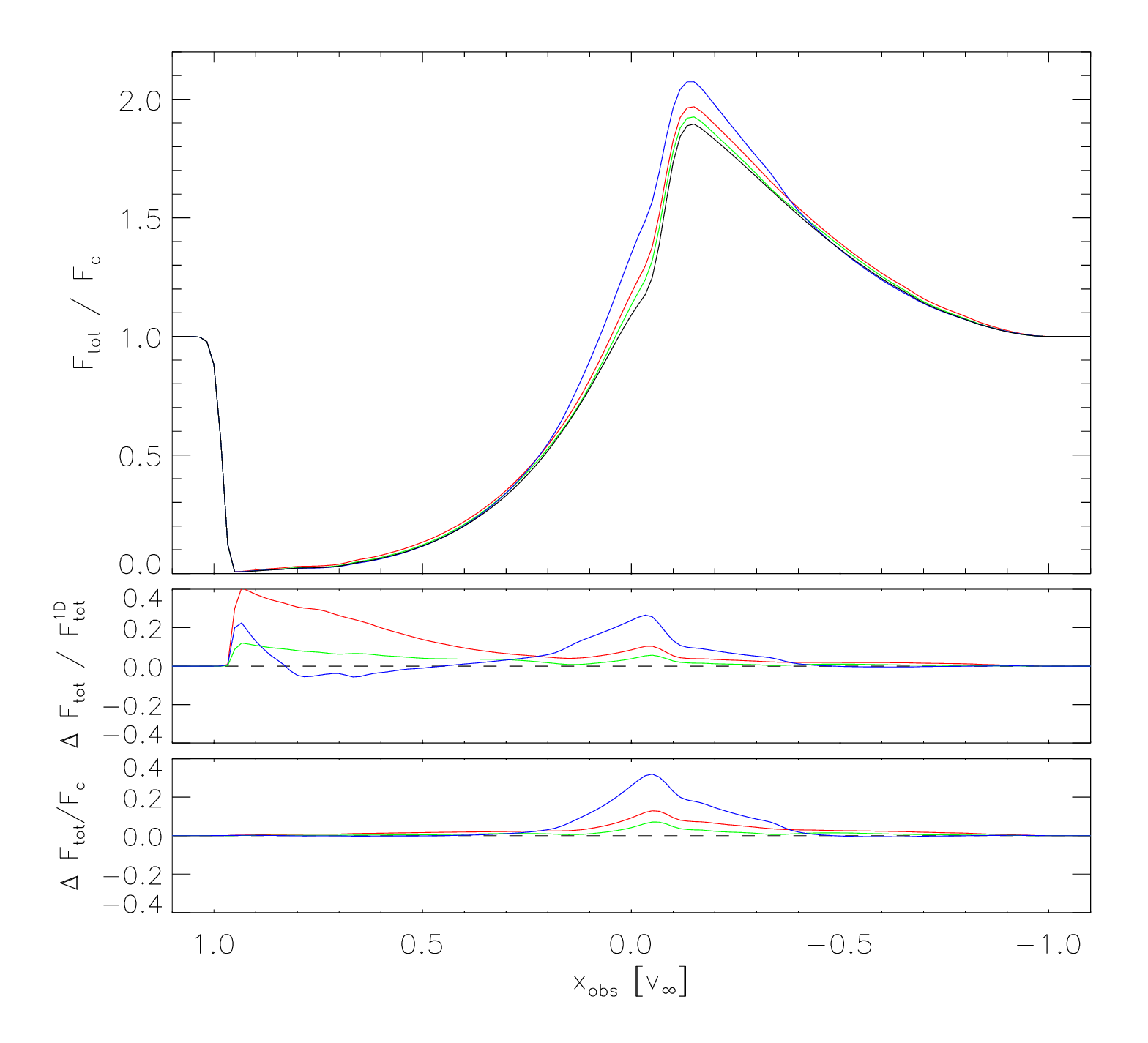}}
%
\caption{Emergent flux profiles of an intermediate ($\kline=10^3$, top panel)
  and strong ($\kline=10^5$, bottom panel) line. The blue, red, and green
  curves correspond to the solution of the FVM, SClin, and SCbez methods,
  respectively. The reference profile (black solid line) has been derived from
  the `exact' 1D source function interpolated onto the 3D Cartesian
  grid. Corresponding relative and absolute errors are shown at the bottom of
  each chart. For all profiles, the continuum level has been determined from a
  zero-opacity model.}
\label{fig:line_profiles}
\end{figure}
The converged source functions are used to calculate the emergent flux profile
using the same postprocessing LC solver as in
\citetalias{Hennicker2018}. Based on \cite{Lamersetal87}, \cite{Busche2005},
\cite{Sundqvist12c}, this method solves the equation of radiative transfer in
cylindrical coordinates with the $z$-axis being aligned with the line of sight
of the observer's direction under consideration. All quantities required on
the rays are found by trilinear interpolation from the 3D grid, and the
equation of radiative transfer is integrated using linear interpolations. To
extract the error resulting from the FVM and SC methods alone, we interpolated
the `exact' 1D solution onto our 3D grid and calculated the reference profile
using the same technique. The continuum has been calculated from a
zero-opacity model given by the unattenuated illumination from the projected
stellar disc. Then, the differences of line profiles are exclusively related
to the differences of line source functions. Fig.~\ref{fig:line_profiles}
shows the line profiles with corresponding absolute and relative errors for
the intermediate ($\kline=10^3$) and strong ($\kline=10^5$) line, obtained
from the converged source functions from above.

The line profiles are in good agreement with the 1D solution for both applied
SC methods, with slight advantages of the SCbez method when compared to the
SClin. Major (relative) deviations arise particularly at large frequency
shifts on the blue side due to the enlarged source functions in corresponding
resonance regions (\ie~at large radii in front of the star). At such frequency
shifts, however, the line profile is mainly controlled by absorption, and the
absolute error remains small. At low frequency shifts, the emission peak
becomes slightly overestimated, particularly when considering the strong
line. The corresponding resonance regions are mainly located near to the star
(at low absolute velocities) and in the whole plane perpendicular to the line
of sight (with low projected velocities). For the intermediate line, the
emission from this plane at large radii only plays a minor role due to
relatively small optical-depth increments along the line of sight. Thus, both
SC methods are in excellent agreement with the 1D reference profile. With
increasing line strength, however, the emission from the outer wind region
contributes significantly to the line formation, and the discrepancies between
the 1D and the SClin/SCbez methods become more pronounced. For all test
calculations, the B\'ezier method performs best, closely followed by the SClin
method, and (far behind) the FVM.

With Fig.~\ref{fig:line_profiles} and the argumentation from above, we
conclude that (at least) a short-characteristics solution scheme is required
to enable a quantitative interpretation of ultra-violet (UV) resonance line
profiles, where both the linear and B\'ezier interpolation techniques perform
similarly well. The less accurate (however computationally cheaper) FVM can
still be applied for qualitative discussions.
%
%
\section{Rotating winds} 
\label{sec:rotation} 
As a first application of the 3D SC method to non-spherical atmospheres, we
consider the UV resonance line formation in the winds of (fast) rotating O
stars. Fast rotation has two immediate consequences on the stellar geometry
and wind structure. Firstly, the surface of any rotating star becomes distorted,
with $R_{\mathrm{eq}}/R_{\mathrm{pole}}$ approaching 3/2 for rotational speeds
near the critical velocity (\citealt{Collins63} assuming a Roche model, and
the critical velocity defined by Eq.~\eqref{eq:omega} for $\Omega=1$).
Secondly, the emergent flux depends on the (local) effective gravity (corrected for the
centrifugal acceleration), and thus, decreases towards the equator (`gravity
darkening', see \citealt{Zeipel24}, and \citealt{MaederIV}, \citealt{MaederVI}
for uniform and shellular rotation, respectively).  The first attempt to model
the winds of fast rotating OB stars was made by \cite{BC93}. These authors
considered a purely radial line force, and neglected gravity darkening and the
surface distortion. Within these approximations, a `wind compressed disc' is
formed in the equatorial plane. \cite{CO95} and \cite{Owocki96} included the
effects of non-radial line-forces into their 2D radiation-hydrodynamic
simulations, and showed that the formation of the disc becomes suppressed due
to a small, but significant polewards acceleration, giving rise to an
associated polar velocity component that prevents the formation of a
disc. When also accounting for gravity darkening (\ie~a decreased radial
acceleration in equatorial regions), \cite{Owocki96} further showed that a
prolate wind structure develops, with decreased equatorial mass loss and
velocity (see also the review by \citealt{Owocki98bmw}).  \cite{MaederIV}
proposed that an oblate wind structure might still be possible, when
accounting for a polar variation of the ionization equilibrium induced by
gravity darkening (the so-called $\kappa$-effect). This effect becomes
particularly important when the local effective temperature drops below the
bi-stability jump temperature\footnote{The jump temperature is theoretically
  motivated by a stronger radiative line-driving due to lower ionization
  stages of iron for $\Teff \lesssim T_{\rm jump} \approx 25\,\rm{kK}$
  (\citealt{Vink1999}). More recently, \cite{Petrov2016} predicted a somewhat
  lower jump temperature, $T_{\rm jump}\approx 20\,\rm{kK}$.}.
\cite{PetrenzPuls00} extended the hydrodynamic calculations from above by
allowing for spatially varying line force multipliers, and showed that no
major differences from the prolate wind structure arise, at least for OB stars
above $\Teff\gtrsim 20\,\rm{kK}$ with an optically thin Lyman
continuum. Recently, \cite{Gagnier2019} reinvestigated the effects of rotation
using 2D \textsc{ESTER} models (see \citealt{Rieutord2016} for a description
of this code). Using a different implementation of gravity darkening
(consistent with the so-called $\omega$-model by \citealt{Espinosa2011}, which
basically results in a slower decrease of effective temperature with
co-latitude than obtained from the von Zeipel theorem), these authors predict
either a `single-wind regime' (with enhanced polar mass loss) or a `two-wind
regime' (with enhanced mass loss at latitudes where the effective temperature
drops below the bi-stability jump temperature). To understand which of the
different models represents reality best (in different temperature regimes),
one needs to compare synthetic profiles with observations. In this respect,
investigating the effects of prolate and oblate wind structures is
particularly important to distinguish between different theories.
 
As a consequence of the distinct wind structure resulting from a particular
model, the wind lines of rotating stars are expected to differ as a function
of rotational speed and inclination. To predict UV resonance line profiles of
fast rotating stars, we calculated the source function of a prototypical
resonance line transition including the effects of gravity darkening and
surface distortion for models with different rotational velocities. As a first
step, we used a wind description that is consistent with the prolate wind
model. For all calculations, we applied the SClin method.
%
%
\subsection{Wind model}
\label{subsec:rotation_model}
\begin{figure}[t]
\resizebox{\hsize}{!}{\includegraphics{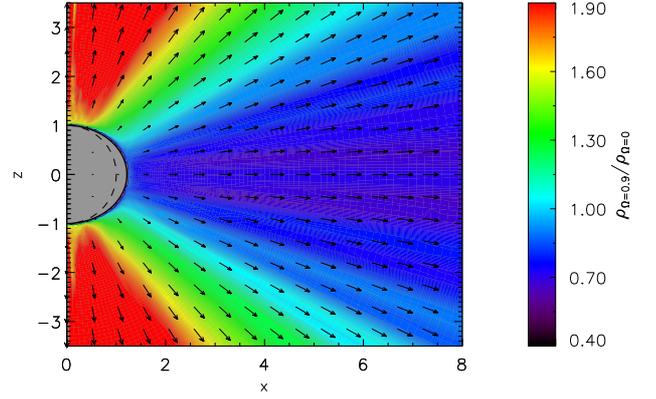}}
\caption{Contours of the density in the $xz$-plane ($z$ being the rotation
  axis) for a prototypical rotating O star with $\Lstar=10^6 \,\lsun$, $\Mstar
  = 52.5\,\msun$, $\Rpole = 19\,\rsun$, and $\vrot=432\,\kms$ (corresponding
  to $\Omega=0.9$). The density has been scaled by values from the
  non-rotating model with $\Omega=0$. While the thick solid line corresponds
  to the surface of the (distorted) star, the dashed line would correspond to
  a spherical surface. Additionally, the velocity vectors are displayed.}
\label{fig:rotating_wind}
\end{figure}
\begin{figure*}[t]
\resizebox{\hsize}{!}{
   \begin{minipage}{0.25\hsize}
      \resizebox{\hsize}{!}{\includegraphics{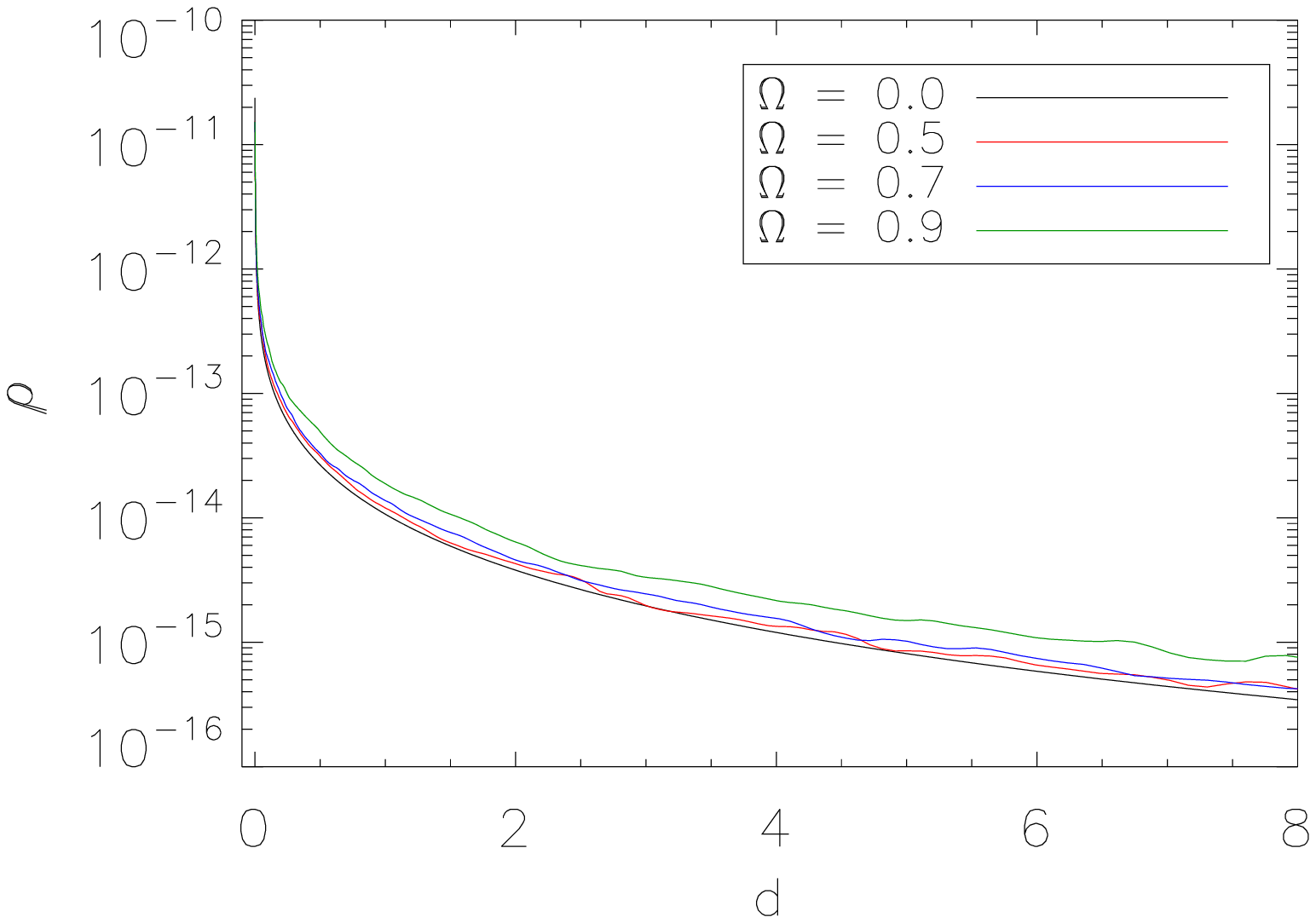}}
   \end{minipage}
   \begin{minipage}{0.25\hsize}
      \resizebox{\hsize}{!}{\includegraphics{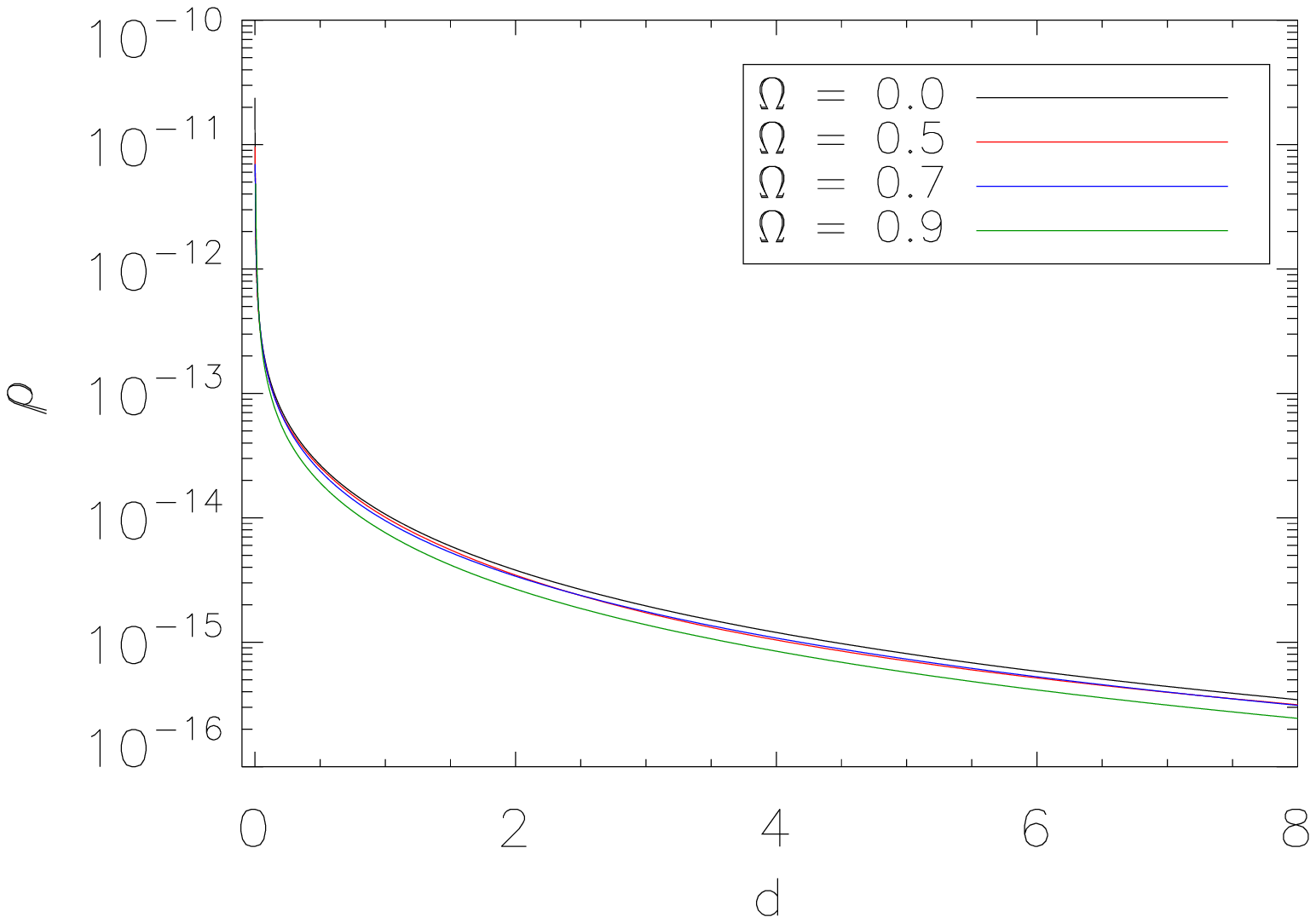}}
   \end{minipage}
   \begin{minipage}{0.25\hsize}
      \resizebox{\hsize}{!}{\includegraphics{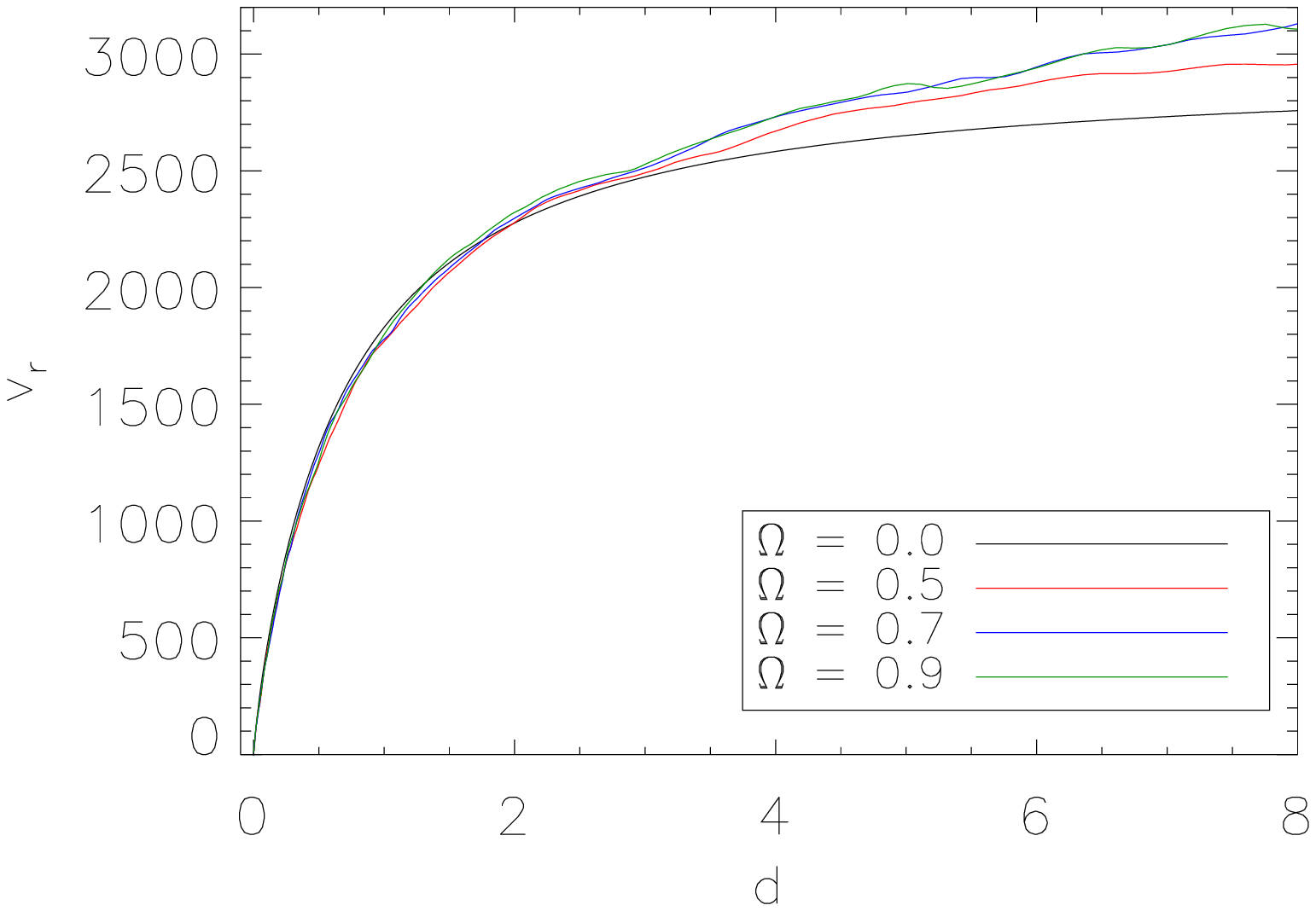}}
   \end{minipage}
   \begin{minipage}{0.25\hsize}
      \resizebox{\hsize}{!}{\includegraphics{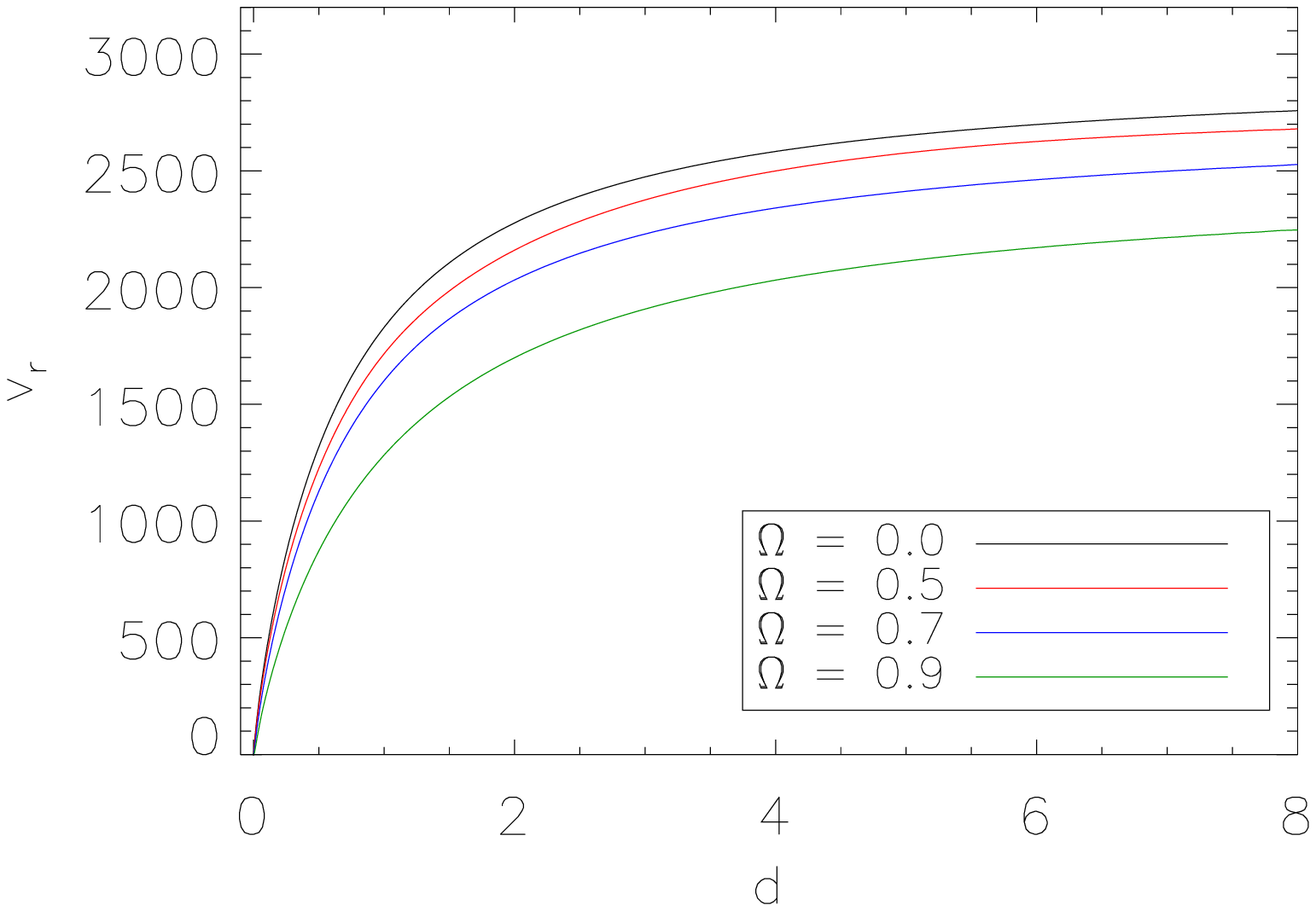}}
   \end{minipage}
}
%
\caption{Density and radial velocity as a function of distance from the
  stellar surface in polar (first and third panel) and equatorial (second and
  fourth panel) regions, for different rotation parameters $\Omega$.}
\label{fig:rotating_wind2}
\end{figure*}
\begin{table}
\begin{center}
\caption{Specific parameters used and obtained for the rotating wind
  models. For a given stellar luminosity $\Lstar=10^6 \,\lsun$, stellar mass
  $\Mstar=52.5 \,\msun$, and polar radius $\Rpole=19 \,\rsun$, rows two to
  eight display the rotation parameter $\Omega$, the equatorial radius $\Req$,
  the polar and equatorial effective temperature $\Teffpole$, $\Teffeq$, the
  total mass loss rate $\mdot$, and the polar and equatorial terminal velocity
  $v_{\infty, \rm p}$, $v_{\infty, \rm eq}$, for different equatorial rotation
  speeds $\vrot$.}
\label{tab:rotating_wind}
\begin{tabular}{ccccc}
\vrot~[\kms] & 0 & 210 & 294 & 432 \\\hline\hline
\noalign{\vskip 0.5mm}
$\Omega$ & 0 & 0.5 & 0.7 & 0.9 \\
\noalign{\vskip 0.5mm}
\Req~[\Rpole] & 1 & 1.04 & 1.09 & 1.22 \\
\noalign{\vskip 0.5mm}
\Teffpole~[kK] & 41.84 & 42.44 & 43.07 & 44.66 \\
\noalign{\vskip 0.5mm}
\Teffeq~[kK] & 41.84 & 40.61 & 39.28 & 35.20 \\
\noalign{\vskip 0.5mm}
\mdot~[\mdu] & 2.70 & 2.73 & 2.79 & 2.93 \\
\noalign{\vskip 0.5mm}
$v_{\infty, \rm p}$ [\kms] & 2781 & 2989 & 3255 & 3159 \\
\noalign{\vskip 0.5mm}
$v_{\infty, \rm eq}$ [\kms] & 2781 & 2651 & 2556 & 2273
\end{tabular}
\end{center}
\end{table}
To obtain a model for the structure of rotating winds, we applied a 2D
version of the \textsc{VH-1}
code\footnote{\url{http://wonka.physics.ncsu.edu/pub/VH-1/}} developed by
J. M. Blondin and co-workers. Our model includes the effects of gravity
darkening and surface distortion (see below). Using a 1D input model derived
from radiation driven wind theory including finite cone angle corrections
(\citealt{CAK} and \citealt{PPK}) for the first time step, the radiation
hydrodynamic equations (accounting for non-radial line forces) are solved
until a (quasi) stationary solution is obtained (see \citealt{CO95} and
\citealt{Owocki96} for the description of the line force). Assuming azimuthal
symmetry, the resulting 2D density and velocity structure is then used as
input for our 3D SC code. Table \ref{tab:rotating_wind} summarizes specific
parameters used and obtained for our model calculations. While the surface
integrated mass flux, \mdot, becomes only slightly increased with increasing
rotational speed, the polar (equatorial) terminal velocities are significantly
enhanced (reduced). For the fastest rotating model ($\vrot = 432 \, \kms$),
Fig.~\ref{fig:rotating_wind} shows corresponding density contours in the
$xz$-plane. The $z$-axis is aligned with the rotation axis. To explicitly show
the prolate wind structure, we have scaled the density by the density
resulting from the non-rotating (spherically symmetric) model, as a function
of distance from the stellar surface. For different rotational speeds,
Fig.~\ref{fig:rotating_wind2} displays the density and radial velocity along
the polar axis and along an (arbitrarily defined) axis in the equatorial
plane. When compared with the spherically symmetric wind, the densities of the
rotating models are enhanced in polar regions, and become reduced along the
equator. Further, the radial velocity along the polar axis remains nearly the
same, except in regions far from the star, where the terminal velocity of all
rotating models becomes enhanced with increasing $v_{\rm rot}$. We note that
one would expect clearer differences of the (radial) velocity fields for
different rotation rates, due to different accelerations induced,
particularly, by the different radiative fluxes resulting from gravity
darkening, and, though to a lesser extent, by the specific density structure
and rotational velocities. Such differences can be barely observed within our
simulations, most presumably because the wind structure in polar regions has
not completely settled to a stationary state at the last time steps. In
contrast, the radial velocity in equatorial regions is significantly reduced
at all distances, when compared to the non-rotating wind, and the deviations
from spherical symmetry become more pronounced with increasing rotational
velocity.  Although we have averaged the hydrodynamic simulations over the
last 20 time steps, the atmospheric structure still suffers from small
numerical artefacts.

To calculate the stellar surface distortion, we consider the gravitational
potential of the star accounting for the effects of centrifugal forces. Under
reasonable assumptions, we can approximate this potential by a Roche model
(\eg~\citealt{Collins63}, see also \citealt{CO95}):
\beq
\Phi(r,\clatitude) = - \dfrac{G\Mstar}{r} - \dfrac{\omega^2r^2\sin^2(\clatitude)}{2} \,,
\eeq
with angular velocity $\omega$. The surface of the star is defined on
equipotential lines and can be calculated by setting
$\Phi(\Rpole,\clatitude=0)=\Phi(\Rstar(\clatitude),\clatitude)$, with $\Rpole$
the polar radius. Solving the resulting cubic equation, one finds:
\beq
\label{eq:rstar}
\Rstar(\Omega, \clatitude) = \dfrac{3 \Rpole}{\Omega \sin(\clatitude)}
  \cos \Bigl( \dfrac{\pi +
    \cos^{-1}\bigl(\Omega\sin(\clatitude)\bigr)}{3}\Bigr) \,,
\eeq
with $\Omega = \omega/\omega_{\rm crit}$ the ratio of the actual to the
critical (`breakup') angular velocity. Defining the rotational speed at the
equator $\vrot$ as input parameter, one easily obtains (cf.~\citealt[their
  Eq. 27]{CO95}):
\begin{eqnarray}
\label{eq:omega}
\Omega &=& \dfrac{\vrot}{\Req} \dfrac{1}{\omega_{\rm crit}} \\
\Req &=& \dfrac{\Rpole}{1-\vrot^2 \Rpole/2G\Mstar} \,,
\end{eqnarray}
with equatorial radius $\Req$. Following \cite{MaederVI}, we use the actual
stellar mass to calculate the equatorial radius and critical velocity without
correcting for Thomson-accelerations. Additionally, we note that our stellar
models are well below the Eddington limit ($\Gamma = 0.5$). Thus, the critical
angular velocity is simply given by $\omega_{\rm crit} = (8G\Mstar/27
\Rpole^3)^{1/2}$. Instead of using Eq.~\eqref{eq:rstar} in our final
implementation, we approximated the stellar surface by a spheroid with
semi-major axes $a=b=\Req$ and semi-minor axis $c=\Rpole$. Such a formulation
greatly simplifies the calculation of the intersection of a given ray with the
stellar surface (required for the boundary conditions, see
Sect.~\ref{subsec:boundary}). For the most extreme test case considered here
($\Omega=0.9$), the maximum error on $\Rstar(\clatitude)$ due to this
approximation is well below the 2\%-level, and rapidly decreases for lower
rotational velocities.

To calculate the intensity emerging from the stellar core, we set $I_{\rm
  c}^+(\clatitude)=B_\nu(\Teff(\clatitude))$, with the effective temperature
as a function of co-latitude. For a given luminosity of the star, $\Lstar$, we
obtain (see also \citealt{PetrenzPuls96}):
\begin{eqnarray}
\Teff(\clatitude) &=& \Bigl[ \dfrac{\Lstar}{2 \pi
    \sigmab \Sigma}\lvert \vecown{g} \rvert^{4\beta_{\rm Z}} \Bigr]^{1/4} \\\nonumber
\Sigma &=& \int_0^\pi \lvert \vecown{g} \rvert^{4\beta_{\rm Z}}
\dfrac{\Rstar^2(\clatitude)\sin(\clatitude)}{- g_r/\lvert \vecown{g} \rvert}
\dd \clatitude \,,
\end{eqnarray}
with $\sigmab$ the Stefan Boltzmann constant, and the surface integrated
effective gravity $\Sigma$ derived from $\vecown{g}(\clatitude) =
-\vecown{\nabla} \Phi(\Rstar(\clatitude),\clatitude)$. The parameter
$\beta_{\rm Z}$ describes the gravity darkening law in terms of
$\Teff(\clatitude)\propto \lvert \vecown{g}(\clatitude) \rvert^{\beta_{\rm
    Z}}$. As originally formulated by \cite{Zeipel24}, $\beta_{\rm
  Z}=1/4$. Though $\beta_{\rm Z}$ might be significantly lower
(\eg~\citealt{deSouza2014}, \citealt{Gagnier2019}), for simplicity we
nevertheless used $\beta_{\rm Z}=1/4$. As long as we assume constant
ionization fractions, the effect of $\beta_{\rm Z}$ on the line profiles will
be minor anyway.
%
%
\subsection{Line formation}
\label{subsec:rotation_profile}
\begin{figure*}[t]
\resizebox{\hsize}{!}{
   \begin{minipage}{0.33\hsize}
      \resizebox{\hsize}{!}{\includegraphics{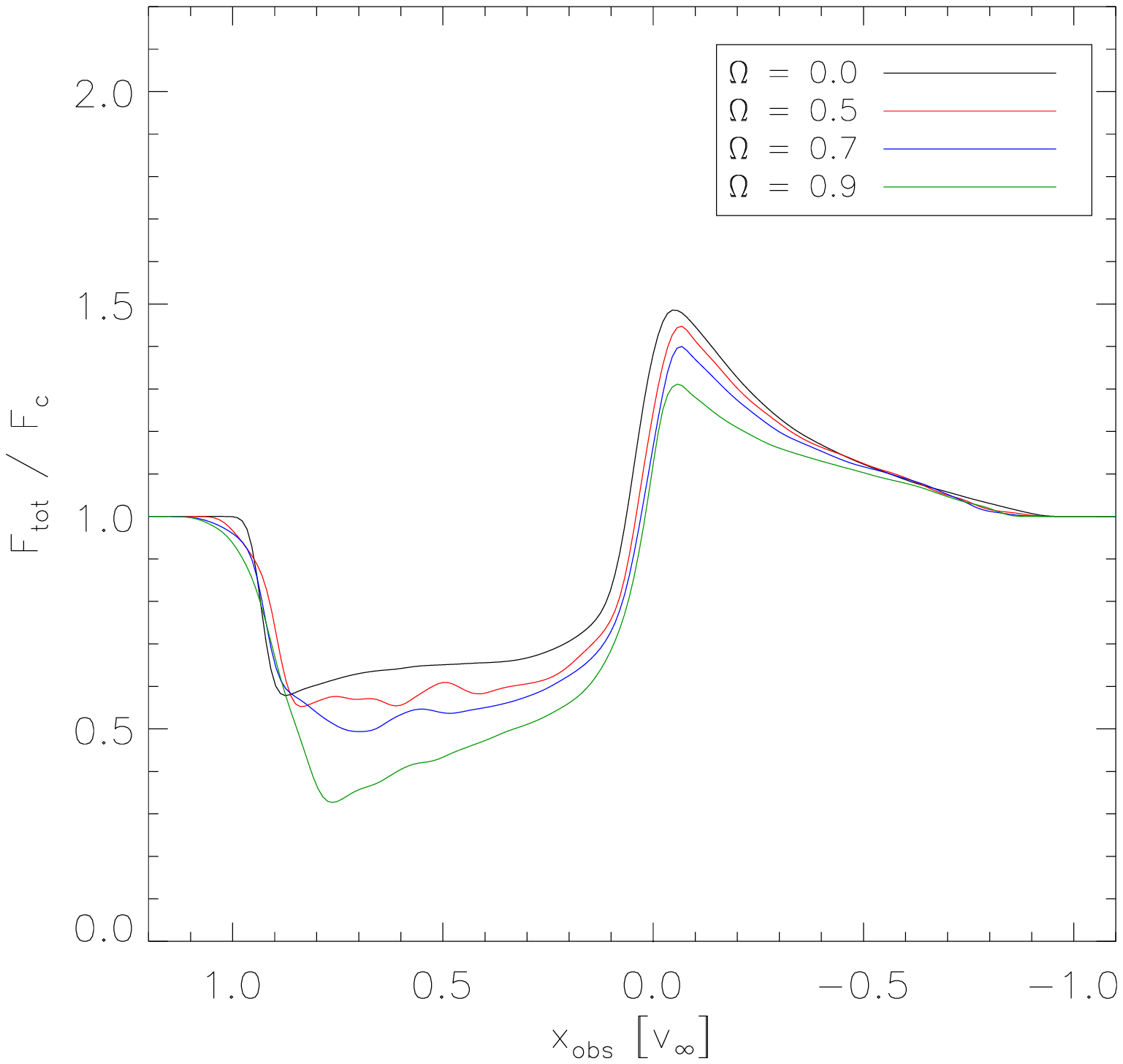}}
   \end{minipage}
   \begin{minipage}{0.33\hsize}
      \resizebox{\hsize}{!}{\includegraphics{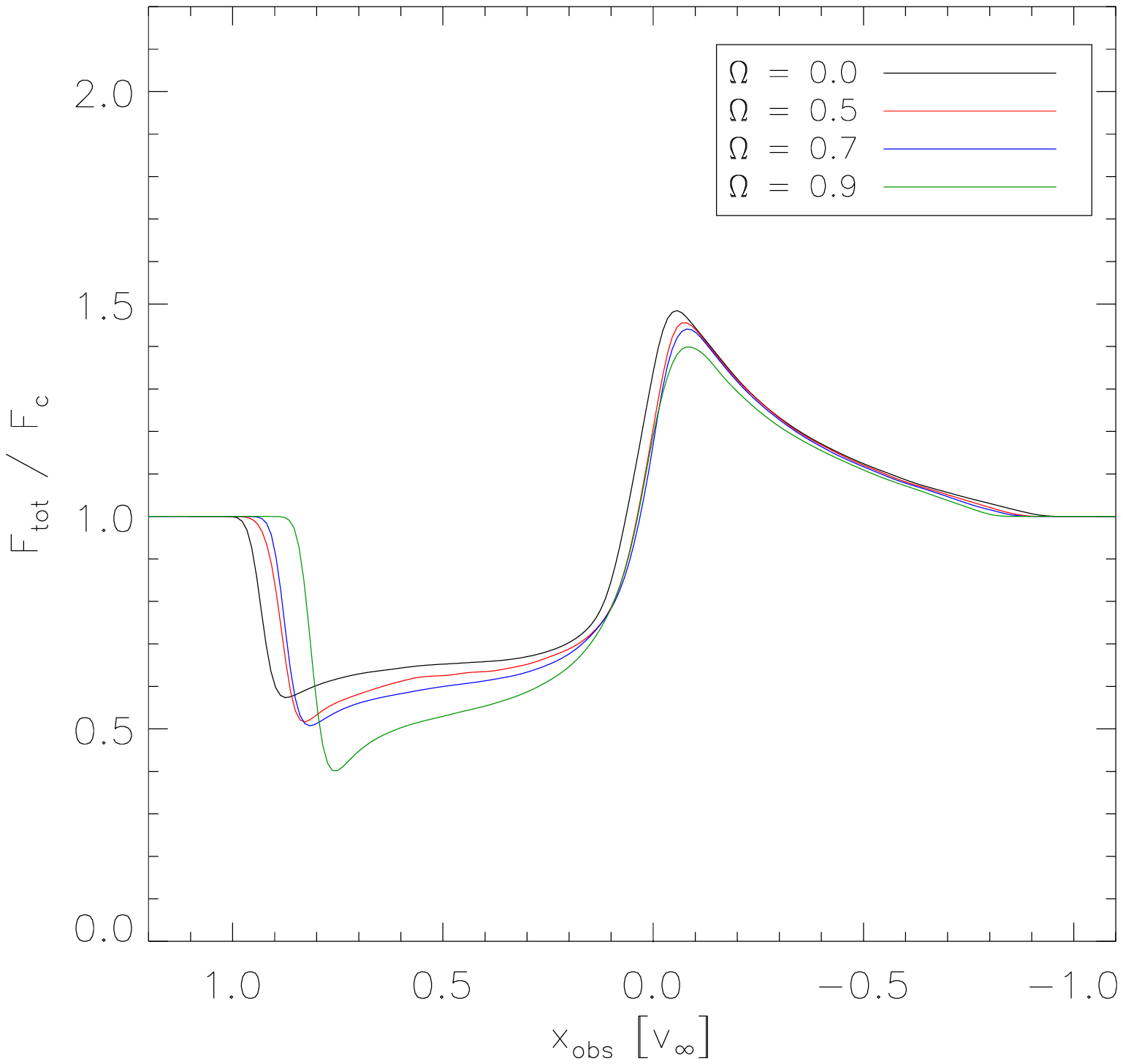}}
   \end{minipage}
   \begin{minipage}{0.33\hsize}
      \resizebox{\hsize}{!}{\includegraphics{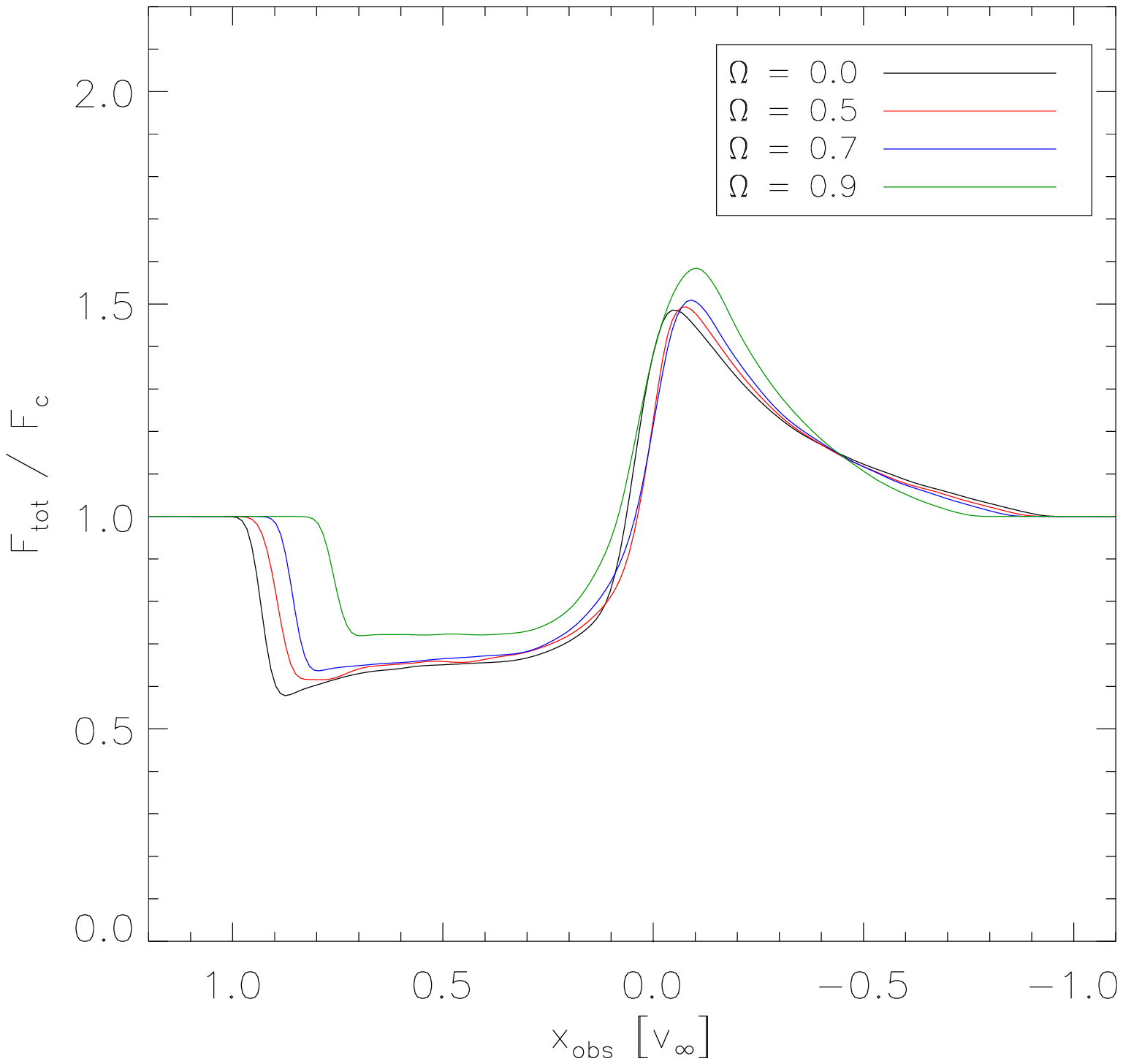}}
   \end{minipage}
}
\\
\resizebox{\hsize}{!}{
   \begin{minipage}{0.33\hsize}
      \resizebox{\hsize}{!}{\includegraphics{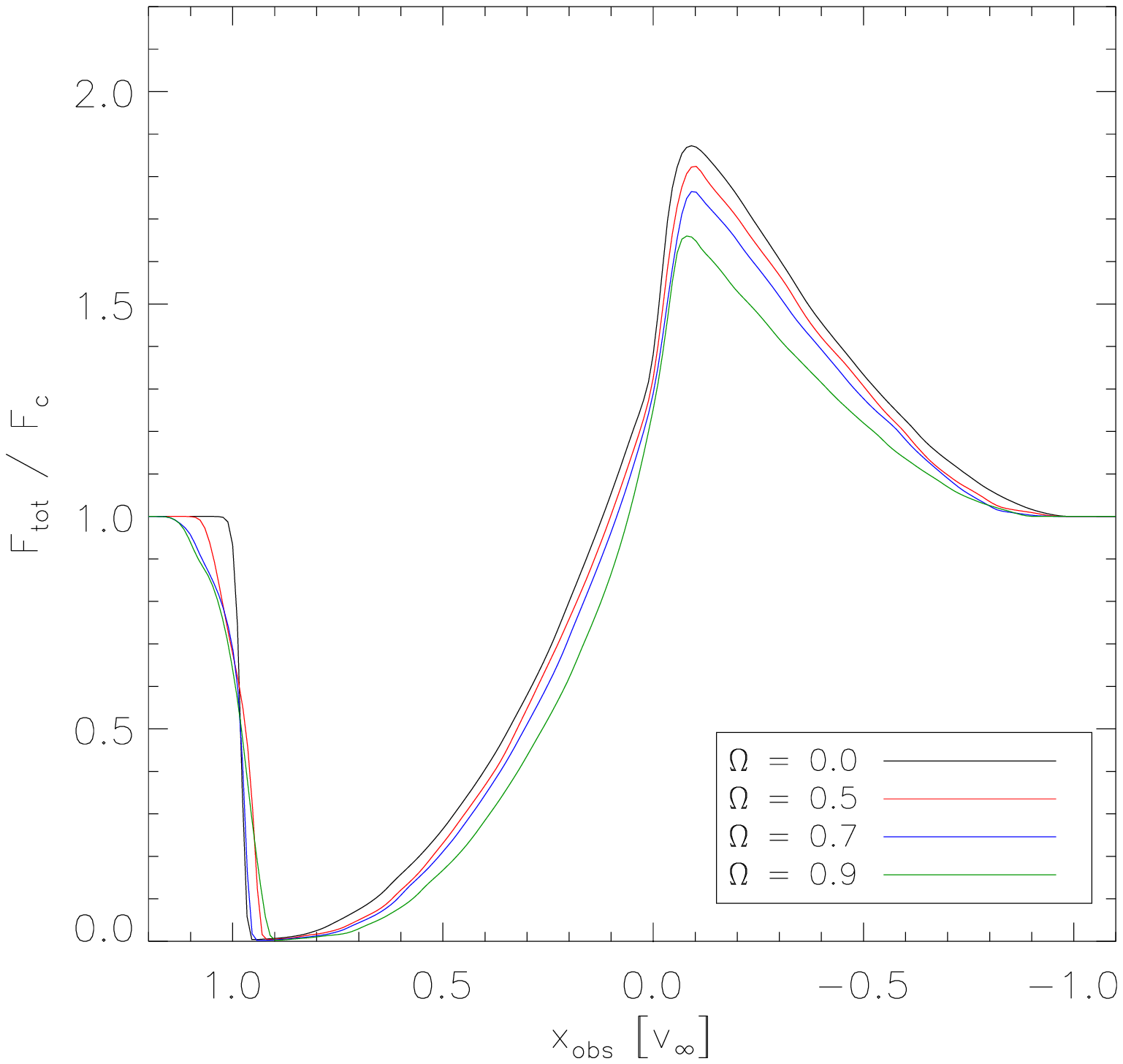}}
   \end{minipage}
   \begin{minipage}{0.33\hsize}
      \resizebox{\hsize}{!}{\includegraphics{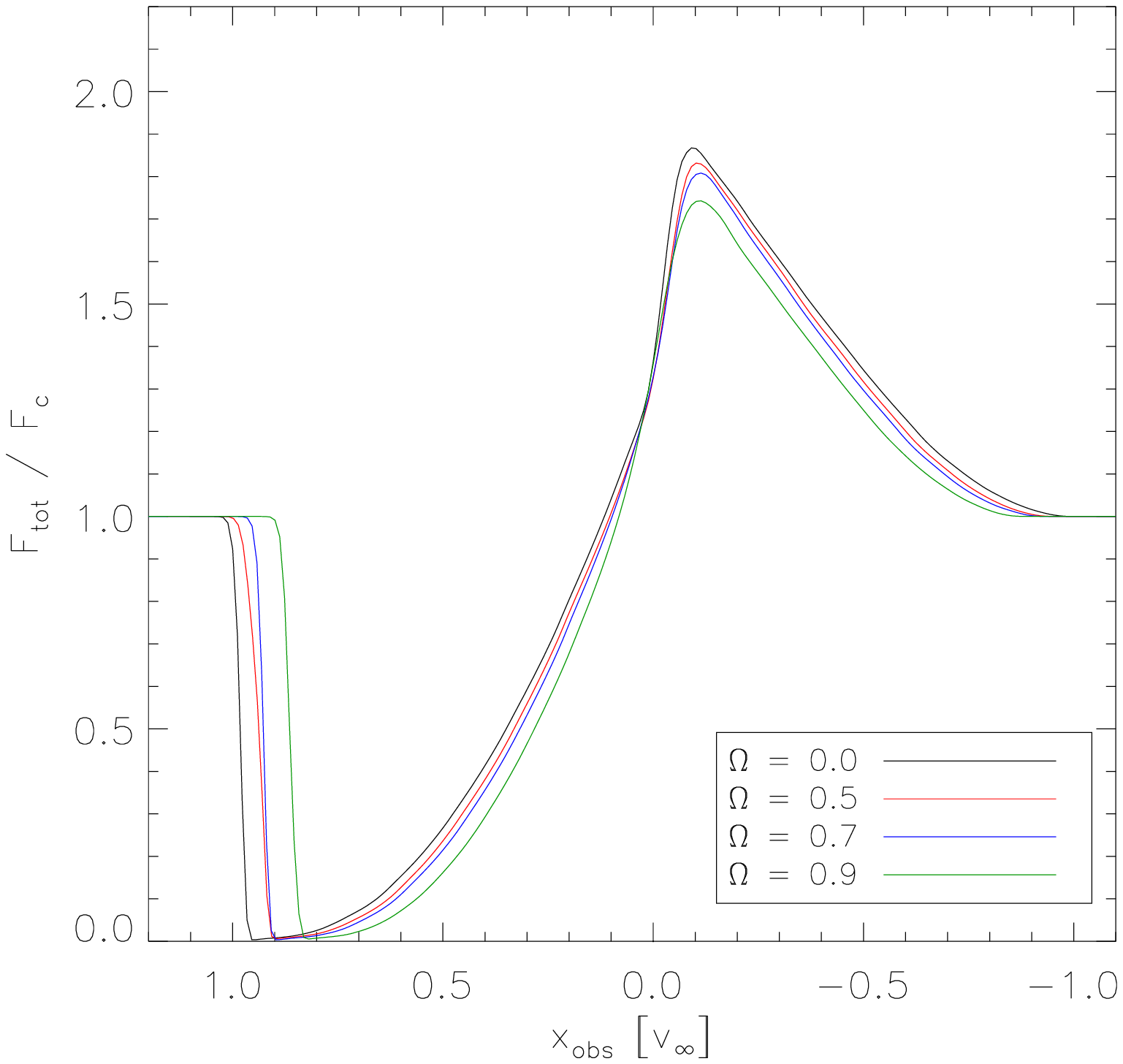}}
   \end{minipage}
   \begin{minipage}{0.33\hsize}
      \resizebox{\hsize}{!}{\includegraphics{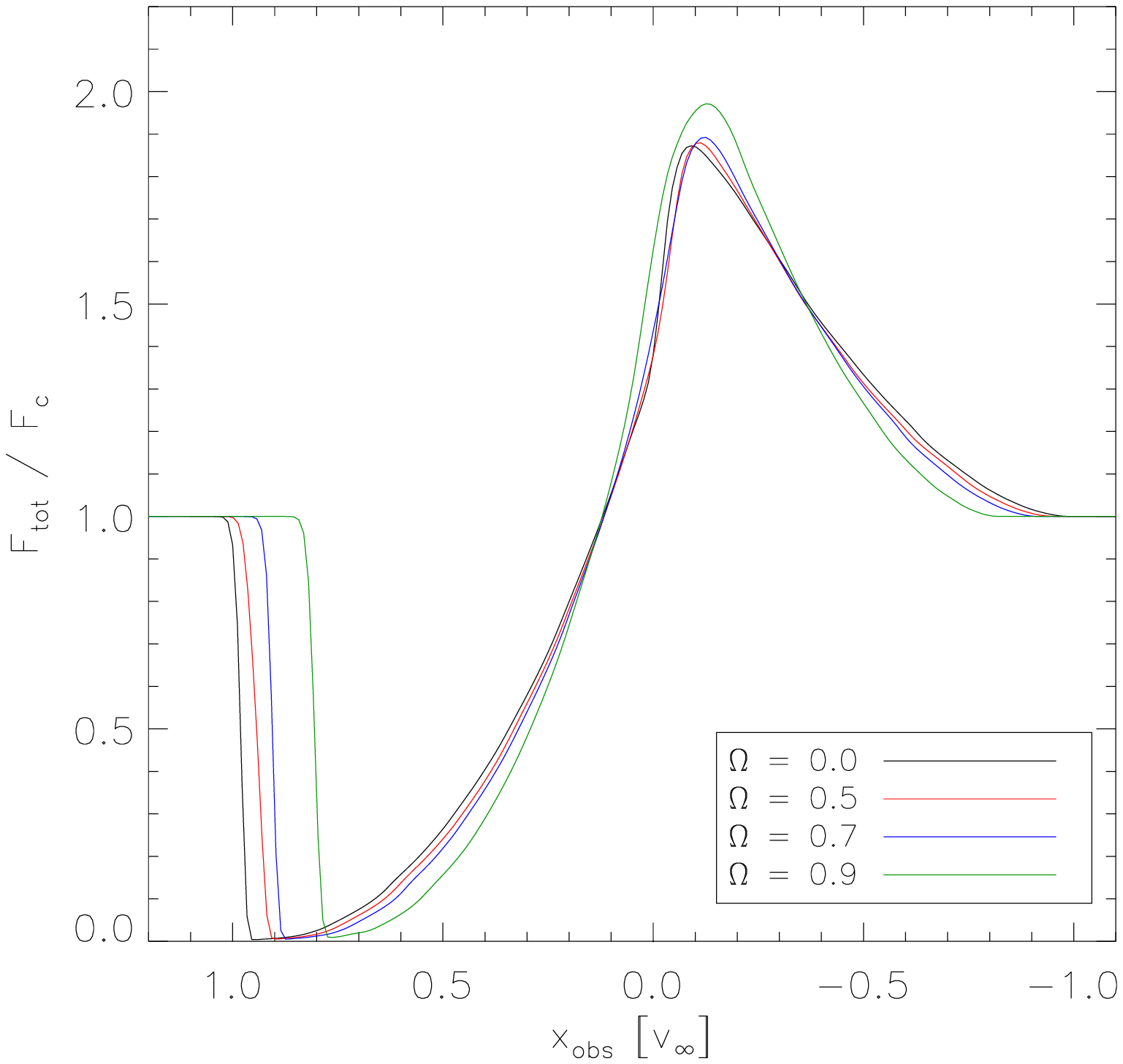}}
   \end{minipage}
}
%
\caption{Predicted emergent flux profiles for the rotating star models with
  $\Omega = [0,0.5,0.7,0.9]$ (see Table~\ref{tab:rotating_wind}). The upper
  and lower panels show the intermediate and strong line with $\kline=10^3$
  and $\kline=10^5$, respectively. The inclination angle has been set to
  $\sin(i)=[0,0.707,1]$ from left to right.}
\label{fig:profiles_rotation}
\end{figure*}
\begin{figure}[t]
\resizebox{\hsize}{!}
{\includegraphics{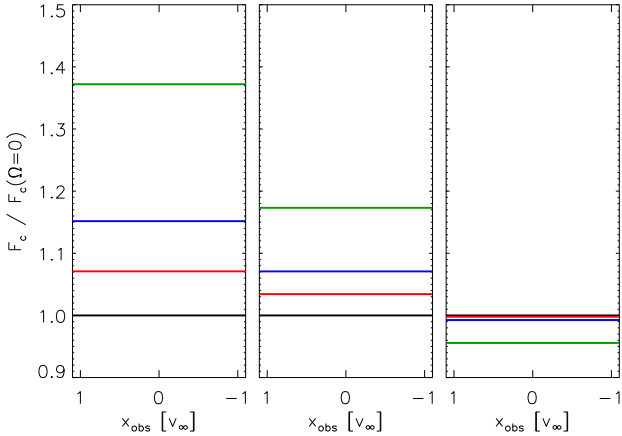}}
\caption{Continuum fluxes for different inclinations $\sin(i)=[0,0.707,1]$
  from left to right, and different rotation parameters (using the same colour
  coding as in Fig.~\ref{fig:profiles_rotation}). The continuum fluxes have
  been scaled by the corresponding flux obtained from the non-rotating model.}
\label{fig:profiles_rotation_cont}
\end{figure}
\begin{figure}[t]
\resizebox{\hsize}{!}{\includegraphics{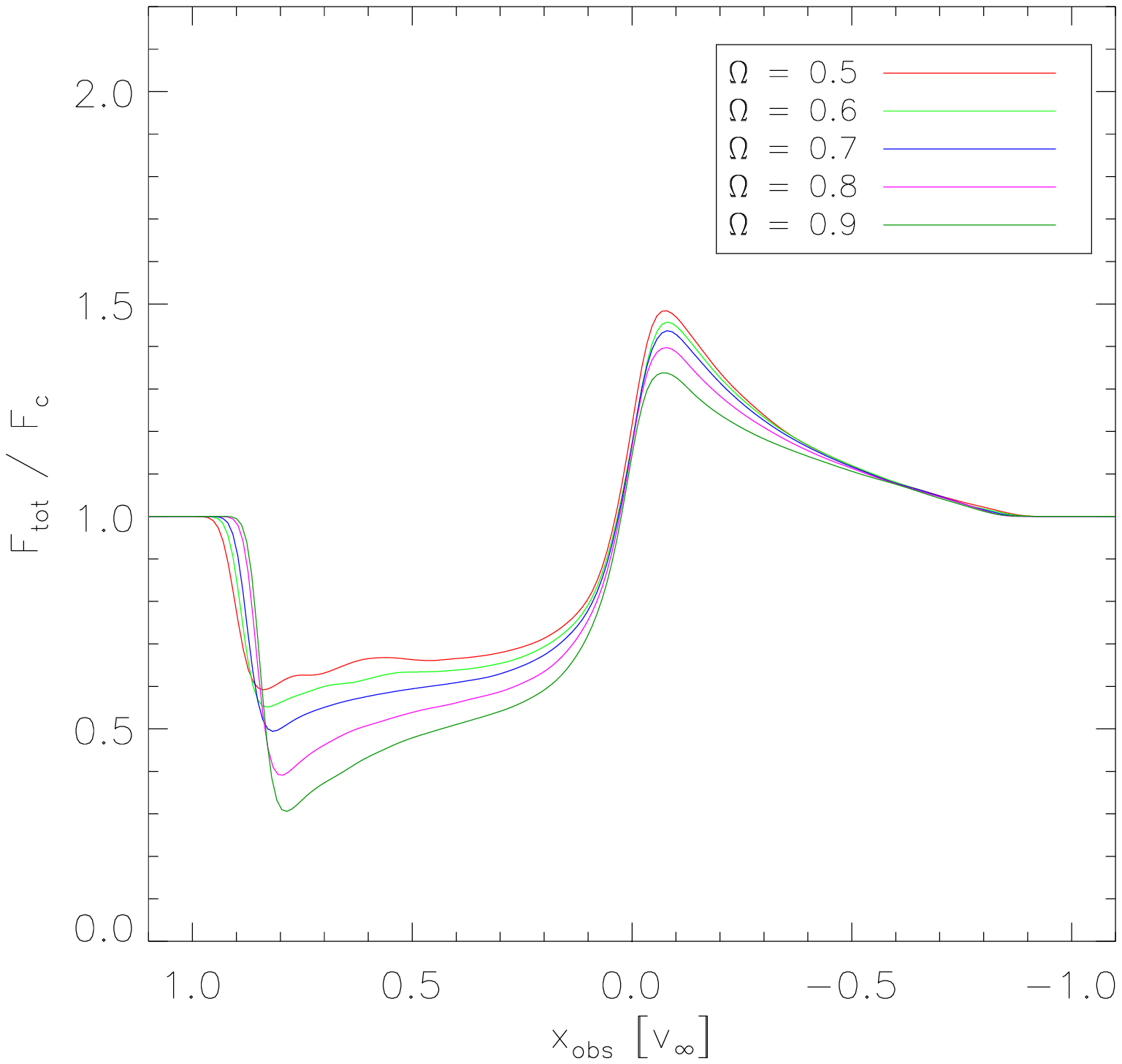}}
\\
\resizebox{\hsize}{!}{\includegraphics{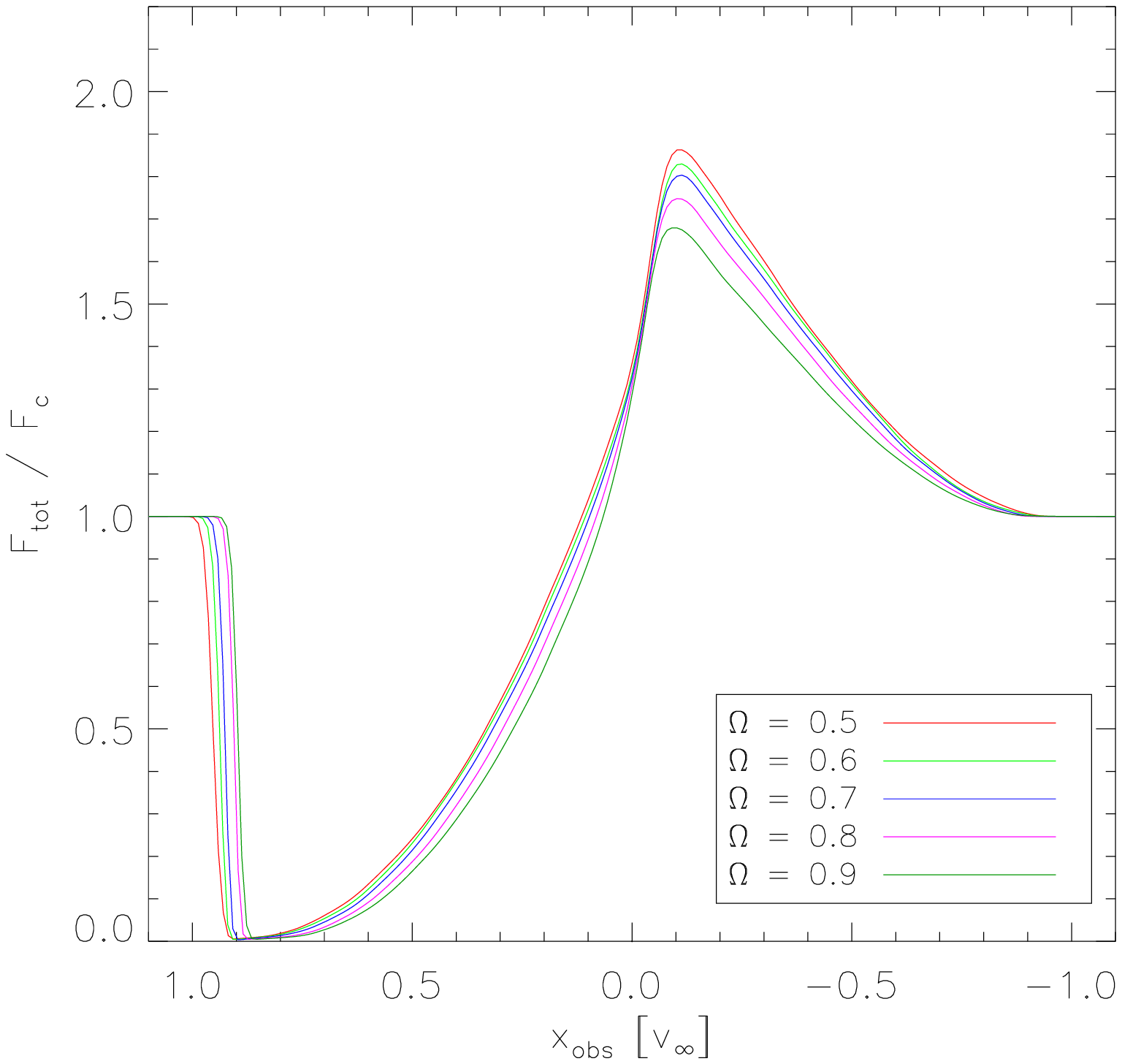}}
%
\caption{Predicted emergent flux profiles for the rotating star models with
  $\Omega = [0.5,0.6,0.7,0.8,0.9]$, and different inclination angles such that
  $v\sin(i)=200\,\kms$ for all models. The upper and lower panel display the
  intermediate ($\kline=10^3$) and strong ($\kline=10^5$) line, respectively.}
\label{fig:profiles_rotation2}
\end{figure}
For our test models, we used $v_{\rm micro} = 100\,\kms$, and calculated the
frequency integrated opacity from Eq.~\ref{eq:opal} for an intermediate and a
strong line with line-strength parameter $\kline=10^3$ and $\kline=10^5$. To
obtain the source function, we applied the 3D SClin method and set
$\epsl=10^{-6}$. The resulting (normalized) line profiles are shown in
Fig.~\ref{fig:profiles_rotation} for different rotational velocities and
inclination angles. Additionally, we display the continuum flux used for
normalization in Fig.~\ref{fig:profiles_rotation_cont}. Due to gravity
darkening and the surface distortion, the continuum depends on the rotation
rate and inclination, with largest fluxes found for high rotation rates and
low inclinations (resulting from the higher temperatures in polar regions and
a larger projected stellar disc). In Figs.~\ref{fig:profiles_rotation} and
\ref{fig:profiles_rotation_cont}, the $x$-axes have been normalized to an
(arbitrarily chosen) terminal velocity $\vinf=3000\,\kms$.
The behaviour of the line profiles can be qualitatively explained with the
hydrodynamic structure:

(i) For pole on observers ($\sin(i)=0$, left panel of
  Fig.~\ref{fig:profiles_rotation}), the absorption column in front of the
  star is enhanced with increasing rotational velocity due to the larger
  densities (and opacities) in polar regions. Thus, the absorption trough (of
  unsaturated lines) becomes more pronounced. The absorption edge of the
  intermediate lines is found at slightly lower velocities than expected from
  the hydrodynamic simulations, because the optical depths of the
  corresponding resonance regions are too low to efficiently contribute to the
  absorption. When considering the strong lines, the optical depth is larger,
  and the absorption edge is more consistent with the actual terminal
  velocity. For both applied line strength parameters, the emission peak
  becomes weaker with increasing rotation rate, particularly at low projected
  velocities on the red side of the line centre (for negative $\xobs$). This
  effect can be partly explained by the reduced emission from the equatorial
  plane, due to the lower densities in these regions. More importantly,
  however, is the increased continuum flux that mainly determines the
  (normalized) height of the emission peak.

(ii) When increasing the inclination towards equator-on observers
  ($\sin(i)=1$, right panel of Fig.~\ref{fig:profiles_rotation}), the
  behaviour is reversed. For such directions, the continuum plays an only
  minor role, since the lower (equatorial) effective temperatures of the
  rotating models are nearly compensated by the enlarged projected stellar
  disc. With increasing rotation parameter, the absorption trough of the
  intermediate line becomes reduced and shifted towards lower terminal
  velocities, consistent with the hydrodynamical model. When considering the
  strong line, the absorption becomes saturated, and only the impact of the
  different terminal velocities can be observed. Additionally, and for both
  line strengths, the absorption slightly extends towards the red side,
  because of (negative) projected line of sight velocities near the stellar
  surface induced by rotation. For the fastest rotating model with
  $\Omega=0.9$, this effect becomes suppressed due to an increased emission
  from the (dense) prolate wind structure. This latter effect is only moderate
  for lower rotational speeds.
Based on the current hydrodynamic wind structure, we would therefore expect to
observe either rather low terminal velocities or relatively deep absorption
troughs for fast rotating stars, and we are able, at least in
  principle, to check the theory by comparing our synthetic spectra with
(past or future) UV observations. This point, however, is beyond the scope of
this paper.

Finally, if the projected rotational velocity is known (\eg~from photospheric
lines), one might even estimate the actual rotational velocity from UV
resonance lines. This latter point becomes clear from
Fig.~\ref{fig:profiles_rotation2}, where we predict the line profiles of
models with different rotational speed for a given $v\sin(i)$ (set here to
$200\,\kms$). Since, at least for the intermediate line, the profile shapes
differ, $\sin(i)$ could be derived if $v\sin(i)$ was known. Of course, such
constraints will become feasible only if the underlying models correctly
describe the wind structure (including possibly varying ionization stages) of
rotating stars.
%
%
\section{Summary and conclusions} \label{sec:conclusions}
In this study, we have presented a 3D short-characteristics method tailored
for the solution of continuum- and line-scattering problems in the winds of
hot stars. To obtain the formal solution, we have implemented a purely linear
interpolation scheme (for calculating upwind quantities and for the solution
of the radiative transfer equation along a ray), as well as a second order,
monotonic, B\'ezier technique. We use Cartesian coordinates with a non-uniform
grid spacing to ensure a reasonable spatial resolution in important regions
(\ie~where velocity and/or density gradients are large).  As a first step
towards full NLTE radiative transfer models, we consider a single
resonance-line transition (approximated by a two-level-atom) assuming an
optically thin background continuum, whereas for pure continuum problems we
use the thermalization parameter, \epsc, and split the source function into a
scattering and a thermal part. A generalization (including multi-level atoms)
is planned for future applications.

To calculate strong scattering lines and optically thick, scattering dominated
continua, we have implemented an accelerated $\Lambda$-iteration scheme using
different non-local approximate $\Lambda$-operators (ALOs), together with
applying the Ng-extrapolation method for subsequent iterations. With
increasing complexity of the ALO (\ie~from a purely diagonal ALO to a
nearest-neighbour ALO including also the 26 neighbouring terms), the rate of
convergence is improved. When applying the NN-ALO, the converged solution is
generally found within $\lesssim 20$ iteration steps even for the most
challenging test cases.

We have estimated the error of the applied methods in different regimes by
calculating spherically symmetric test models within our 3D SC framework, and
with a 3D finite-volume method. To our knowledge, this is the first study,
where different 3D solution schemes for spherical problems have been compared,
and their precision explored. When rated against the solution obtained from
(accurate) 1D solvers, we found a mean relative error for the converged
continuum source function of roughly $5-10\,\%$ and $5-20 \,\%$ when using
B\'ezier and linear interpolations, respectively. Particularly for optically
thick continua, the (first order) FVM method breaks down, and a (high order)
SC or LC method is required to accurately solve the radiative transfer. When
considering the solution of the line source function for different
line-strength parameters, the mean relative errors of both SC methods are on
the $10\,\%$-level and below, with slight advantages of the B\'ezier technique
compared to purely linear interpolations. The resulting synthetic line
profiles are calculated with a long-characteristics postprocessing routine
using the previously calculated converged source functions. The SC method
using B\'ezier interpolations almost perfectly matches the 1D reference
profiles for all our models (\ie~for weak and strong lines). When linear
interpolations are used, we obtain slight deviations originating mainly in the
outer wind regions. In contrast, the 3D FVM always overestimates the
emission. Nevertheless, all methods have their own advantages and
disadvantages, particularly when also accounting for the computation time
(with fastest turn-around times for the FVM method).  Thus, the 3D FVM method
should be used for qualitative interpretations and for finding (rough)
estimates of the parameters of interest, while the SC methods are to be
preferred when aiming at a quantitative analysis of line profiles, and for
optically thick continua.

As a first application of the 3D SC code to non-spherical problems, we
considered the resonance line formation in the winds of (fast) rotating O
stars. Assuming azimuthal symmetry, the hydrodynamic structure for a
prototypical O star with different rotation rates has been calculated by means
of the 2D \textsc{VH-1} code. We have included the effects of surface
distortion and gravity darkening into our 3D radiative transfer
framework. Given the hydrodynamic models, we are able to predict the shape of
line profiles for different rotational speeds and inclination angles. When
compared with a (non-rotating) spherically symmetric wind (obtained using the
same stellar parameters), rotating stars should either show relatively low
terminal velocities (for equator-on observers) or deeper absorption troughs
(for pole-on observers). The latter effect, however, would only be observable
when considering intermediate (\ie~unsaturated) lines. Additionally, we showed
that the line profile shapes vary as a function of rotational speed at a given
$v\sin(i)$. Thus, assuming that $v\sin(i)$ was known (\eg~from photospheric
line modelling), one could estimate the rotational speed, though with a rather
large uncertainty, since the differences of the line profiles are only
moderate. We emphasize that other effects (such as clumping or a flatter
gravity darkening law) may additionally shape the line profiles. When
analysing UV observations of fast rotating stars, the 3D SC code developed in
this work certainly will help to understand the manifestations of various
(aforementioned) effects, and to distinguish between different theoretical
predictions (\eg~prolate vs. oblate wind structures). Ideal testbeds for
future investigations of fast rotating winds are the stars VFTS102 (O9 Vnnne,
\citealt{Dufton11}) and VFTS285 (O7.5 Vnnn, \citealt{Walborn2012}), both
rotating at nearly their critical velocity.

Finally, we note that our tools are, of course, not limited to rotating stars.
Indeed, almost any kind of stellar wind that deviates from spherical
  symmetry (with non-relativistic velocity fields), such as magnetic winds or
  colliding winds in close binaries, can be investigated.
\begin{acknowledgements}
We thank our referee, Dr. Achim Feldmeier, for many helpful comments and
suggestions, that improved the first version of this manuscript considerably.
LH gratefully acknowledges support from the German Research Foundation, DFG,
under grant PU 117/9-1. NDK acknowledges funding from the KU Leuven C1 grant
MAESTRO C16/17/007. JOS acknowledges support from FWO Odysseus grant
D4545-GH9218. Finally, we thank J. M. Blondin and co-workers for providing the
VH-1 code.
\end{acknowledgements}

\bibliographystyle{aa}
\bibliography{bib_levin}

\appendix
%
%
\section{Grid construction}
\label{subsec:grid_construction}
\begin{figure}[t]
\resizebox{\hsize}{!}{\includegraphics{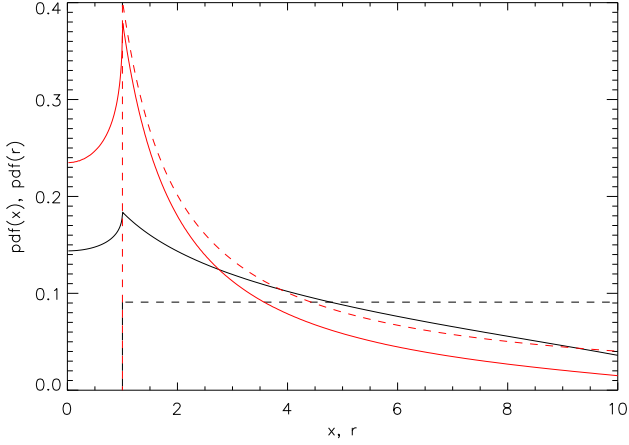}}
\caption{Probability density functions of radial (dashed) and $x$ (solid)
  coordinates for different spherical and Cartesian grids. In this example,
  two spherical grids are given in 2D as input to our 3D code, with uniformly
  (black) or logarithmically (red) distributed $r$-coordinates, and a
  uniformly distributed polar angle. The corresponding distributions of
  $x$-coordinates are calculated within our grid construction procedure (see
  text). Large values of the probability density functions correspond to a
  high resolution of $x$ and $r$-coordinates.}
%
\label{fig:gridpdf}
\end{figure}
In this section, we describe the grid construction procedure used within our
3D code. Generally, we assume the wind structure (\ie~density, velocity field,
and temperature) to be given by an input model obtained from hydrodynamic
simulations or external (semi)-analytic calculations. Since the input grid is
not necessarily compatible with our 3D SC solver, and to minimize
interpolation errors when calculating upwind and downwind quantities, we
construct an own grid that uses the distribution of the input-grid coordinates
in an optimum way. When the input grid uses spherical coordinates
$(r,\clatitude,\azimuth)$, we define a joint probability distribution
\beq
h(x,z) = f(r)g(\clatitude) \lvert J \rvert \,,
\eeq
where $f(r)$ and $g(\clatitude)$ are the probability density functions derived
from the distribution of the input coordinates, and $\lvert J \rvert =
\sqrt{x^2+z^2}$ is the Jacobian determinant. Since we consider only
axisymmetric atmospheres in this paper, we use the $x$-coordinate distribution
also for the y-coordinates.  To calculate the probability density functions
for $x$ and $z$, we simply marginalize $h(x,z)$ over $z$ and $x$,
respectively. The discretized coordinates are finally determined by demanding
that the probabilities of selecting a (continuous) coordinate in each
(discrete) interval shall be the same.  Fig.~\ref{fig:gridpdf} shows the
probability density functions of the $x$-coordinates for two different input
distributions of the radial grid. Here, the polar angle $\clatitude$ has been
assumed to be uniformly distributed for both examples.

Since the final number of core and non-core points depends on the slope of the
probability density of the radial grid, yielding in worst cases a much larger
number of core points than non-core points, and because the total number of
used points is memory-limited, we define two input parameters $N_{\rm core}$
and $N_{\rm non-core}$ to keep control on the final grid. For all test
calculations (including those presented in Sect.~\ref{subsec:spherical_wind}),
the best solution has always been found for a number of $N_{\rm core}/N_{\rm
  non-core} \in [0.25,0.5]$. An explicit choice of $N_{\rm core}$ and $N_{\rm
  non-core}$ corresponds to a re-normalization of the probability density
function in the regimes $x,z \in[0,\Rstar]$ and $x,z \in[\Rstar,R_{\rm max}]$,
where $R_{\rm max}$ defines the border of the computational domain.  We note
that the same procedure can be used for an input grid given in Cartesian
coordinates, with the probability density function of the input-grid
coordinates derived directly from the corresponding (discrete) distribution.
%
%
\section{1D B\'ezier interpolation}
\label{app:bez}
\begin{figure}[t]
\resizebox{\hsize}{!}{\includegraphics{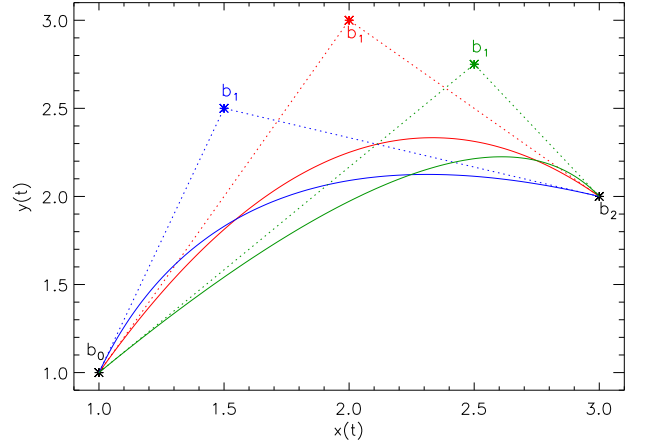}}
%
\caption{B\'ezier curves (solid lines) for three given points $\vecown{b}_0$,
  $\vecown{b}_1$, $\vecown{b}_2$. The blue, red, and green lines represent the
  resulting curves for different control points $\vecown{b}_1$. The straight
  connections of the control points with the data points are indicated by the
  dotted lines.}
\label{fig:bez_orig}
\end{figure}
\begin{figure}[t]
\resizebox{\hsize}{!}{\includegraphics{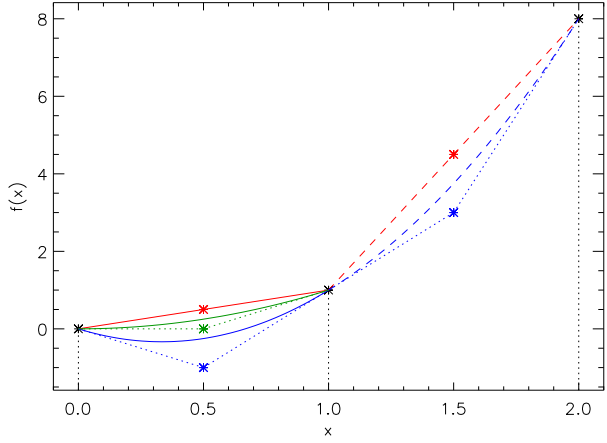}}
%
\caption{Different interpolation techniques for a set of three data points at
  $x$-coordinates indicated by the dotted vertical lines. The solid and dashed
  lines correspond to the interpolation in the different intervals
  $[\ximo,\xii]$ and $[\xii,\xipo]$, respectively. Linear interpolations,
  quadratic interpolations (connecting all three data points), and a
  \textbf{monotonic} B\'ezier curve (with $\paramon$ calculated from
  Eq.~\eqref{eq:alphal1} in the interval $[\ximo,\xii]$) are indicated in red,
  blue, and green. Since the quadratic interpolation is already monotonic in
  the interval $[\xii,\xipo]$, the monotonic B\'ezier curve coincides with the
  dashed, blue line. Control points are indicated with coloured asterisks.}
\label{fig:bez}
\end{figure}
In this section, we discuss an interpolation technique using quadratic
B\'ezier curves (\eg~\citealt{Auer2003}, \citealt{SchwarzNumerics}). Such
curves are generally constructed from three given points $\vecown{b}_0$,
$\vecown{b}_1$, $\vecown{b}_2$ (see Fig.~\ref{fig:bez_orig}), and have the
following useful properties:
\begin{enumerate}
\item[(i)] The boundary points, $\vecown{b}_0$ and $\vecown{b}_2$ are
  reproduced exactly by the B\'ezier curve.
\item[(ii)] The straight connections $\left(\vecown{b}_1-\vecown{b}_0\right)$
  and $\left(\vecown{b}_2-\vecown{b}_1\right)$ define the tangent lines of the
  curve at $\vecown{b}_0$ and $\vecown{b}_2$, respectively.
\item[(iii)] Any point on the B\'ezier curve is located in the convex hull of
  $\vecown{b}_0$, $\vecown{b}_1$, $\vecown{b}_2$.
\end{enumerate}
In a 2D plane described by coordinates $x$ and $y$, the quadratic B\'ezier
curve is parameterized as:
\beq
\label{eq:app_bez_orig}
\vecown{b}(t) = 
\begin{pmatrix}
x(t) \\ y(t)
\end{pmatrix}
=
\left(1-t\right)^2 \vecown{b}_0 +
2t\left(1-t\right)\vecown{b}_1 + t^2 \vecown{b}_2 \,,
\eeq
with $t\in[0,1]$, and $\vecown{b}_0 = (x_0,y_0)$, $\vecown{b}_1 = (x_1,y_1)$,
$\vecown{b}_2=(x_2,y_2)$. With Eq.~\eqref{eq:app_bez_orig}, the properties
(i)-(iii) can be exploited to construct a monotonic interpolation scheme by
identifying $\vecown{b}_0$,$\vecown{b}_2$ with two given data points
$(x_0,f_0)$, $(x_2,f_2)$, and defining $\vecown{b}_1$ as a free (and tunable)
parameter. Thus, $\vecown{b}_1$ is commonly named control point, and is
`only' required to set the slope of the B\'ezier curve. To reproduce the
underlying function best, and to preserve monotonicity of the resulting curve,
the control point should be chosen with care.

In the following, we present a B\'ezier-interpolation technique for an
interval $x\in[\ximo,\xii]$, given three data points, $(\ximo,\fimo)$,
$(\xii,\fii)$, $(\xipo,\fipo)$. The interpolation formulas corresponding to
the interval $x\in[\xii,\xipo]$ are given in subsection \ref{subsec:app_bez2}.
\subsection{Interval $[\ximo,\xii]$}
A quadratic B\'ezier curve in the interval $[\ximo,\xii]$ is given from
Eq.~\eqref{eq:app_bez_orig}:
\beq
\label{eq:app_bez_orig2}
\begin{pmatrix}
x(t) \\
f\left(x\left(t\right)\right)
\end{pmatrix}
= 
\left(1-t\right)^2
\begin{pmatrix}
x_{\indx{i-1}} \\
f_{\indx{i-1}}
\end{pmatrix}
+ 2t\left(1-t\right)
\begin{pmatrix}
x_{\indxc} \\
f_{\indxc}
\end{pmatrix}
+ t^2
\begin{pmatrix}
x_{\indx{i}} \\
f_{\indx{i}}
\end{pmatrix} \,,
\eeq
with $(x_{\indxc},f_{\indxc})$ the control point. The abscissa of the control
point, $x_\indxc$, can be chosen arbitrarily (at least in principal). To
obtain a second-order interpolation scheme, however, $x_\indxc$ needs to be
located at the centre of the data-point's abscissae\footnote{If $x_\indxc$ was
  located at $x_\indxc=\ximo+3/4(\xii-\ximo)$, for instance, one can easily show that
  the resulting B\'ezier curve never reproduces the unit parabola for any
  ordinate value $f_\indxc$.}, and is therefore set to
$x_{\indxc}=(\ximo+\xii)/2$. Then, the quadratic B\'ezier interpolation scheme
is given by:
\beqa
\label{eq:bezl}
f(x) &=& (1-t)^2\fimo + 2t(1-t)f_{\indxc} + t^2 \fii \\\nonumber
t &=&(x-\ximo)/(\xii-\ximo) \,,
\eeqa
where $t$ has been determined from the definition of $x_\indxc$ and
Eq.~\eqref{eq:app_bez_orig2}. Since the straight connection of the control
point $(x_{\indxc},f_{\indxc})$ with the data point $(\xii,\fii)$ defines the
tangent line of the B\'ezier curve at this data point, $f_{\indxc}$ is
calculated as
\beq
\nonumber
f_{\indxc} = \fii - \dfrac{\dd f}{\dd x}\biggr\rvert_{\xii} \dfrac{\dxi}{2} \,,
\eeq
with $\dxi=\xii-\ximo$. The unknown derivative at $\xii$ needs to be
approximated. Using also the information from the next data point,
$(\xipo,\fipo)$, and assigning a weight $\paramon$ to the forward and backward
derivatives (obtained from finite differences), we find
\beq
\label{eq:bezl_fc}
f_{\indxc} = \fii - \dfrac{\dxi}{2}\Biggl( \paramon \dfrac{\fii-\fimo}{\dxi} +
(1-\paramon) \dfrac{\fipo-\fii}{\dxipo} \Biggr) \,,
\eeq
with $\dxipo=\xipo-\xii$. With a proper choice of $\paramon$, we can adjust
the B\'ezier curve to our needs by shifting the control point up or down. For
instance, setting $\paramon = \dxipo/(\dxi+\dxipo)$ results in the unique
parabola connecting the three given data points, while $\paramon=1$ would
yield the linear interpolation. To avoid overshoots and negative function
values, we demand that the B\'ezier curve shall be monotonic in the interval
$[\ximo,\xii]$. Noting that monotonicity is obtained when the control point is
located in the interval $f_{\indxc} \in [\fimo,\fii]$, corresponding
$\paramon$-values should lie in between the following limits:
\beqa
\label{eq:alphal1}
\paramon_{\indx{i-1}}^{[\indx{i-1},\indx{i}]} &:=& \paramon (f_{\indxc}^{[\indx{i-1},\indx{i}]}=\fimo) =
1 + \dfrac{1}{1-\frac{\fipo-\fii}{\fii-\fimo}\frac{\dxi}{\dxipo}} \\
\label{eq:alphal2}
\paramon_{\indx{i}}^{[\indx{i-1},\indx{i}]} &:=& \paramon(f_{\indxc}^{[\indx{i-1},\indx{i}]}=\fii) =
\dfrac{1}{1-\frac{\fii-\fimo}{\fipo-\fii}\frac{\dxipo}{\dxi}} \,,
\eeqa
where the superscript $[\indx{i-1},\indx{i}]$ denotes that $\paramon$
corresponds to the interpolation scheme in the left interval, $[\ximo,\xii]$.
In the final implementation, we avoid the division by zero if $\fii=\fimo$ or
$\fii=\fipo$, of course. Our standard interpolation is then performed as
follows. At first, we calculate $\paramon$ such that we obtain the unique parabola
connecting all three data points. Secondly, if $\paramon$ lies outside the allowed
limits from Eq.~\eqref{eq:alphal1} and \eqref{eq:alphal2}, we adjust
$\paramon$ to yield monotonic interpolations. In Fig.~\ref{fig:bez}, we
display the monotonic B\'ezier curve resulting from a $\paramon$-parameter
calculated by means of Eq.~\eqref{eq:alphal1}, together with linear and
quadratic interpolations (the latter connecting the three data points). Since
monotonicity is always obtained for $\paramon \in
[\paramon_{\indx{i-1}},\paramon_{\indx{i}}]$, we can define even stricter
limits in order to avoid oscillations during the iteration scheme, by setting
$\paramon=1$ to obtain purely linear interpolations, for instance (see
Sect.~\ref{subsec:ali}).

To calculate the elements of the (approximate) $\matown{\Lambda}$-matrix, the
interpolation coefficients are required. Combining Eqs.~\eqref{eq:bezl} and
\eqref{eq:bezl_fc} then gives:
\beq
\label{eq:bezleft_coeff}
f\bigl(x\in[\ximo,\xii]\bigr) = \tilde{a}^{[\indx{i-1},\indx{i}]}\fimo +
\tilde{b}^{[\indx{i-1},\indx{i}]}\fii + \tilde{c}^{[\indx{i-1},\indx{i}]}\fipo \,,
\eeq
with
\begin{eqnarray}
\label{eq:bezleft_coeffa}
\tilde{a}^{[\indx{i-1},\indx{i}]} &=& 1 + (\paramon-2)\dfrac{x-\ximo}{\xii-\ximo} +
(1-\paramon)\Biggl(\dfrac{x-\ximo}{\xii-\ximo}\Biggr)^2 \\\nonumber
\tilde{b}^{[\indx{i-1},\indx{i}]} &=&
\dfrac{(1-\paramon)\dxi+(2-\paramon)\dxipo}{\dxipo}\dfrac{x-\ximo}{\xii-\ximo} 
\\\label{eq:bezleft_coeffb} &+& (\paramon-1) \dfrac{\dxi+\dxipo}{\dxipo}\Biggl( \dfrac{x-\ximo}{\xii-\ximo} 
\Biggr)^2  \\\nonumber
\tilde{c}^{[\indx{i-1},\indx{i}]} &=&
\dfrac{(\paramon-1)\dxi}{\dxipo}\dfrac{x-\ximo}{\xii-\ximo} \\\label{eq:bezleft_coeffc}
&-& \dfrac{(\paramon-1)\dxi}{\dxipo}\Biggl(\dfrac{x-\ximo}{\xii-\ximo}\Biggr)^2 \,.
\end{eqnarray}
%
%
\subsection{Interval $[\xii,\xipo]$}
\label{subsec:app_bez2}
The interpolation formula for the right interval $[\xii,\xipo]$ uses the same
data points as above. Since the value of the control point needs to be
calculated at a different x-coordinate, $x_{\indxc}=(\xipo+\xii)/2$, we cannot
simply substitute indices. Using
\beqa
\label{eq:bezr}
f(x) &=& (1-t)^2\fii + 2t(1-t)f_{\indxc} + t^2 \fipo \\\nonumber
t &:=&(x-\xii)/(\xipo-\xii)  \\\label{eq:bezr_fc}
f_{\indxc} &=& \fii + \dfrac{\dxipo}{2}\Biggl( \paramon \dfrac{\fipo-\fii}{\dxipo} +
(1-\paramon) \dfrac{\fii-\fimo}{\dxi} \Biggr) \,,
\eeqa
we obtain for this interval:
\beq
\label{eq:bezr_coeff}
f(x\in[\xii,\xipo]) = \tilde{a}^{[\indx{i},\indx{i+1}]}\fimo + \tilde{b}^{[\indx{i},\indx{i+1}]}\fii + \tilde{c}^{[\indx{i},\indx{i+1}]}\fipo \,,
\eeq
with
\beqa
\nonumber
\tilde{a}^{[\indx{i},\indx{i+1}]} &=&
\dfrac{(\paramon-1)\dxipo}{\dxi}\dfrac{x-\xii}{\xipo-\xii} \\\label{eq:bezr_coeffa}
&-& \dfrac{(\paramon-1)\dxipo}{\dxi}\Biggl(\dfrac{x-\xii}{\xipo-\xii}\Biggr)^2 \\
\nonumber
\tilde{b}^{[\indx{i},\indx{i+1}]} &=& 1 -
\dfrac{\paramon\dxi+(\paramon-1)\dxipo}{\dxi}\dfrac{x-\xii}{\xipo-\xii} \\\label{eq:bezr_coeffb}
&+& (\paramon-1)\dfrac{\dxi+\dxipo}{\dxi} \Biggl( \dfrac{x-\xii}{\xipo-\xii}
\Biggr)^2  \\
\label{eq:bezr_coeffc}
\tilde{c}^{[\indx{i},\indx{i+1}]} &=& \paramon \dfrac{x-\xii}{\xipo-\xii} + (1-\paramon)\Biggl(
\dfrac{x-\xii}{\xipo-\xii} \,,
\Biggr)^2
\eeqa
and
\beqa
\label{eq:alphar1}
\paramon_{\indx{i}}^{[\indx{i},\indx{i+1}]} &:=& \paramon (f_{\indxc}^{[\indx{i},\indx{i+1}]}=\fii) =
\dfrac{1}{1-\frac{\fipo-\fii}{\fii-\fimo}\frac{\dxi}{\dxipo}} \\
\label{eq:alphar2}
\paramon_{\indx{i+1}}^{[\indx{i},\indx{i+1}]} &:=& \paramon (f_{\indxc}^{[\indx{i},\indx{i+1}]}=\fipo) =
1 + \dfrac{1}{1-\frac{\fii-\fimo}{\fipo-\fii}\frac{\dxipo}{\dxi}} \,.
\eeqa
The corresponding B\'ezier curves for different $\paramon$-parameters
($\paramon=1$ for linear and $\paramon=\dxi/(\dxi+\dxipo)$ for continuous
quadratic interpolations) are also shown in Fig.~\ref{fig:bez}. We note that
the B\'ezier interpolation gives a continuous function in the complete
interval $[\ximo, \xipo]$ only for those $\paramon$-values that define the
parabola connecting all three data points.
%
%
\section{2D B\'ezier interpolation}
\label{app:bez2d}
\begin{figure}[t]
\psfrag{i-1j-1}{$\indx{i-1,j-1}$}
\psfrag{ij-1}{$\indx{i,j-1}$}
\psfrag{i+1j-1}{$\indx{i+1,j-1}$}
\psfrag{i-1j}{$\indx{i-1,j}$}
\psfrag{ij}{$\indx{i,j}$}
\psfrag{i+1j}{$\indx{i+1,j}$}
\psfrag{i-1j+1}{$\indx{i-1,j+1}$}
\psfrag{ij+1}{$\indx{i,j+1}$}
\psfrag{i+1j+1}{$\indx{i+1,j+1}$}
\psfrag{j+1}{$x,\indx{j+1}$}
\psfrag{j}{$x,\indx{j}$}
\psfrag{j-1}{$x,\indx{j-1}$}
\psfrag{xy}{$(x,y)$}
\resizebox{\hsize}{!}{\includegraphics{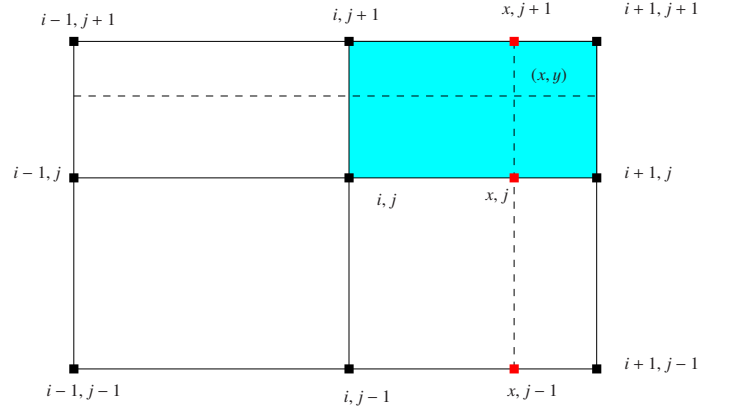}}
\caption{2D interpolation for upwind or downwind quantities required in the
  cyan shaded area. The 2D B\'ezier interpolation consists of three 1D
  interpolations to obtain the values at the desired $x$-coordinate (indicated
  by red dots), followed by a 1D interpolation along the y-coordinate using
  the obtained values at the red dots.}
\label{fig:interp2d}
\end{figure}
To interpolate upwind and downwind quantities, a 2D interpolation scheme is
required. Fig.~\ref{fig:interp2d} displays the geometry for a 2D rectangular
area, with grid points indicated by the black dots. With this setup, we
perform a 2D B\'ezier interpolation by applying three 1D B\'ezier
interpolations along the $x$-axis on each $y$-level at $(\indx{j-1})$,
$(\indx{j})$, $(\indx{j+1})$, followed by another 1D B\'ezier interpolation
along $y$ at the desired $x$-coordinate. Within the cyan shaded interval, we
obtain with the 1D B\'ezier interpolation given by
Eqs.~\eqref{eq:bezr_coeff}-\eqref{eq:bezr_coeffc}:
\begin{align}
\nonumber
f(x,y) &= \tilde{a}_y \tilde{a}_x^{(\indx{j-1})} f_{\indx{i-1,j-1}} + \tilde{a}_y
\tilde{b}_x^{(\indx{j-1})} f_{\indx{i,j-1}} +
\tilde{a}_y\tilde{c}_x^{(\indx{j-1})} f_{\indx{i+1,j-1}} \\\nonumber 
&+
\tilde{b}_y \tilde{a}_x^{(\indx{j})} f_{\indx{i-1,j}} + \tilde{b}_y
\tilde{b}_x^{(\indx{j})} f_{\indx{i,j}} + \tilde{b}_y\tilde{c}_x^{(\indx{j})}
f_{\indx{i+1,j}} \\\label{eq:interp2d} 
&+ \tilde{c}_y \tilde{a}_x^{(\indx{j+1})} f_{\indx{i-1,j+1}} + \tilde{c}_y
\tilde{b}_x^{(\indx{j+1})} f_{\indx{i,j+1}} + \tilde{c}_y\tilde{c}_x^{(\indx{j+1})} f_{\indx{i+1,j+1}} \,,
\end{align}
where the subscripts of the interpolation coefficients indicate the coordinate
used for each 1D interpolation. We note that all upwind and downwind
interpolations are performed in the upper right interval of a given surface,
in order to obtain a simple representation of the $\matown{\Lambda}$-matrix
elements.
%
%
\section{ALO coefficients}
\label{app:alo}
\begin{figure}[t]
\psfrag{iim2jm1}{\Large{$I_{\indx{i-2\alpha,j-\beta}}=0$}}
\psfrag{iim2jm2}{\Large{$I_{\indx{i-2\alpha,j-2\beta}}=0$}}
\psfrag{iuim1jm1}{\Large{$I_{\indxu}^{\indx{i-\alpha,j-\beta}}=0$}}
\psfrag{suim1jm1}{\Large{$S_{\indxu}^{\indx{i-\alpha,j-\beta}}=0$}}
\psfrag{ipim1jm1}{\Large{$\Lambda_{\indx{i-\alpha,j-\beta}}^{\indx{ij}}=I_{\indx{i-\alpha,j-\beta}}$}}
\psfrag{spim1jm1}{\Large{$S_{\indxp}^{\indx{i-\alpha,j-\beta}}=S_{\indx{i-\alpha, j-\beta}}=0$}}
\psfrag{sdim1jm1}{\Large{$S_{\indxd}^{\indx{i-\alpha,j-\beta}}\neq0$}}
\psfrag{sij}{\Large{$S_{\indx{i,j}}=1$}}
\psfrag{iuijm1}{\Large{$I_{\indxu}^{\indx{i,j-\beta}}\neq 0$}}
\psfrag{suijm1}{\Large{$S_{\indxu}^{\indx{i,j-\beta}}=0$}}
\psfrag{ipijm1}{\Large{$\Lambda_{\indx{i,j-\beta}}^{\indx{ij}}=I_{\indx{i,j-\beta}}$}}
\psfrag{spijm1}{\Large{$S_{\indxp}^{\indx{i,j-\beta}}=S_{\indx{i, j-\beta}}=0$}}
\psfrag{sdijm1}{\Large{$S_{\indxd}^{\indx{i-\alpha,j-\beta}}\neq0$}}
\psfrag{sij}{\Large{$S_{\indx{i,j}}=1$}}
\psfrag{iim1jm1}{\Large{$I_{\indx{i-\alpha,j-\beta}}=\Lambda_{\indx{i-\alpha,j-\beta}}^{\indx{ij}}$}}
\psfrag{iim1jm2}{\Large{$I_{\indx{i-\alpha,j-2\beta}}=0$}}
\psfrag{n}{\Large{$\vecown{n}$}}
\psfrag{upanel}{
   \begin{minipage}{2cm}
      \large{(see upper \\panel)}
   \end{minipage}
}
\resizebox{\hsize}{!}{\includegraphics{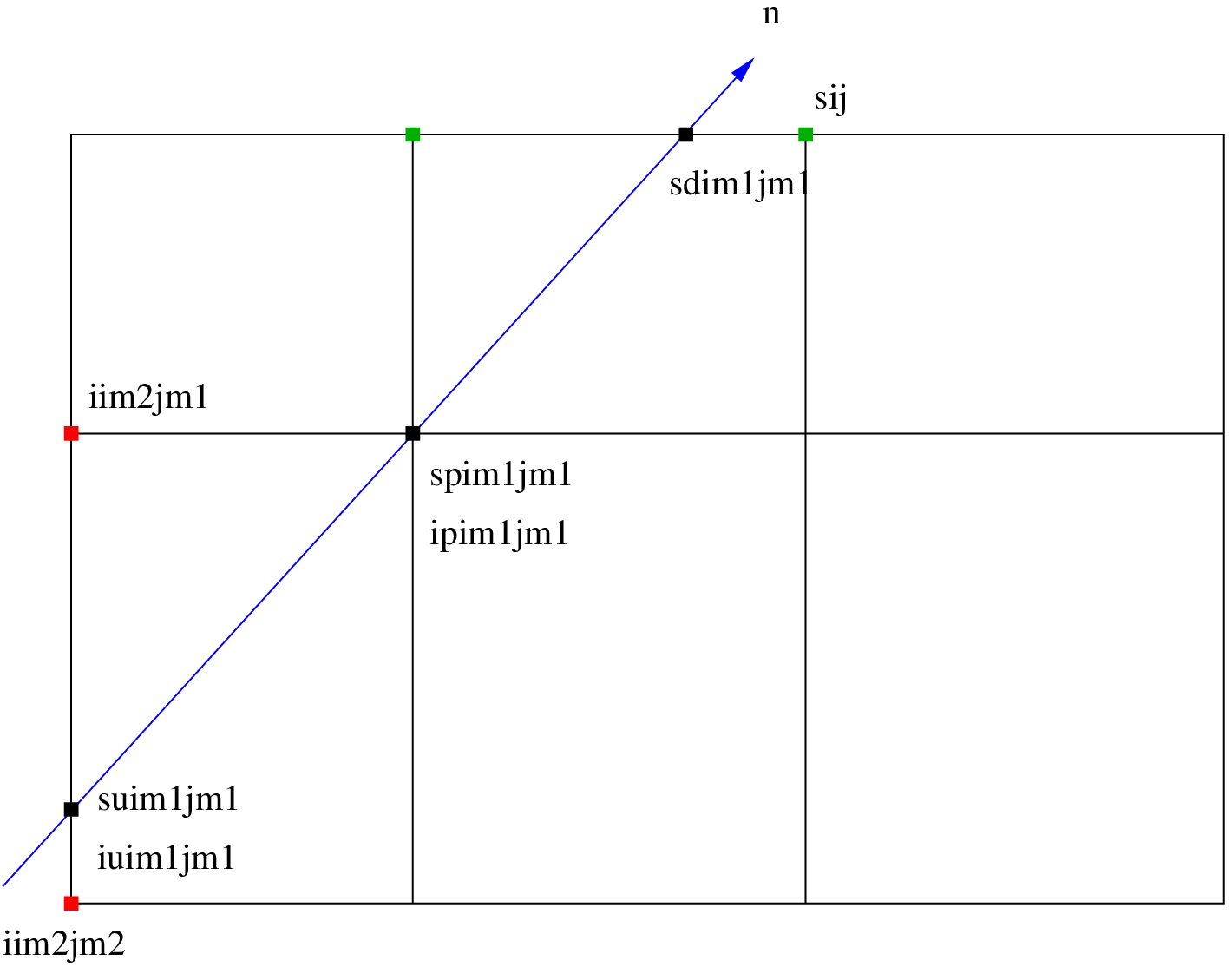}}
\\
\resizebox{\hsize}{!}{\includegraphics{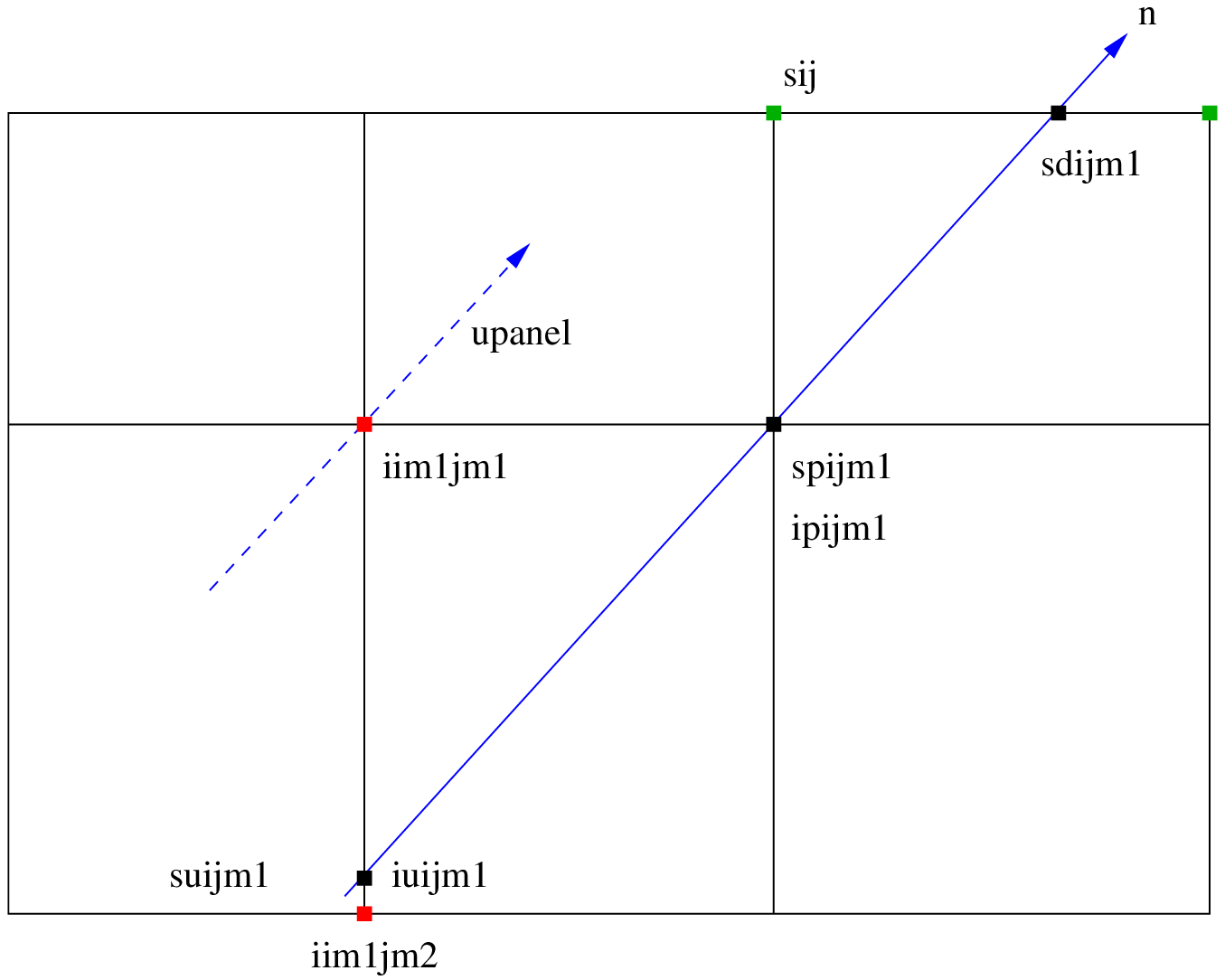}}
\caption{2D example for calculating the $\matown{\Lambda}_{\Omega,\nu}$-matrix
  elements at a grid point $(\indx{i-\alpha,j-\beta})$ (upper panel) and
  $(\indx{i,j-\beta})$ (lower panel). The matrix elements correspond to the
  intensity at the considered grid points calculated for a source function
  $S_{\indx{ij}}=1$ and zero everywhere else. For such a configuration, the
  downwind source function is interpolated from grid points indicated with the
  green dots, while the upwind source function and upwind intensity are
  obtained from the red dots (for simplicity we here assume linear
  interpolations for determining upwind and downwind quantities). We emphasize
  that the upwind intensity vanishes only when considering the grid point
  $(\indx{i-\alpha,j-\beta})$.}
\label{fig:alocoeff}
\end{figure}
In this section, we derive the $\matown{\Lambda}$-matrix coefficients used to
construct the approximate $\Lambda$-operator. We note that the obtained matrix
elements can also be used for any other (2nd or lower order) interpolation
scheme using the same geometry, with different interpolation coefficients
though.

For a source function set to unity at grid point $(\indx{ijk})$ and zero
everywhere else, we consider all 27 points ranging from
$(\indx{i-\alpha,j-\beta,k-\gamma})$ to
$(\indx{i+\alpha,j+\beta,k+\gamma})$. The corresponding matrix coefficients
are derived from Eq.~\eqref{eq:lambda_elements}, using the discretized
equation of radiative transfer,
Eqs.~\eqref{eq:eqrt_disc}/\eqref{eq:eqrt_disc2}, with upwind and downwind
interpolations defined by Eqs.~\eqref{eq:interp2du} and
\eqref{eq:interp2dd}. We further consider only the
$\Lambda_{\Omega\nu}$-operator, since the integration over frequency and/or
solid angle is straightforward. Each $\Lambda$-matrix element then corresponds
to the intensity (resulting from $S_{\indx{ijk}}=1$) at a considered grid
point $\indxp$ (not necessarily identical to $(\indx{ijk})$), and consists of
an emission term (defined by the interpolated source functions and
optical-depth steps at the corresponding upwind, current, and downwind points),
and the irradiation from the upwind point (defined by the upwind intensity and
upwind optical-depth step). The upper and lower panel of
Fig.~\ref{fig:alocoeff} show an example in 2D considering the points
$(\indx{i-\alpha,j-\beta})$ and $(\indx{i,j-\beta})$ for a source function
$S_{\indx{ij}}=1$.

In the following, we sketch the derivation of the (3D) matrix element for the
first neighbour, and only present the solution for the remaining ones.  To
save space, we skip the indices $\Omega,\nu$. The $m,n$-th $\Lambda$-element
is written as $\Lambda_{m}^{n}$, with matrix indices $n, m$ calculated from
Eq.~\eqref{eq:index_conversion}. While $n$ corresponds to the 3D indices of
the local grid point ($S_{\indx{ijk}}=S_{n}=1$), $m$ represents the
neighbouring point, $(\indx{i-\alpha,j-\beta,k-\gamma})$.  Applying
Eq.~\eqref{eq:lambda_elements} to the specific intensity at point $m$, we
obtain:
\beqa
\nonumber
\Lambda_{\indx{m}}^{\indx{n}} &=& I_{\indx{m}} \left(\vecown{S}=\vecown{e}_{\indx{n}},
\vecown{\Phi}_{\rm B}=0 \right) 
\\\nonumber
=\Lambda_{\indx{i-\alpha,j-\beta,k-\gamma}}^{\indx{ijk}} &=&
I_{\indx{i-\alpha,j-\beta,k-\gamma}}\left(S_{\indx{ijk}}=\delta_{\tilde{i},i}\delta_{\tilde{j},j}\delta_{\tilde{k},k}\right) \\\nonumber
&=& a_{\indx{i-\alpha,j-\beta,k-\gamma}}S_{\indxu}^{(\indx{i-\alpha,j-\beta,k-\gamma})}\left(S_{\indx{ijk}}=\delta_{\tilde{i},i}\delta_{\tilde{j},j}\delta_{\tilde{k},k}\right) \\\nonumber
&+& b_{\indx{i-\alpha,j-\beta,k-\gamma}}S_{\indxp}^{(\indx{i-\alpha,j-\beta,k-\gamma})}\left(S_{\indx{ijk}}=\delta_{\tilde{i},i}\delta_{\tilde{j},j}\delta_{\tilde{k},k}\right) \\\nonumber
&+& c_{\indx{i-\alpha,j-\beta,k-\gamma}}S_{\indxd}^{(\indx{i-\alpha,j-\beta,k-\gamma})}\left(S_{\indx{ijk}}=\delta_{\tilde{i},i}\delta_{\tilde{j},j}\delta_{\tilde{k},k}\right) \\\nonumber 
&+& d_{\indx{i-\alpha,j-\beta,k-\gamma}}I_{\indxu}^{(\indx{i-\alpha,j-\beta,k-\gamma})}\left(S_{\indx{ijk}}=\delta_{\tilde{i},i}\delta_{\tilde{j},j}\delta_{\tilde{k},k}\right) \,,
\eeqa
with boundary contribution $\vecown{\Phi}_{\rm B}$, $n$-th unit vector
$\vecown{e}_{\indx{n}}$, and $\delta_{\tilde{i},i}$, $\delta_{\tilde{j},j}$,
$\delta_{\tilde{k},k}$ the Kronecker-$\delta$ for all possible
$x_{\tilde{i}}$, $y_{\tilde{j}}$, and $z_{\tilde{k}}$ coordinates,
respectively. $S_{\indxu}$ and $S_{\indxd}$ are the upwind and downwind source
functions corresponding to a considered short characteristic at grid point
$\indxp \leftrightarrow (\indx{i-\alpha,j-\beta,k-\gamma})$, $S_{\indxp}$ is
the source function at the grid point\footnote{$S_{\indxp}\neq 0$ only when
  considering the grid point $\indxp \leftrightarrow (\indx{ijk})$. Then,
  $S_{\indxd}^{(\indx{ijk})}=0$, and $S_{\indxu}^{(\indx{ijk})}\neq 0$ only
  when the upwind point is located on the stellar surface
  (Eq.~\eqref{eq:alocoeff14}).}, $I_{\indxu}$ is the upwind intensity, and
$a,b,c,d$ are the integration coefficients for this particular short
characteristic. All upwind and downwind quantities are to be interpolated from
neighbouring grid points. We use the notation $w$, $\hat{w}$, $\tilde{w}$, to
identify different interpolation coefficients corresponding to the upwind
source function, upwind intensity, and downwind source function,
respectively. Using Eqs.~\eqref{eq:interp2du} and \eqref{eq:interp2dd} to
interpolate upwind and downwind quantities, we find:
\beqa
\nonumber
\Lambda_{\indx{i-\alpha,j-\beta,k-\gamma}}^{\indx{ijk}} &=& 
a_{\indx{i-\alpha,j-\beta,k-\gamma}}\cdot \bigl[ w_{\rm A} S_{\indx{i-3\alpha,j-2\beta,k-3\gamma}} \\\nonumber
&+&\quad w_{\rm B} S_{\indx{i-2\alpha,j-2\beta,k-3\gamma}}
+w_{\rm C} S_{\indx{i-\alpha,j-2\beta,k-3\gamma}} \\\nonumber
&+&\quad w_{\rm D} S_{\indx{i-3\alpha,j-2\beta,k-2\gamma}}
+w_{\rm E} S_{\indx{i-2\alpha,j-2\beta,k-2\gamma}}\\\nonumber
&+&\quad w_{\rm F} S_{\indx{i-\alpha,j-2\beta,k-2\gamma}}
+w_{\rm G} S_{\indx{i-3\alpha,j-2\beta,k-\gamma}}\\\nonumber
&+&\quad w_{\rm H} S_{\indx{i-2\alpha,j-2\beta,k-\gamma}}
+w_{\rm I} S_{\indx{i-\alpha,j-2\beta,k-\gamma}}\\\nonumber
&+&\quad w_{\rm J} S_{\indx{i-3\alpha,j-3\beta,k-2\gamma}}
+w_{\rm K} S_{\indx{i-2\alpha,j-3\beta,k-2\gamma}}\\\nonumber
&+&\quad w_{\rm L} S_{\indx{i-\alpha,j-3\beta,k-2\gamma}}
+w_{\rm M} S_{\indx{i-3\alpha,j-\beta,k-2\gamma}}\\\nonumber
&+&\quad w_{\rm N} S_{\indx{i-2\alpha,j-\beta,k-2\gamma}}
+w_{\rm O} S_{\indx{i-\alpha,j-\beta,k-2\gamma}}\\\nonumber
&+&\quad w_{\rm P} S_{\indx{i-2\alpha,j-3\beta,k-3\gamma}}
+w_{\rm Q} S_{\indx{i-2\alpha,j-\beta,k-3\gamma}}\\\nonumber
&+&\quad w_{\rm R} S_{\indx{i-2\alpha,j-3\beta,k-\gamma}}
+w_{\rm S} S_{\indx{i-2\alpha,j-\beta,k-\gamma}}\\\nonumber
&+&\quad w_{\indx{i-\alpha,j-\beta,k-\gamma}} S_{\indx{ i-\alpha,j-\beta,k-\gamma}}
 \bigr] \\\nonumber
&+& b_{\indx{i-\alpha,j-\beta,k-\gamma}} S_{\indx{ i-\alpha, j-\beta,k-\gamma}}
\\\nonumber
&+& c_{\indx{i-\alpha,j-\beta,k-\gamma}} \cdot \bigl[ 
\tilde{w}_{\rm A} S_{\indx{ i-2\alpha,j,k-2\gamma}}\\\nonumber
&+&\quad \tilde{w}_{\rm B} S_{\indx{ i-\alpha,j,k-2\gamma}}
+\tilde{w}_{\rm C} S_{\indx{ i,j,k-2\gamma}}\\\nonumber
&+&\quad \tilde{w}_{\rm D} S_{\indx{ i-2\alpha,j,k-\gamma}}
+\tilde{w}_{\rm E} S_{\indx{ i-\alpha,j,k-\gamma}}\\\nonumber
&+&\quad \tilde{w}_{\rm F} S_{\indx{ i,j,k-\gamma}}
+\tilde{w}_{\rm G} S_{\indx{ i-2\alpha,j,k}}\\\nonumber
&+&\quad \tilde{w}_{\rm H} S_{\indx{ i-\alpha,j,k}}
+\tilde{w}_{\rm I} S_{\indx{ i,j,k}}\\\nonumber
&+&\quad \tilde{w}_{\rm J} S_{\indx{ i-2\alpha,j-2\beta,k}}
+\tilde{w}_{\rm K} S_{\indx{ i-\alpha,j-2\beta,k}}\\\nonumber
&+&\quad \tilde{w}_{\rm L} S_{\indx{ i,j-2\beta,k}}
+\tilde{w}_{\rm M} S_{\indx{ i-2\alpha,j-\beta,k}}\\\nonumber
&+&\quad \tilde{w}_{\rm N} S_{\indx{ i-\alpha,j-\beta,k}}
+\tilde{w}_{\rm O} S_{\indx{ i,j-\beta,k}}\\\nonumber
&+&\quad \tilde{w}_{\rm P} S_{\indx{ i,j-2\beta,k-2\gamma}}
+\tilde{w}_{\rm Q} S_{\indx{ i,j-\beta,k-2\gamma}}\\\nonumber
&+&\quad \tilde{w}_{\rm R} S_{\indx{ i,j-2\beta,k-\gamma}}
+\tilde{w}_{\rm S} S_{\indx{ i,j-\beta,k-\gamma}} \bigr] \\\nonumber
&+& d_{\indx{i-\alpha,j-\beta,k-\gamma}} \bigl[\hat{w}_{\rm A}
  I_{\indx{i-3\alpha,j-2\beta,k-3\gamma}} (S_{\indx{ijk}}=1) + \cdots
  \\\nonumber 
&+& \quad \hat{w}_{\rm S} I_{\indx{i-2\alpha, j-\beta, k-\gamma}}
  (S_{\indx{ ijk}} = 1)\bigr] \,,
\eeqa
with the upwind intensity interpolated from the same points as the upwind
source function, and a compact notation for the interpolation coefficients
(with skipped superscripts). Since only $S_{\indx{ijk}}=1$ (and zero
everywhere else), and because the upwind intensity vanishes (for this
particular grid point, see Fig.~\ref{fig:alocoeff} for an example in 2D), we
finally obtain:
\beq\label{eq:alocoeff01}
\Lambda_{\indx{i-\alpha,j-\beta,k-\gamma}}^{\indx{ijk}} =
c_{\indx{i-\alpha,j-\beta,k-\gamma}} \tilde{w}_{\rm
  I}^{(\indx{i-\alpha,j-\beta,k-\gamma})} \,.
\eeq
The matrix element for a point $(\indx{i-\alpha,j-\beta,k-\gamma})$ with a
non-vanishing source function at point $(\indx{ijk})$ is thus solely given by
the integration coefficient $c_{\indx{i-\alpha,j-\beta,k-\gamma}}$ from the
discretized equation of radiative transfer multiplied with the interpolation
coefficient for the downwind source function of point I (corresponding to grid
point $(\indx{ijk})$, see Fig.~\ref{fig:sc_cell_interp}). The other neighbours
are obtained analogously, without vanishing incident intensities,
however. Accounting also for the interpolation of upwind source functions and
intensities when necessary, we find:
\beqa
\nonumber
\Lambda_{\indx{i,j-\beta,k-\gamma}}^{\indx{ijk}} &=& c_{\indx{i,j-\beta,k-\gamma}}\tilde{w}_{\rm
  H}^{\indx{i,j-\beta,k-\gamma}} \\ \label{eq:alocoeff02}
 &+& d_{\indx{i,j-\beta,k-\gamma}}\hat{w}_{\rm S}^{\indx{i,j-\beta,k-\gamma}}
\Lambda_{\indx{i-\alpha,j-\beta,k-\gamma}}^{\indx{ijk}}
\eeqa
\beqa
\nonumber
\Lambda_{\indx{i+\alpha,j-\beta,k-\gamma}}^{\indx{ijk}} &=&
c_{\indx{i+\alpha,j-\beta,k-\gamma}}\tilde{w}_{\rm G}^{\indx{i+\alpha,j-\beta,k-\gamma}} \\ \label{eq:alocoeff03}
 &+& d_{\indx{i+\alpha,j-\beta,k-\gamma}}\hat{w}_{\rm
  S}^{\indx{i+\alpha,j-\beta,k-\gamma}} \Lambda_{\indx{i,j-\beta,k-\gamma}}^{\indx{ijk}}
\eeqa
\beqa
\nonumber
\Lambda_{\indx{i-\alpha,j,k-\gamma}}^{\indx{ijk}} &=& c_{\indx{i-\alpha,j,k-\gamma}}\tilde{w}_{\rm
  O}^{\indx{i-\alpha,j,k-\gamma}} \\\label{eq:alocoeff04}
 &+& d_{\indx{i-\alpha,j,k-\gamma}}\hat{w}_{\rm
  I}^{\indx{i-\alpha,j,k-\gamma}} \Lambda_{\indx{i-\alpha,j-\beta,k-\gamma}}^{\indx{ijk}}
\eeqa
\beqa
\nonumber
\Lambda_{\indx{i,j,k-\gamma}}^{\indx{ijk}} &=& c_{\indx{i,j,k-\gamma}}\tilde{w}_{\rm
  N}^{\indx{i,j,k-\gamma}}  + d_{\indx{i,j,k-\gamma}} \cdot \Bigl[
 \hat{w}_{\rm H}^{\indx{i,j,k-\gamma}} \Lambda_{\indx{i-\alpha,j-\beta,k-\gamma}}^{\indx{ijk}} \\\label{eq:alocoeff05}
&+& 
 \hat{w}_{\rm I}^{\indx{i,j,k-\gamma}} \Lambda_{\indx{i,j-\beta,k-\gamma}}^{\indx{ijk}}
 + \hat{w}_{\rm S}^{\indx{i,j,k-\gamma}} \Lambda_{\indx{i-\alpha,j,k-\gamma}}^{\indx{ijk}} \Bigr]
\eeqa
\beqa
\nonumber
\Lambda_{\indx{i+\alpha,j,k-\gamma}}^{\indx{ijk}} &=& c_{\indx{i+\alpha,j,k-\gamma}}\tilde{w}_{\rm
  M}^{\indx{i+\alpha,j,k-\gamma}}  +  d_{\indx{i+\alpha,j,k-\gamma}} \\\nonumber
&\cdot& \Bigl[ \hat{w}_{\rm D}^{\indx{i+\alpha,j,k-\gamma}}
  \Lambda_{\indx{i-\alpha,j-\beta,k-\gamma}}^{\indx{ijk}} + \hat{w}_{\rm
    H}^{\indx{i+\alpha,j,k-\gamma}} \Lambda_{\indx{i,j-\beta,k-\gamma}}^{\indx{ijk}} \\\label{eq:alocoeff06}
&+& \hat{w}_{\rm I}^{\indx{i+\alpha,j,k-\gamma}}
\Lambda_{\indx{i+\alpha,j-\beta,k-\gamma}}^{\indx{ijk}} + \hat{w}_{\rm
  S}^{\indx{i+\alpha,j,k-\gamma}} \Lambda_{\indx{i,j,k-\gamma}}^{\indx{ijk}} \Bigr]
\eeqa
\beqa
\nonumber
\Lambda_{\indx{i-\alpha,j+\beta,k-\gamma}}^{\indx{ijk}} &=& c_{\indx{i-\alpha,j+\beta,k-\gamma}}\tilde{w}_{\rm
  L}^{\indx{i-\alpha,j+\beta,k-\gamma}}  \\\label{eq:alocoeff07}
&+& d_{\indx{i-\alpha,j+\beta,k-\gamma}} \hat{w}_{\rm I}^{\indx{i-\alpha,j+\beta,k-\gamma}}
 \Lambda_{\indx{i-\alpha,j,k-\gamma}}^{\indx{ijk}}
\eeqa
\beqa
\nonumber
\Lambda_{\indx{i,j+\beta,k-\gamma}}^{\indx{ijk}} &=& c_{\indx{i,j+\beta,k-\gamma}}\tilde{w}_{\rm
  K}^{\indx{i,j+\beta,k-\gamma}}  + d_{\indx{i,j+\beta,k-\gamma}} \\\nonumber
&\cdot& \Bigl[
\hat{w}_{\rm R}^{\indx{i,j+\beta,k-\gamma}}\Lambda_{\indx{i-\alpha,j-\beta,k-\gamma}}^{\indx{ijk}}+
\hat{w}_{\rm H}^{\indx{i,j+\beta,k-\gamma}}\Lambda_{\indx{i-\alpha,j,k-\gamma}}^{\indx{ijk}} \\\label{eq:alocoeff08}
&+&
\hat{w}_{\rm I}^{\indx{i,j+\beta,k-\gamma}}\Lambda_{\indx{i,j,k-\gamma}}^{\indx{ijk}}+
\hat{w}_{\rm S}^{\indx{i,j+\beta,k-\gamma}}\Lambda_{\indx{i-\alpha,j+\beta,k-\gamma}}^{\indx{ijk}} \Bigr] 
\eeqa
\beqa
\nonumber
\Lambda_{\indx{i+\alpha,j+\beta,k-\gamma}}^{\indx{ijk}} &=& c_{\indx{i+\alpha,j+\beta,k-\gamma}}\tilde{w}_{\rm
  J}^{\indx{i+\alpha,j+\beta,k-\gamma}}  + d_{\indx{i+\alpha,j+\beta,k-\gamma}}  \\\nonumber
&\cdot& \Bigl[ \hat{w}_{\rm R}^{\indx{i+\alpha,j+\beta,k-\gamma}}\Lambda_{\indx{i,j-\beta,k-\gamma}}^{\indx{ijk}}+
\hat{w}_{\rm G}^{\indx{i+\alpha,j+\beta,k-\gamma}}\Lambda_{\indx{i-\alpha,j,k-\gamma}}^{\indx{ijk}} \\\nonumber
&+&
\hat{w}_{\rm H}^{\indx{i+\alpha,j+\beta,k-\gamma}}\Lambda_{\indx{i,j,k-\gamma}}^{\indx{ijk}}+
\hat{w}_{\rm I}^{\indx{i+\alpha,j+\beta,k-\gamma}}\Lambda_{\indx{i+\alpha,j,k-\gamma}}^{\indx{ijk}}
\\\label{eq:alocoeff09}
&+&\hat{w}_{\rm S}^{\indx{i+\alpha,j+\beta,k-\gamma}}\Lambda_{\indx{i,j+\beta,k-\gamma}}^{\indx{ijk}}\Bigr]  
\eeqa
\beqa
\nonumber
\Lambda_{\indx{i-\alpha,j-\beta,k}}^{\indx{ijk}} &=& c_{\indx{i-\alpha,j-\beta,k}}\tilde{w}_{\rm
  F}^{\indx{i-\alpha,j-\beta,k}}  \\\label{eq:alocoeff10}
&+& d_{\indx{i-\alpha,j-\beta,k}} \hat{w}_{\rm O}^{\indx{i-\alpha,j-\beta,k}}\Lambda_{\indx{i-\alpha,j-\beta,k-\gamma}}^{\indx{ijk}}
\eeqa
\beqa
\nonumber
\Lambda_{\indx{i,j-\beta,k}}^{\indx{ijk}} &=& c_{\indx{i,j-\beta,k}}\tilde{w}_{\rm
  E}^{\indx{i,j-\beta,k}} + d_{\indx{i,j-\beta,k}} \cdot \Bigl[
\hat{w}_{\rm N}^{\indx{i,j-\beta,k}}\Lambda_{\indx{i-\alpha,j-\beta,k-\gamma}}^{\indx{ijk}}
\\\label{eq:alocoeff11} &+&
\hat{w}_{\rm O}^{\indx{i,j-\beta,k}}\Lambda_{\indx{i,j-\beta,k-\gamma}}^{\indx{ijk}} +
\hat{w}_{\rm S}^{\indx{i,j-\beta,k}}\Lambda_{\indx{i-\alpha,j-\beta,k}}^{\indx{ijk}} \Bigr]
\eeqa
\beqa
\nonumber
\Lambda_{\indx{i+\alpha,j-\beta,k}}^{\indx{ijk}} &=& c_{\indx{i+\alpha,j-\beta,k}}\tilde{w}_{\rm
  D}^{\indx{i+\alpha,j-\beta,k}} + d_{\indx{i+\alpha,j-\beta,k}} \\\nonumber
&\cdot& \Bigl[
\hat{w}_{\rm M}^{\indx{i+\alpha,j-\beta,k}}\Lambda_{\indx{i-\alpha,j-\beta,k-\gamma}}^{\indx{ijk}}+
\hat{w}_{\rm N}^{\indx{i+\alpha,j-\beta,k}}\Lambda_{\indx{i,j-\beta,k-\gamma}}^{\indx{ijk}} \\\label{eq:alocoeff12}
&+&
\hat{w}_{\rm O}^{\indx{i+\alpha,j-\beta,k}}\Lambda_{\indx{i+\alpha,j-\beta,k-\gamma}}^{\indx{ijk}}+
\hat{w}_{\rm S}^{\indx{i+\alpha,j-\beta,k}}\Lambda_{\indx{i,j-\beta,k}}^{\indx{ijk}} \Bigr] 
\eeqa
\beqa
\nonumber
\Lambda_{\indx{i-\alpha,j,k}}^{\indx{ijk}} &=& c_{\indx{i-\alpha,j,k}}\tilde{w}_{\rm
  S}^{\indx{i-\alpha,j,k}}  + d_{\indx{i-\alpha,j,k}} \cdot \Bigl[
 \hat{w}_{\rm F}^{\indx{i-\alpha,j,k}} \Lambda_{\indx{i-\alpha,j-\beta,k-\gamma}}^{\indx{ijk}} \\\label{eq:alocoeff13}
&+& 
 \hat{w}_{\rm O}^{\indx{i-\alpha,j,k}} \Lambda_{\indx{i-\alpha,j,k-\gamma}}^{\indx{ijk}} +
 \hat{w}_{\rm I}^{\indx{i-\alpha,j,k}} \Lambda_{\indx{i-\alpha,j-\beta,k}}^{\indx{ijk}} \Bigr]
\eeqa
\beqa
\nonumber
\Lambda_{\indx{ijk}}^{\indx{ijk}} &=& a_{\indx{ijk}}w_{\indx{ijk}} + b_{\indx{ijk}} + d_{\indx{ijk}} \\\nonumber
&\cdot& \Bigl[
 \hat{w}_{\rm E}^{\indx{ijk}} \Lambda_{\indx{i-\alpha,j-\beta,k-\gamma}}^{\indx{ijk}} +
 \hat{w}_{\rm F}^{\indx{ijk}} \Lambda_{\indx{i,j-\beta,k-\gamma}}^{\indx{ijk}} \\\nonumber
&+&
 \hat{w}_{\rm N}^{\indx{ijk}} \Lambda_{\indx{i-\alpha,j,k-\gamma}}^{\indx{ijk}} +
 \hat{w}_{\rm O}^{\indx{ijk}} \Lambda_{\indx{i,j,k-\gamma}}^{\indx{ijk}} +
 \hat{w}_{\rm H}^{\indx{ijk}} \Lambda_{\indx{i-\alpha,j-\beta,k}}^{\indx{ijk}} \\\label{eq:alocoeff14}
&+&
 \hat{w}_{\rm I}^{\indx{ijk}} \Lambda_{\indx{i,j-\beta,k}}^{\indx{ijk}} +
 \hat{w}_{\rm S}^{\indx{ijk}} \Lambda_{\indx{i-\alpha,j,k}}^{\indx{ijk}} \Bigr]
\eeqa
\beqa
\nonumber
\Lambda_{\indx{i+\alpha,j,k}}^{\indx{ijk}} &=& a_{\indx{i+\alpha,j,k}}w_{\rm S}^{\indx{i+\alpha,j,k}} +
d_{\indx{i+\alpha,j,k}} \\\nonumber
&\cdot& \Bigl[
 \hat{w}_{\rm D}^{\indx{i+\alpha,j,k}} \Lambda_{\indx{i-\alpha,j-\beta,k-\gamma}}^{\indx{ijk}} +
 \hat{w}_{\rm E}^{\indx{i+\alpha,j,k}} \Lambda_{\indx{i,j-\beta,k-\gamma}}^{\indx{ijk}} \\\nonumber
&+&
 \hat{w}_{\rm F}^{\indx{i+\alpha,j,k}} \Lambda_{\indx{i+\alpha,j-\beta,k-\gamma}}^{\indx{ijk}} +
 \hat{w}_{\rm M}^{\indx{i+\alpha,j,k}} \Lambda_{\indx{i-\alpha,j,k-\gamma}}^{\indx{ijk}}
 \\\nonumber
&+& 
\hat{w}_{\rm N}^{\indx{i+\alpha,j,k}} \Lambda_{\indx{i,j,k-\gamma}}^{\indx{ijk}} +
 \hat{w}_{\rm O}^{\indx{i+\alpha,j,k}} \Lambda_{\indx{i+\alpha,j,k-\gamma}}^{\indx{ijk}}
 \\\nonumber
&+&
 \hat{w}_{\rm G}^{\indx{i+\alpha,j,k}} \Lambda_{\indx{i-\alpha,j-\beta,k}}^{\indx{ijk}} +
 \hat{w}_{\rm H}^{\indx{i+\alpha,j,k}} \Lambda_{\indx{i,j-\beta,k}}^{\indx{ijk}} \\\label{eq:alocoeff15}
&+&
 \hat{w}_{\rm I}^{\indx{i+\alpha,j,k}} \Lambda_{\indx{i+\alpha,j-\beta,k}}^{\indx{ijk}} +
 \hat{w}_{\rm S}^{\indx{i+\alpha,j,k}} \Lambda_{\indx{ijk}}^{\indx{ijk}} \Bigr]
\eeqa
\beqa
\nonumber
\Lambda_{\indx{i-\alpha,j+\beta,k}}^{\indx{ijk}} &=& c_{\indx{i-\alpha,j+\beta,k}}\tilde{w}_{\rm
  R}^{\indx{i-\alpha,j+\beta,k}} + d_{\indx{i-\alpha,j+\beta,k}} \\\nonumber
&\cdot& \Bigl[
 \hat{w}_{\rm L}^{\indx{i-\alpha,j+\beta,k}} \Lambda_{\indx{i-\alpha,j-\beta,k-\gamma}}^{\indx{ijk}} +
 \hat{w}_{\rm F}^{\indx{i-\alpha,j+\beta,k}} \Lambda_{\indx{i-\alpha,j,k-\gamma}}^{\indx{ijk}}
 \\\label{eq:alocoeff16}
&+&
 \hat{w}_{\rm O}^{\indx{i-\alpha,j+\beta,k}} \Lambda_{\indx{i-\alpha,j+\beta,k-\gamma}}^{\indx{ijk}} +
 \hat{w}_{\rm I}^{\indx{i-\alpha,j+\beta,k}} \Lambda_{\indx{i-\alpha,j,k}}^{\indx{ijk}}  \Bigr]
\eeqa
\beqa
\nonumber
\Lambda_{\indx{i,j+\beta,k}}^{\indx{ijk}} &=& a_{\indx{i,j+\beta,k}}w_{\rm I}^{\indx{i,j+\beta,k}} +
d_{\indx{i,j+\beta,k}} \\\nonumber
&\cdot& \Bigl[
 \hat{w}_{\rm E}^{\indx{i,j+\beta,k}} \Lambda_{\indx{i-\alpha,j,k-\gamma}}^{\indx{ijk}} +
 \hat{w}_{\rm F}^{\indx{i,j+\beta,k}} \Lambda_{\indx{i,j,k-\gamma}}^{\indx{ijk}}  \\\nonumber
&+&
 \hat{w}_{\rm H}^{\indx{i,j+\beta,k}} \Lambda_{\indx{i-\alpha,j,k}}^{\indx{ijk}} +
 \hat{w}_{\rm I}^{\indx{i,j+\beta,k}} \Lambda_{\indx{ijk}}^{\indx{ijk}} \\\nonumber
&+& 
\hat{w}_{\rm K}^{\indx{i,j+\beta,k}} \Lambda_{\indx{i-\alpha,j-\beta,k-\gamma}}^{\indx{ijk}} +
\hat{w}_{\rm L}^{\indx{i,j+\beta,k}} \Lambda_{\indx{i,j-\beta,k-\gamma}}^{\indx{ijk}} \\\nonumber
&+&
 \hat{w}_{\rm N}^{\indx{i,j+\beta,k}} \Lambda_{\indx{i-\alpha,j+\beta,k-\gamma}}^{\indx{ijk}} +
 \hat{w}_{\rm O}^{\indx{i,j+\beta,k}} \Lambda_{\indx{i,j+\beta,k-\gamma}}^{\indx{ijk}} \\\label{eq:alocoeff17}
&+&
 \hat{w}_{\rm R}^{\indx{i,j+\beta,k}} \Lambda_{\indx{i-\alpha,j-\beta,k}}^{\indx{ijk}} +
 \hat{w}_{\rm S}^{\indx{i,j+\beta,k}} \Lambda_{\indx{i-\alpha,j+\beta,k}}^{\indx{ijk}} +
  \Bigr]
\eeqa
\beqa
\nonumber
\Lambda_{\indx{i+\alpha,j+\beta,k}}^{\indx{ijk}} &=& a_{\indx{i+\alpha,j+\beta,k}}w_{\rm H}^{\indx{i+\alpha,j+\beta,k}} +
d_{\indx{i+\alpha,j+\beta,k}} \\\nonumber
&\cdot& \Bigl[
 \hat{w}_{\rm D}^{\indx{i+\alpha,j+\beta,k}} \Lambda_{\indx{i-\alpha,j,k-\gamma}}^{\indx{ijk}} +
 \hat{w}_{\rm E}^{\indx{i+\alpha,j+\beta,k}} \Lambda_{\indx{i,j,k-\gamma}}^{\indx{ijk}}  \\\nonumber
&+&
 \hat{w}_{\rm F}^{\indx{i+\alpha,j+\beta,k}} \Lambda_{\indx{i+\alpha,j,k-\gamma}}^{\indx{ijk}} +
 \hat{w}_{\rm G}^{\indx{i+\alpha,j+\beta,k}} \Lambda_{\indx{i-\alpha,j,k}}^{\indx{ijk}} \\\nonumber
&+&
 \hat{w}_{\rm H}^{\indx{i+\alpha,j+\beta,k}} \Lambda_{\indx{ijk}}^{\indx{ijk}} +
 \hat{w}_{\rm I}^{\indx{i+\alpha,j+\beta,k}} \Lambda_{\indx{i+alpha,j,k}}^{\indx{ijk}} \\\nonumber
&+&
 \hat{w}_{\rm J}^{\indx{i+\alpha,j+\beta,k}} \Lambda_{\indx{i-\alpha,j-\beta,k-\gamma}}^{\indx{ijk}} +
 \hat{w}_{\rm K}^{\indx{i+\alpha,j+\beta,k}} \Lambda_{\indx{i,j-\beta,k-\gamma}}^{\indx{ijk}} \\\nonumber
&+&
 \hat{w}_{\rm L}^{\indx{i+\alpha,j+\beta,k}} \Lambda_{\indx{i+\alpha,j-\beta,k-\gamma}}^{\indx{ijk}} +
 \hat{w}_{\rm M}^{\indx{i+\alpha,j+\beta,k}} \Lambda_{\indx{i-\alpha,j+\beta,k-\gamma}}^{\indx{ijk}} \\\nonumber
&+&
 \hat{w}_{\rm O}^{\indx{i+\alpha,j+\beta,k}} \Lambda_{\indx{i+\alpha,j+\beta,k-\gamma}}^{\indx{ijk}} +
 \hat{w}_{\rm R}^{\indx{i+\alpha,j+\beta,k}} \Lambda_{\indx{i,j-\beta,k}}^{\indx{ijk}} \\\label{eq:alocoeff18}
&+&
  \hat{w}_{\rm S}^{\indx{i+\alpha,j+\beta,k}} \Lambda_{\indx{i,j+\beta,k}}^{\indx{ijk}}
  \Bigr]
\eeqa
\beqa
\nonumber
\Lambda_{\indx{i-\alpha,j-\beta,k+\gamma}}^{\indx{ijk}} &=&
c_{\indx{i-\alpha,j-\beta,k+\gamma}}\tilde{w}_{\rm C}^{\indx{i-\alpha,j-\beta,k+\gamma}}
\\\label{eq:alocoeff19}
&+& d_{\indx{i-\alpha,j-\beta,k+\gamma}} 
 \hat{w}_{\rm O}^{\indx{i-\alpha,j-\beta,k+\gamma}} \Lambda_{\indx{i-\alpha,j-\beta,k}}^{\indx{ijk}}
\eeqa
\beqa
\nonumber
\Lambda_{\indx{i,j-\beta,k+\gamma}}^{\indx{ijk}} &=& c_{\indx{i,j-\beta,k+\gamma}}\tilde{w}_{\rm B}^{\indx{i,j-\beta,k+\gamma}} +
d_{\indx{i,j-\beta,k+\gamma}} \\\nonumber
&\cdot& \Bigl[
 \hat{w}_{\rm N}^{\indx{i,j-\beta,k+\gamma}} \Lambda_{\indx{i-\alpha,j-\beta,k}}^{\indx{ijk}} +
 \hat{w}_{\rm O}^{\indx{i,j-\beta,k+\gamma}} \Lambda_{\indx{i,j-\beta,k}}^{\indx{ijk}}  \\\nonumber
&+&
 \hat{w}_{\rm Q}^{\indx{i,j-\beta,k+\gamma}} \Lambda_{\indx{i-\alpha,j-\beta,k-\gamma}}^{\indx{ijk}} \\\label{eq:alocoeff20}
&+&
 \hat{w}_{\rm S}^{\indx{i,j-\beta,k+\gamma}} \Lambda_{\indx{i-\alpha,j-\beta,k+\gamma}}^{\indx{ijk}} 
  \Bigr]
\eeqa
\beqa
\nonumber
\Lambda_{\indx{i+\alpha,j-\beta,k+\gamma}}^{\indx{ijk}} &=& c_{\indx{i+\alpha,j-\beta,k+\gamma}}\tilde{w}_{\rm A}^{\indx{i+\alpha,j-\beta,k+\gamma}} +
d_{\indx{i+\alpha,j-\beta,k+\gamma}} \\\nonumber
&\cdot& \Bigl[
 \hat{w}_{\rm M}^{\indx{i+\alpha,j-\beta,k+\gamma}} \Lambda_{\indx{i-\alpha,j-\beta,k}}^{\indx{ijk}} +
 \hat{w}_{\rm N}^{\indx{i+\alpha,j-\beta,k+\gamma}} \Lambda_{\indx{i,j-\beta,k}}^{\indx{ijk}}
 \\\nonumber
&+&
\hat{w}_{\rm O}^{\indx{i+\alpha,j-\beta,k+\gamma}} \Lambda_{\indx{i+\alpha,j-\beta,k}}^{\indx{ijk}} +
\hat{w}_{\rm Q}^{\indx{i+\alpha,j-\beta,k+\gamma}} \Lambda_{\indx{i,j-\beta,k-\gamma}}^{\indx{ijk}}
\\\label{eq:alocoeff21}
&+&
 \hat{w}_{\rm S}^{\indx{i+\alpha,j-\beta,k+\gamma}} \Lambda_{\indx{i,j-\beta,k+\gamma}}^{\indx{ijk}}
  \Bigr]
\eeqa
\beqa
\nonumber
\Lambda_{\indx{i-\alpha,j,k+\gamma}}^{\indx{ijk}} &=& c_{\indx{i-\alpha,j,k+\gamma}}\tilde{w}_{\rm Q}^{\indx{i-\alpha,j,k+\gamma}} +
d_{\indx{i-\alpha,j,k+\gamma}} \\\nonumber
&\cdot& \Bigl[
 \hat{w}_{\rm C}^{\indx{i-\alpha,j,k+\gamma}} \Lambda_{\indx{i-\alpha,j-\beta,k-\gamma}}^{\indx{ijk}} +
 \hat{w}_{\rm F}^{\indx{i-\alpha,j,k+\gamma}} \Lambda_{\indx{i-\alpha,j-\beta,k}}^{\indx{ijk}} \\\label{eq:alocoeff22}
&+&
 \hat{w}_{\rm I}^{\indx{i-\alpha,j,k+\gamma}} \Lambda_{\indx{i-\alpha,j-\beta,k+\gamma}}^{\indx{ijk}} +
 \hat{w}_{\rm O}^{\indx{i-\alpha,j,k+\gamma}} \Lambda_{\indx{i-\alpha,j,k}}^{\indx{ijk}}
  \Bigr]
\eeqa
\beqa
\nonumber
\Lambda_{\indx{i,j,k+\gamma}}^{\indx{ijk}} &=& a_{\indx{i,j,k+\gamma}}w_{\rm O}^{\indx{i,j,k+\gamma}} +
d_{\indx{i,j,k+\gamma}} \\\nonumber
&\cdot& \Bigl[
 \hat{w}_{\rm B}^{\indx{i,j,k+\gamma}} \Lambda_{\indx{i-\alpha,j-\beta,k-\gamma}}^{\indx{ijk}} +
 \hat{w}_{\rm C}^{\indx{i,j,k+\gamma}} \Lambda_{\indx{i,j-\beta,k-\gamma}}^{\indx{ijk}} 
 \\\nonumber
&+&
 \hat{w}_{\rm E}^{\indx{i,j,k+\gamma}} \Lambda_{\indx{i-\alpha,j-\beta,k}}^{\indx{ijk}} +
 \hat{w}_{\rm F}^{\indx{i,j,k+\gamma}} \Lambda_{\indx{i,j-\beta,k}}^{\indx{ijk}} \\\nonumber
&+&
 \hat{w}_{\rm H}^{\indx{i,j,k+\gamma}} \Lambda_{\indx{i-\alpha,j-\beta,k+\gamma}}^{\indx{ijk}} +
 \hat{w}_{\rm I}^{\indx{i,j,k+\gamma}} \Lambda_{\indx{i,j-\beta,k+\gamma}}^{\indx{ijk}} \\\nonumber
&+&
 \hat{w}_{\rm N}^{\indx{i,j,k+\gamma}} \Lambda_{\indx{i-\alpha,j,k}}^{\indx{ijk}} +
 \hat{w}_{\rm O}^{\indx{i,j,k+\gamma}} \Lambda_{\indx{ijk}}^{\indx{ijk}} \\\label{eq:alocoeff23}
&+&
 \hat{w}_{\rm Q}^{\indx{i,j,k+\gamma}} \Lambda_{\indx{i-\alpha,j,k-\gamma}}^{\indx{ijk}} +
 \hat{w}_{\rm S}^{\indx{i,j,k+\gamma}} \Lambda_{\indx{i-\alpha,j,k+\gamma}}^{\indx{ijk}} 
  \Bigr]
\eeqa
\beqa
\nonumber
\Lambda_{\indx{i+\alpha,j,k+\gamma}}^{\indx{ijk}} &=& a_{\indx{i+\alpha,j,k+\gamma}}w_{\rm N}^{\indx{i+\alpha,j,k+\gamma}} +
d_{\indx{i+\alpha,j,k+\gamma}} \\\nonumber
&\cdot& \Bigl[
 \hat{w}_{\rm A}^{\indx{i+\alpha,j,k+\gamma}} \Lambda_{\indx{i-\alpha,j-\beta,k-\gamma}}^{\indx{ijk}} +
 \hat{w}_{\rm B}^{\indx{i+\alpha,j,k+\gamma}} \Lambda_{\indx{i,j-\beta,k-\gamma}}^{\indx{ijk}}
 \\\nonumber
&+&
 \hat{w}_{\rm C}^{\indx{i+\alpha,j,k+\gamma}} \Lambda_{\indx{i+\alpha,j-\beta,k-\gamma}}^{\indx{ijk}} +
 \hat{w}_{\rm D}^{\indx{i+\alpha,j,k+\gamma}} \Lambda_{\indx{i-\alpha,j-\beta,k}}^{\indx{ijk}}
 \\\nonumber
&+&
 \hat{w}_{\rm E}^{\indx{i+\alpha,j,k+\gamma}} \Lambda_{\indx{i,j-\beta,k}}^{\indx{ijk}} +
 \hat{w}_{\rm F}^{\indx{i+\alpha,j,k+\gamma}} \Lambda_{\indx{i+alpha,j,k}}^{\indx{ijk}}
 \\\nonumber
&+&
 \hat{w}_{\rm G}^{\indx{i+\alpha,j,k+\gamma}} \Lambda_{\indx{i-\alpha,j-\beta,k+\gamma}}^{\indx{ijk}} +
 \hat{w}_{\rm H}^{\indx{i+\alpha,j,k+\gamma}} \Lambda_{\indx{i,j-\beta,k+\gamma}}^{\indx{ijk}}
 \\\nonumber
&+&
 \hat{w}_{\rm I}^{\indx{i+\alpha,j,k+\gamma}} \Lambda_{\indx{i+\alpha,j-\beta,k+\gamma}}^{\indx{ijk}} +
 \hat{w}_{\rm M}^{\indx{i+\alpha,j,k+\gamma}} \Lambda_{\indx{i-\alpha,j,k}}^{\indx{ijk}}
 \\\nonumber
&+&
 \hat{w}_{\rm N}^{\indx{i+\alpha,j,k+\gamma}} \Lambda_{\indx{ijk}}^{\indx{ijk}} +
 \hat{w}_{\rm O}^{\indx{i+\alpha,j,k+\gamma}} \Lambda_{\indx{i+alpha,j,k}}^{\indx{ijk}} \\\label{eq:alocoeff24}
&+&
 \hat{w}_{\rm Q}^{\indx{i+\alpha,j,k+\gamma}} \Lambda_{\indx{i,j,k-\gamma}}^{\indx{ijk}} +
 \hat{w}_{\rm S}^{\indx{i+\alpha,j,k+\gamma}} \Lambda_{\indx{i,j,k+\gamma}}^{\indx{ijk}} 
  \Bigr]
\eeqa
\beqa
\nonumber
\Lambda_{\indx{i-\alpha,j+\beta,k+\gamma}}^{\indx{ijk}} &=& c_{\indx{i-\alpha,j+\beta,k+\gamma}}\tilde{w}_{\rm P}^{\indx{i-\alpha,j+\beta,k+\gamma}} +
d_{\indx{i-\alpha,j+\beta,k+\gamma}} \\\nonumber
&\cdot& \Bigl[
 \hat{w}_{\rm C}^{\indx{i-\alpha,j+\beta,k+\gamma}} \Lambda_{\indx{i-\alpha,j,k-\gamma}}^{\indx{ijk}} +
 \hat{w}_{\rm F}^{\indx{i-\alpha,j+\beta,k+\gamma}} \Lambda_{\indx{i-\alpha,j,k}}^{\indx{ijk}}
 \\\nonumber
&+&
 \hat{w}_{\rm I}^{\indx{i-\alpha,j+\beta,k+\gamma}} \Lambda_{\indx{i-\alpha,j,k+\gamma}}^{\indx{ijk}} +
 \hat{w}_{\rm L}^{\indx{i-\alpha,j+\beta,k+\gamma}} \Lambda_{\indx{i-\alpha,j-\beta,k}}^{\indx{ijk}} \\\label{eq:alocoeff25}
&+&
 \hat{w}_{\rm O}^{\indx{i-\alpha,j+\beta,k+\gamma}} \Lambda_{\indx{i-\alpha,j+\beta,k}}^{\indx{ijk}}
  \Bigr]
\eeqa
\beqa
\nonumber
\Lambda_{\indx{i,j+\beta,k+\gamma}}^{\indx{ijk}} &=& a_{\indx{i,j+\beta,k+\gamma}}w_{\rm F}^{\indx{i,j+\beta,k+\gamma}} +
d_{\indx{i,j+\beta,k+\gamma}} \\\nonumber
&\cdot& \Bigl[
 \hat{w}_{\rm B}^{\indx{i,j+\beta,k+\gamma}} \Lambda_{\indx{i-\alpha,j,k-\gamma}}^{\indx{ijk}} +
 \hat{w}_{\rm C}^{\indx{i,j+\beta,k+\gamma}} \Lambda_{\indx{i,j,k-\gamma}}^{\indx{ijk}} \\\nonumber
&+&
 \hat{w}_{\rm E}^{\indx{i,j+\beta,k+\gamma}} \Lambda_{\indx{i-\alpha,j,k}}^{\indx{ijk}} +
 \hat{w}_{\rm F}^{\indx{i,j+\beta,k+\gamma}} \Lambda_{\indx{ijk}}^{\indx{ijk}} \\\nonumber
&+&
 \hat{w}_{\rm H}^{\indx{i,j+\beta,k+\gamma}} \Lambda_{\indx{i-\alpha,j,k+\gamma}}^{\indx{ijk}} +
 \hat{w}_{\rm I}^{\indx{i,j+\beta,k+\gamma}} \Lambda_{\indx{i,j,k+\gamma}}^{\indx{ijk}} \\\nonumber
&+&
 \hat{w}_{\rm K}^{\indx{i,j+\beta,k+\gamma}} \Lambda_{\indx{i-\alpha,j-\beta,k}}^{\indx{ijk}} +
 \hat{w}_{\rm L}^{\indx{i,j+\beta,k+\gamma}} \Lambda_{\indx{i,j-\beta,k}}^{\indx{ijk}} \\\nonumber
&+&
 \hat{w}_{\rm N}^{\indx{i,j+\beta,k+\gamma}} \Lambda_{\indx{i-\alpha,j+\beta,k}}^{\indx{ijk}} +
 \hat{w}_{\rm O}^{\indx{i,j+\beta,k+\gamma}} \Lambda_{\indx{i,j+\beta,k}}^{\indx{ijk}} \\\nonumber
&+&
 \hat{w}_{\rm P}^{\indx{i,j+\beta,k+\gamma}} \Lambda_{\indx{i-\alpha,j-\beta,k-\gamma}}^{\indx{ijk}} +
 \hat{w}_{\rm Q}^{\indx{i,j+\beta,k+\gamma}} \Lambda_{\indx{i-\alpha,j+\beta,k-\gamma}}^{\indx{ijk}} \\\nonumber
&+&
 \hat{w}_{\rm R}^{\indx{i,j+\beta,k+\gamma}} \Lambda_{\indx{i-\alpha,j-\beta,k+\gamma}}^{\indx{ijk}} \\\label{eq:alocoeff26}
&+&
 \hat{w}_{\rm S}^{\indx{i,j+\beta,k+\gamma}} \Lambda_{\indx{i-\alpha,j+\beta,k-\gamma}}^{\indx{ijk}}
  \Bigr]
\eeqa
\beqa
\nonumber
\Lambda_{\indx{i+\alpha,j+\beta,k+\gamma}}^{\indx{ijk}} &=& a_{\indx{i+\alpha,j+\beta,k+\gamma}}w_{\rm E}^{\indx{i+\alpha,j+\beta,k+\gamma}} +
d_{\indx{i+\alpha,j+\beta,k+\gamma}} \\\nonumber
&\cdot& \Bigl[
 \hat{w}_{\rm A}^{\indx{i+\alpha,j+\beta,k+\gamma}} \Lambda_{\indx{i-\alpha,j,k-\gamma}}^{\indx{ijk}} +
 \hat{w}_{\rm B}^{\indx{i+\alpha,j+\beta,k+\gamma}} \Lambda_{\indx{i,j,k-\gamma}}^{\indx{ijk}}
 \\\nonumber
&+& 
 \hat{w}_{\rm C}^{\indx{i+\alpha,j+\beta,k+\gamma}} \Lambda_{\indx{i+\alpha,j,k-\gamma}}^{\indx{ijk}} +
 \hat{w}_{\rm D}^{\indx{i+\alpha,j+\beta,k+\gamma}} \Lambda_{\indx{i-\alpha,j,k}}^{\indx{ijk}}
 \\\nonumber
&+&
 \hat{w}_{\rm E}^{\indx{i+\alpha,j+\beta,k+\gamma}} \Lambda_{\indx{ijk}}^{\indx{ijk}} +
 \hat{w}_{\rm F}^{\indx{i+\alpha,j+\beta,k+\gamma}} \Lambda_{\indx{i+alpha,j,k}}^{\indx{ijk}}
 \\\nonumber
&+&
 \hat{w}_{\rm G}^{\indx{i+\alpha,j+\beta,k+\gamma}} \Lambda_{\indx{i-\alpha,j,k+\gamma}}^{\indx{ijk}} +
 \hat{w}_{\rm H}^{\indx{i+\alpha,j+\beta,k+\gamma}} \Lambda_{\indx{i,j,k+\gamma}}^{\indx{ijk}}
 \\\nonumber
&+&
 \hat{w}_{\rm I}^{\indx{i+\alpha,j+\beta,k+\gamma}} \Lambda_{\indx{i+\alpha,j,k+\gamma}}^{\indx{ijk}} +
 \hat{w}_{\rm J}^{\indx{i+\alpha,j+\beta,k+\gamma}} \Lambda_{\indx{i-\alpha,j-\beta,k}}^{\indx{ijk}} \\\nonumber
&+&
 \hat{w}_{\rm K}^{\indx{i+\alpha,j+\beta,k+\gamma}} \Lambda_{\indx{i,j-\beta,k}}^{\indx{ijk}} +
 \hat{w}_{\rm L}^{\indx{i+\alpha,j+\beta,k+\gamma}} \Lambda_{\indx{i+\alpha,j-\beta,k}}^{\indx{ijk}} \\\nonumber
&+&
 \hat{w}_{\rm M}^{\indx{i+\alpha,j+\beta,k+\gamma}} \Lambda_{\indx{i-\alpha,j+\beta,k}}^{\indx{ijk}} +
 \hat{w}_{\rm N}^{\indx{i+\alpha,j+\beta,k+\gamma}} \Lambda_{\indx{i,j+\beta,k}}^{\indx{ijk}}
 \\\nonumber
&+&
 \hat{w}_{\rm O}^{\indx{i+\alpha,j+\beta,k+\gamma}} \Lambda_{\indx{i+\alpha,j+\beta,k}}^{\indx{ijk}} +
 \hat{w}_{\rm P}^{\indx{i+\alpha,j+\beta,k+\gamma}} \Lambda_{\indx{i,j-\beta,k-\gamma}}^{\indx{ijk}} \\\nonumber
&+&
 \hat{w}_{\rm Q}^{\indx{i+\alpha,j+\beta,k+\gamma}} \Lambda_{\indx{i,j+\beta,k-\gamma}}^{\indx{ijk}} +
 \hat{w}_{\rm R}^{\indx{i+\alpha,j+\beta,k+\gamma}} \Lambda_{\indx{i,j-\beta,k+\gamma}}^{\indx{ijk}} \\\label{eq:alocoeff27}
&+&
 \hat{w}_{\rm S}^{\indx{i+\alpha,j+\beta,k+\gamma}} \Lambda_{\indx{i,j+\beta,k+\gamma}}^{\indx{ijk}}
  \Bigr]
\eeqa
Since the integration and interpolation coefficients need to be calculated
only once at each considered grid point (here denoted by $(\indx{u,v,w})$, to
avoid confusion), we obtain the $\matown{\Lambda}$-matrix coefficients by
substituting indices. For Eq.~\eqref{eq:alocoeff01}, we find:
\beq
\Lambda_{\indx{i-\alpha,j-\beta,k-\gamma}}^{\indx{ijk}} =
\Lambda_{\indx{uvw}}^{\indx{u+\alpha,v+\beta,w+\gamma}} = c_{\indx{uvw}}\tilde{w}_{\rm
  I}^{\indx{uvw}} \,,
\eeq
and proceed analogously for all other elements in
Eqs.~\eqref{eq:alocoeff02}-\eqref{eq:alocoeff27}. Thus, the ALO can be
calculated in parallel to the formal solution scheme.
\end{document}